\documentclass[aps,prl,twocolumn,amsmath,amssymb,nofootinbib,superscriptaddress,floatfix,reprint,longbibliography]{revtex4-1}
\usepackage{amsmath}
\usepackage{amsfonts}
\usepackage{amsthm}
\usepackage{amssymb}
\usepackage{lipsum}
\usepackage{graphicx}
\usepackage{nicefrac}
\usepackage[usenames,dvipsnames]{color}
\usepackage{float}
\usepackage{multirow}
\usepackage{soul}
\usepackage{appendix}
\usepackage[colorlinks=true, urlcolor=blue, linkcolor=blue, citecolor=blue, pdftex]{hyperref}
\usepackage{dcolumn}
\usepackage{bm}
\usepackage{url}
\usepackage{braket}
\usepackage[range-units=single,separate-uncertainty]{siunitx}
\usepackage{verbatim} 
\usepackage[version=4]{mhchem}
\usepackage{xcolor}
\usepackage{comment}
\usepackage[normalem]{ulem}

\newcommand{\cre}{{\dag}}
\newcommand{\ann}{{\vphantom{\dag}}}
\newcommand{\added}[1]{{#1}}

\newcommand{\supplement}[1]{%
  \clearpage%
  \title{#1}%
  \maketitle%
  \setcounter{equation}{0}%
  \setcounter{figure}{0}%
  \setcounter{table}{0}%
  \setcounter{page}{1}%
  \makeatletter%
  \renewcommand{\thesection}{S\arabic{section}}%
  \renewcommand{\thesubsection}{\Alph{subsection}}%
  \renewcommand{\theequation}{S\arabic{equation}}%
  \renewcommand{\thefigure}{S\arabic{figure}}%
  \renewcommand{\thetable}{S\Roman{table}}%
  \renewcommand{\thepage}{S\arabic{page}}%
  \makeatother%
}
\makeatletter
\def\maketitle{
\@author@finish
\title@column\titleblock@produce
\suppressfloats[t]}
\makeatother

\begin{document}

\author{Anja Wenger}\affiliation{Institut f\"{u}r Theoretische Physik und Astrophysik and W\"{u}rzburg-Dresden Cluster of Excellence ct.qmat, Universit\"{a}t W\"{u}rzburg, 97074 W\"{u}rzburg, Germany}
\author{Armando Consiglio}
\affiliation{Istituto Officina dei Materiali, Consiglio Nazionale delle Ricerche, Trieste I-34149, Italy}
\author{Hendrik Hohmann}\affiliation{Institut f\"{u}r Theoretische Physik und Astrophysik and W\"{u}rzburg-Dresden Cluster of Excellence ct.qmat, Universit\"{a}t W\"{u}rzburg, 97074 W\"{u}rzburg, Germany}
\author{Matteo D\"urrnagel}\email{matteo.duerrnagel@uni-wuerzburg.de}\affiliation{Institut f\"{u}r Theoretische Physik und Astrophysik and W\"{u}rzburg-Dresden Cluster of Excellence ct.qmat, Universit\"{a}t W\"{u}rzburg, 97074 W\"{u}rzburg, Germany}\affiliation{Institute for Theoretical Physics, ETH Z\"{u}rich, 8093 Z\"{u}rich, Switzerland}
\author{Fabian O. von Rohr}
\affiliation{Department of Quantum Matter Physics, University of Geneva, CH-1211 Geneva, Switzerland}
\author{Harley D. Scammell}
\affiliation{School of Mathematical and Physical Sciences, University of Technology Sydney, Ultimo, NSW 2007, Australia}
\author{Julian Ingham}
\affiliation{Department of Physics, Columbia University, New York, NY, 10027, USA}
\author{Domenico Di Sante}
\affiliation{Department of Physics and Astronomy, University of Bologna, 40127 Bologna, Italy}
\author{Ronny Thomale}\email{ronny.thomale@uni-wuerzburg.de}\affiliation{Institut f\"{u}r Theoretische Physik und Astrophysik and W\"{u}rzburg-Dresden Cluster of Excellence ct.qmat, Universit\"{a}t W\"{u}rzburg, 97074 W\"{u}rzburg, Germany}

\date{\today}

\title{Theory of unconventional magnetism in a Cu-based kagome metal}

\begin{abstract}
Kagome metals have established a new arena for correlated electron physics. To date, the predominant experimental evidence centers around unconventional charge order, nematicity, and superconductivity, while magnetic fluctuations due to electronic interactions, i.e., beyond local atomic magnetism, have largely been elusive.
We find the challenge of locating the appropriate parameter regime for such exotic order to center around two aspects. First, the correlations implied by low-energy orbitals have to be sufficiently large to yield a dominance of magnetic fluctuations and weak to retain an itinerant parent state. Second, the kinematic kagome profile at the Fermi level demands an efficient mitigation of sublattice interference causing the suppression of magnetic fluctuations descending from electronic on-site repulsion.
We elucidate our methodology by analyzing the potential copper-based kagome compound CsCu$_3$Cl$_5$: \added{From \textit{ab initio} design and many-body analysis, we develop a model framework of realistic Cu-based kagome materials the simulations of which reveal unconventional magnetic order in a kagome metal.} 

\end{abstract}
\maketitle

\paragraph{Introduction}
The kagome lattice, characterized by its repeating pattern of corner-sharing triangles, forms a hexagonal network with three distinct sublattices. This unique geometry gives rise to exotic quantum phenomena, rendering it an exclusive host for correlated and topologically nontrivial electronic states.
Depending on the balance between electronic interactions and kinetic energy, electronic models on the kagome lattice can yield myriad distinct physical phases. 
Kagome compounds with a metallic parent state generically feature intricate non-magnetic phases such as exotic charge orders, nematicity and superconductivity~\cite{Jiang2021, Mielke2022, li2023electronic, jiang2023flat, nag2024pomeranchuk, jiang2024van}.
Most extensively discussed is the $A$V$_3$Sb$_5$ family ($A$=K,Rb,Cs), in which the lack of magnetic order is best explained by both electron-phonon coupling effects~\cite{Zhong2023, Xie2022} and 
the suppression of on-site Coulomb repulsion due to the sublattice interference (SI) mechanism, preventing local scattering channels between the van Hove singularities (vHS) 
~\cite{Kiesel2012, Hu2022, Kang2022}.
This leads to a rich zoology of unconventional phases like charge bond (CBO) and loop current orders (LCO), which are largely not found in alternative correlated electron material domains such as cuprate or iron-pnictide compounds~\cite{Kiesel2013, Wang2013, Zhan2024}. Due to the predominance of non-magnetic phases, however, magnetic instabilities that might be unique to kagome kinematics have remained largely unexplored at the microscopic and even at the phenomenological level.

The emergence of magnetic instabilities requires a sufficient degree of electronic correlations.
For strong electron-electron interactions, the electrons become Mott-localized on atomic sites, where unpaired spins create magnetic moments. 
The inherent geometric frustration of the kagome lattice leads to unconventional magnetic orders such as spin liquid phases.
The most prominent example is Herbertsmithite (ZnCu$_3$(OH)$_6$Cl$_2$)~\cite{Bert2009, Norman2016}, which lies deep in the Mott-insulating regime, rendering the unique features of the kagome band structure like vHS and Fermi surface nesting -- which drive unconventional charge order in the itinerant case -- irrelevant.

In this \added{article}, we aim to fuse the intricacies of electronic kagome kinematics with the potential emergence of magnetic order.
\added{We identify intermediately correlated kagome compounds close to the mixed $m$-type van Hove filling as a suitable system to stabilize  exotic magnetic orders, that inherit substantial non-local spin expectation values from the partial sublattice polarization of the Fermi surface. 
%
From detailed electronic structure calculations we find that Cu-based 135-kagome compounds, i.e. \ce{CsCu3Cl5}, provide a playground for exploring the potential realization of unconventional magnetic order in a realistic material setting.
The low-energy physics of \ce{CsCu3Cl5} is well described by an isolated set of three bands with an $m$-type vHS close to the Fermi level and exhibits}
%
correlations intermediate between Herbertsmithite and the weakly correlated $A$V$_3$Sb$_5$ compounds. While Cu-O complexes are likely to be located in the Mott-localized regime, we propose Cu-Cl complexes strike the aspired compromise between maximum interaction strength while preserving metallicity.
\added{Our analysis of the magnetic phase diagram unveils \ce{CsCu3Cl5} as a realistic material platform to host correlation-driven spin-bond order and highlights the broader implications of Fermi surface topology for exotic magnetic phases in kagome systems.}

\begin{figure}[tb]
    \centering
    \includegraphics[width=\linewidth]{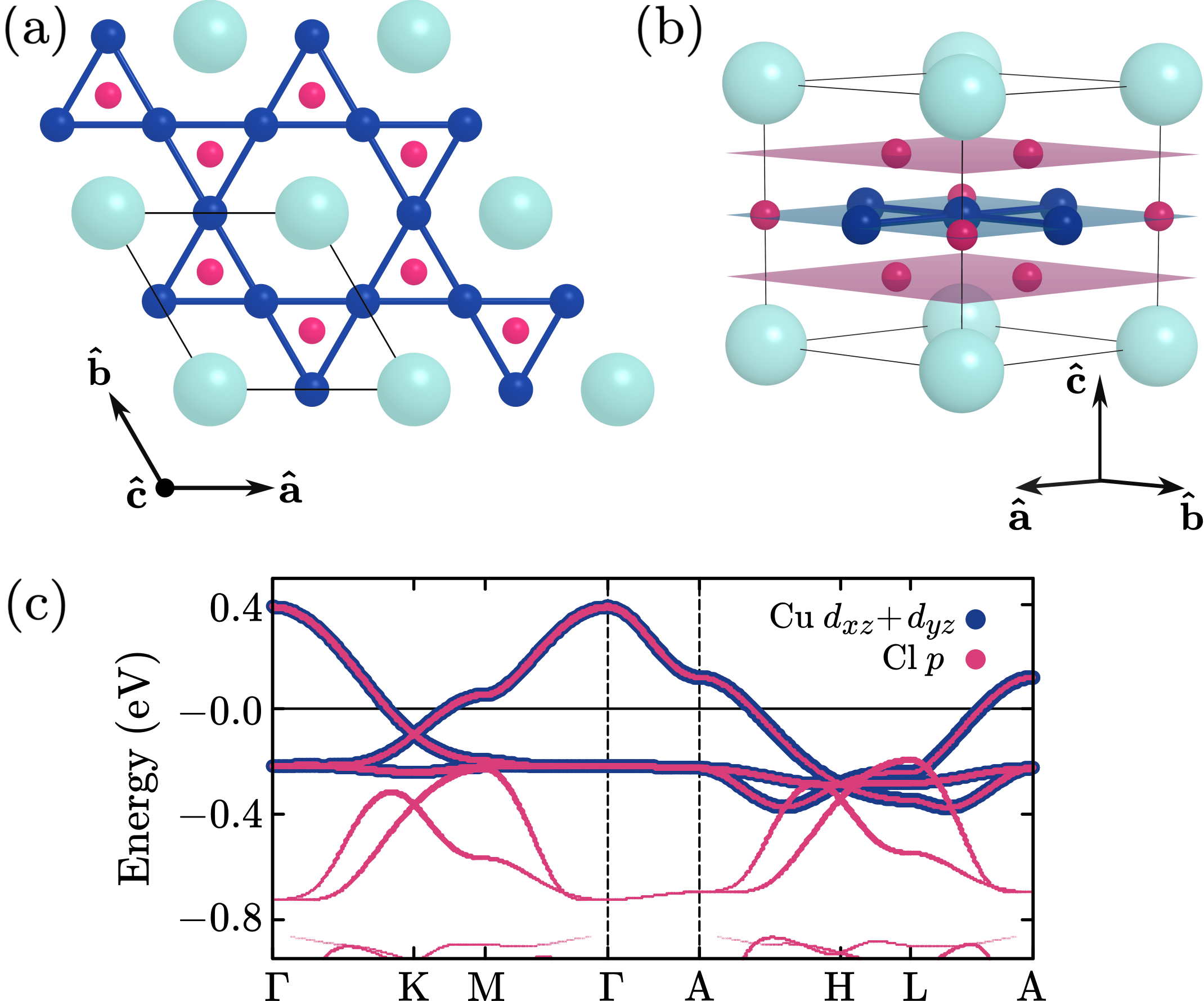}
    \caption{
    Crystal structure of pristine \ce{CsCu3Cl5}.
    (a) Top view showing the Cu kagome lattice highlighted by blue bonds. Turquoise, blue and pink spheres represent Cs, Cu and Cl atoms. The unit cell is delimited by black lines.
    (b) Side view of the unit cell. The blue plane contains the Cu kagome lattice and additional Cl atoms, while the pink planes above and below contain a hexagonal lattice of Cl atoms.
    (c) Electronic band structure with projected orbital weights featuring distinct kagome bands and an $m$-type vHS close to $E_F$.
    }
    \label{fig:pristine}
    \vspace{-0.3cm}
\end{figure}
\paragraph{Model Realization}
%
Our theoretical platform \ce{CsCu3Cl5} is isostructural with other members of the 135 family, such as \ce{$A$V3Sb5}, and crystallizes in the hexagonal space group $P6/mmm$ (No. 191). It features an in-plane kagome network of copper atoms, coordinated octahedrally by chlorine atoms, forming layered \ce{Cu3Cl5} sheets that are separated along the $\hat{c}$-axis by a triangular net of cesium ions. \ce{CsCu3Cl5} exhibits a mixed valence state, where copper ions are present in both Cu(I) and Cu(II) oxidation states, with an average oxidation state of +1.33.

While Cu generally prefers the +II oxidation state, as observed in the chemically related phases \ce{Cs2CuCl4} \cite{mcginnety1972cesium} and \ce{CsCuCl3} \cite{schlueter1966redetermination}, the presence of Cu(I) in compounds such as \ce{Cs3Cu2Cl5} \cite{hull2004crystal} supports the feasibility of this mixed valence state. A notable point is the possibility of a mixed charge on a crystallographic site. Whether this is statistically distributed in the structural model or whether some form of ordering is present remains an open question. The ionic radii of Cu(I) and Cu(II) in octahedral coordination do not differ significantly, making this mixed occupation proposed here certainly plausible \cite{shannon1976revised}.
Due to its full $d^{10}$ electron configuration, Cu(I) lacks the electronic degeneracy needed for Jahn-Teller distortions, making it less prone to such distortions compared to Cu(II). In contrast, Cu(II), with its $d^{9}$ electron configuration, is a well-known Jahn-Teller ion and typically undergoes bond elongations or compressions in its octahedral coordination environment to lower its energy \cite{veidis1969jahn}. It is also worth noting that Cu(I), at least in chloride environments, does not always form regular coordination polyhedra. This irregularity might stem from the nature of Cs-Cl compounds, where the large size of Cs ions can sometimes influence the coordination environments of surrounding atoms, though such effects are less likely in layered structures like those in \ce{CsCu3Cl5}.
\added{A thorough analysis of the compound's stability reveals the pristine configuration as a metastable state. However, the analysis of single particle and derived many-body effects in the energetically favored twisted configuration matches qualitatively the results for the pristine case. To maintain generality, we explicate in the following a study of the pristine structure and refer the interested reader to the SM for the analogous treatment of the twisted configuration.}
%
%
%

Despite being isostructural to other members of the 135 family like \ce{$A$V3Sb5}, the calculated electronic band structure, shown in Fig.~\ref{fig:pristine}\textcolor{blue}{(c)}, reveals striking differences to previous reports on known kagome compounds.
The characteristic kagome band manifold -- including a flat band, a Dirac point located at the K-point, and two vHSs at each of the three inequivalent M-points -- appears well separated from other bands at low energies.
The orbital projection of the bandstructure illustrates that the states at $E_F$ emerge only from the hybridization of the out-of-plane Cu $d_{xz}$ and $d_{yz}$ orbitals with Cl $p$ orbitals.
This can be attributed to a collaborative effect:
The enhanced valence of Cu ($d^9$) suggests a single kagome band manifold close to the Fermi level in analogy to \ce{$A$Ti3Bi5} ($A$ = Cs, Rb)~\cite{Werhahn2022, Yang2023, Bigi2024p}, where the $d^1$ configuration of Ti constitutes the analogue of Cu.
In \ce{CsCu3Cl5}, however, the strong polar Cu-Cl bonding results in an enhanced crystal field splitting, that separates partially filled valence bonds from fully filled bonding and empty anti-bonding states. 

This is directly reflected in the real space arrangement of chlorine atoms around each copper atom, forming a distorted octahedron, elongated by 0.21\;\r{A} towards the center of the hexagonal plaquette (see the SM). 
Within the octahedra, the spatial proximity of the Cu \(d\)-orbitals and Cl \(p\)-orbitals induces a pronounced splitting in the $d$-orbital energy levels. 
Both effects produce a uniquely isolated and undistorted kagome bandstructure close to the Fermi level. 
In addition, \ce{CsCu3Cl5} is distinct from related 135 compounds by its inverted band ordering, with the flat band at the bottom (Fig.~\ref{fig:pristine}\textcolor{blue}{(c)}). This sets  an $m$-type vHS in the vicinity of the Fermi level, opposed to the widely studied $p$-type variant, with important consequences upon the inclusion of interaction effects.

\paragraph{Electronic correlations across kagome compounds}
The members of the 135 family are generally considered weakly correlated, where phonons play a crucial role in charge ordering mechanism in collaboration or competition with electronic interactions~\cite{Ortiz2021, He2024}.
\\
\added{
To estimate interaction parameters from first principles, we perform constrained random phase approximation (cRPA) calculations \cite{PhysRevB.70.195104, PhysRevB.80.155134, PhysRevB.86.165105, DiSante2023} for pristine \ce{CsCu3Cl5} (see SM for details and results for the twisted configuration), 
using as the target space a set of Wannier functions derived from the Wannier model described later in this paper.
The resulting screened on-site effective Coulomb interaction is $U = 3.6\,$eV (bare on-site Coulomb interaction $U_{bare} = 14.7\,$eV). 
With the quite narrow bandwith $t$, this yields a ratio $U/t \approx 4.7$, placing \ce{CsCu3Cl5} above the weakly correlated $A$\ce{V3Sb5} compounds with $U/t < 1$ \cite{DiSante2023} and well below the strongly correlated Herbertmithite in the region of $U_{bare}/t \approx 20$~\cite{Mazin2014}.
The qualitative findings are consistent with a bond length analysis~\cite{Schoop2022}, that estimates the relative degree of correlation across various kagome compounds by the transition metal to ligand distance as explicated in the SM.}
This classification aligns well with the picture, that a larger filling fraction of the 3$d$ shell in copper compared to Ti ($d^1$)~\cite{Yang2023}, V ($d^3$)~\cite{OrtizWilson} or Cr ($d^4$)~\cite{Liu2024a} as the central block of the kagome network can enhance local correlations.

\added{A notable exception to this rule is presented by \ce{CsCr3Sb5}, that displays an admixture of charge and magnetically ordered phases at low temperatures~\cite{Liu2024a, Sangiovanni2024}:
Opposed to other members of the 135 family, this compound features a flat band in vicinity of the Fermi level~\cite{Li2025e}. The large spectral weight close to the Fermi level significantly enhances magnetic fluctuations and drives the system to strong coupling irrespective of the $U$ over bandwidth ratio~\cite{Wu2025f, Xie2025e, Wang2025h}.
Consequently, not only the bare interaction strength inside the correlated manifold plays a crucial role in determining the degree of correlations in a material. Likewise, the available low-lying states for electronic scattering processes play a decisive role beyond crystal field effects. In particular, pronounced Fermi surface nesting can drive the system to stronger coupling as canonically encountered close to vH filling.}
\added{In kagome compounds, the non-trivial quantum geometry of the electronic eigenstates around the vHS points adds an additional layer of inference to the estimation of correlation strengths:
In V-based 135 kagome compounds, the low lying vHS is of pure $p$-type, \textit{i.e.} each vHS point is exclusively supported by electronic states on one of the three kagome sublattice sites~\cite{Hu2022, Kang2022}.
This reduces the effectiveness of local Coulomb interactions in driving Fermi surface instability via the celebrated sublattice interference (SI) mechanism~\cite{Kiesel2012, Kiesel2013} with two important corollaries for the \ce{$A$V3Sb5} family:
Firstly, the enhanced relevance of non-local Coulomb interactions leads to the absence of phases involving local particle-hole pairs at intermediate coupling~\cite{Neupert2021, Profe2024}.
Secondly, phonon coupling arising from a delocalized charge density across the sublattices gains significant importance, making it difficult to discern whether the observed instabilities are driven by phonons or electronic correlations \cite{wang2023quantum, Enzner2025p}.}

\paragraph{Van Hove scattering at m-type filling}

\begin{figure}
    \centering
    \includegraphics[width=\linewidth]{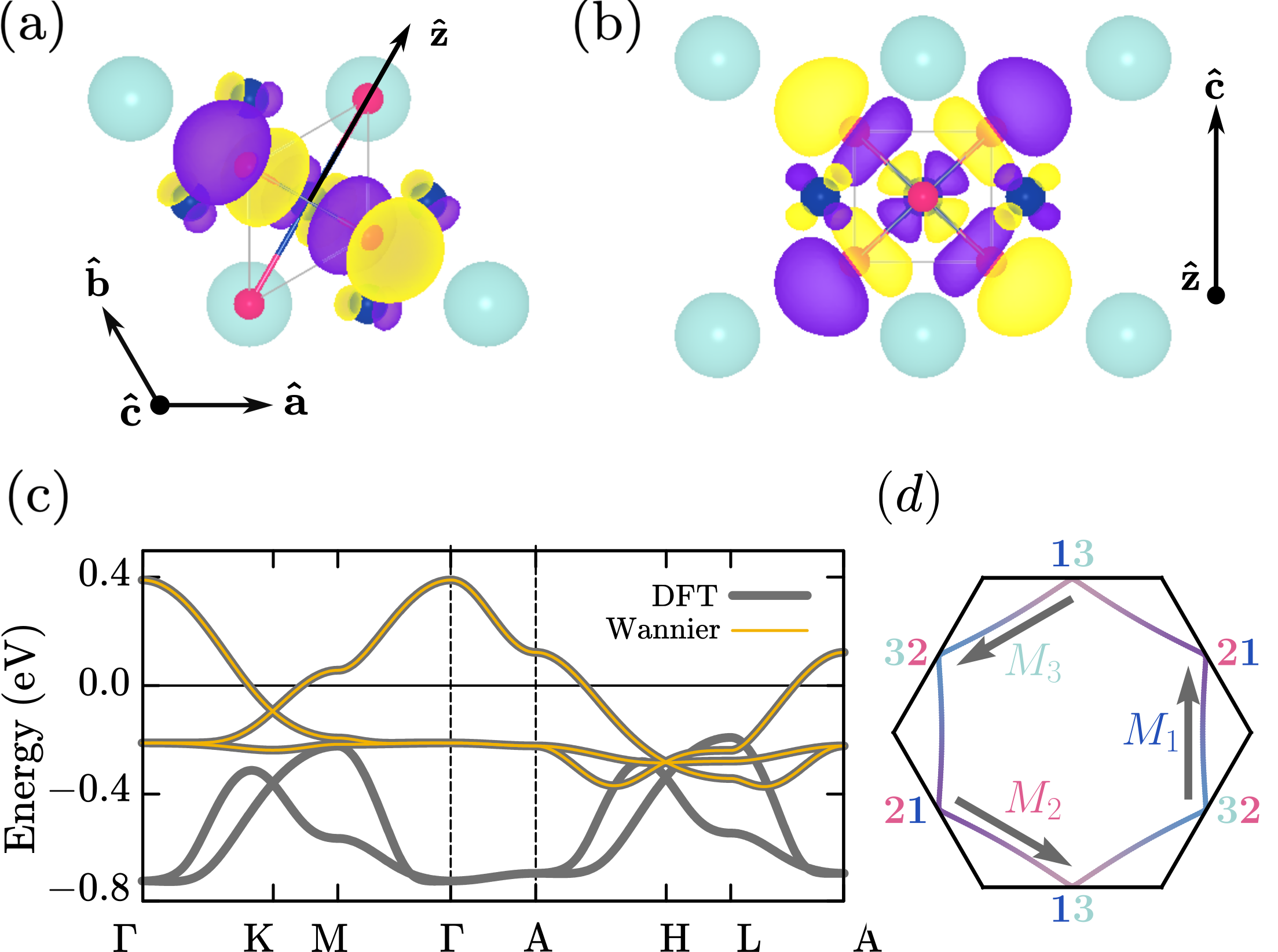}
    \caption{Wannier orbital centered at the Cu site seen from (a) top and (b) side view of the pristine structure. The site-local reference axis $\mathbf{\hat{z}}$ is pointing towards the center of the hexagonal plaquettes. Purple/yellow orbital lobes indicate the phase of the MLWF.
    (c) Comparison of the DFT band dispersion (gray) with bands obtained from the Wannier three-orbital tight-binding model (yellow).
    \added{(d) Fermi surface in the $k_z = 0$ plane for the non-interacting bandstructure at the upper vH filling and dominant nesting vectors. The colors indicate the eigenstate contributions on the three different sublattices. At the three $M$ points, \textit{i.e.} the vH points, the state is equally composed of two sublattices given by the sublattice labels.}}
    \label{fig:wannierorbitals}
    \vspace{-0.3cm}
\end{figure}

\added{
In \ce{CsCu3Cl5}, the much less studied $m$-type vHS promotes a qualitatively different behavior. The sublattice occupation at the upper van Hove filling of \ce{CsCu3Cl5} is indicated in Fig.~\ref{fig:wannierorbitals}\textcolor{blue}{(d)}. At the three inequivalent van Hove points $M_\gamma$, the electronic states are equally distributed across two of the three sublattices with vanishing contribution on the third one. This configuration is known as a mixed $m$-type van Hove singularity~\cite{Kiesel2012}.
Under scattering with $\mathbf{q}$ = M, the electronic eigenstates at a vHS point transition to states comprised of a different combination of two sublattices. This ensures that one sublattice is occupied both before and after scattering and allows on-site interactions to mediate vHS nesting processes.
While this weakens the effect of SI compared to the $p$-type case, on-site scattering is still partially restricted, as the local Hubbard interaction $U$ acts only on one of the three sublattices at each inequivalent M-point.
In the same way the nearest-neighbor (NN) interaction $V$ can mediate scattering between one defined pair of sublattices.
Hence, scattering events with the nesting vector $M_\gamma$ with $M_\alpha = M_\beta + M_\gamma$ between the vH points can be decomposed into a site local and non-local component
\begin{equation}
\begin{split}
    \bra{u(M_\alpha)} & \bra{u(M_\alpha)} \Gamma_{o_1 o_2 o_3 o_4}(M_\gamma) \ket{u(M_\beta)} \ket{u(M_\beta)} \\
    & = \Gamma_{\gamma \gamma \gamma \gamma}(M_\gamma) + \Gamma_{\alpha \alpha \beta \beta}(M_\gamma) \ .
\end{split}
\end{equation}
Here, $u(M_\alpha)$ is the eigenstate of the non-interacting Hamiltonian at the vH point $M_\alpha$ and $\Gamma$ the two-particle interaction.
Thereby, both interaction scales, $U$ and $V$, operate at similar scales due to the effect of the partial sublattice polarization of the vH points.
This leads to a regime where both local and non-local interactions contribute significantly, and $U$ and $V$ are of comparable importance. Moreover, the $m$-type vHS corresponds to a more localized electronic structure in momentum space ~\cite{Kiesel2012, Kiesel2013, Wang2013}. As a result, electrons couple less efficiently to phonons, suppressing phonon-driven instabilities and making electronic correlations more effective.

The correlation strength within the low energy manifold of CsCu$_3$Cl$_5$ is therefore anticipated to reside in a sweet spot between that of the weakly correlated CsV$_3$Sb$_5$, and the strongly correlated Herbertsmithite.}
%

\paragraph{Wannier model}
The isolated nature of the three characteristic kagome bands allows to capture the electronic structure of CsCu$_3$Cl$_5$ around $E_F$ with a single orbital per kagome site.
To construct a tight-binding model with maximally localized Wannier functions (MLWF), we begin with an initial approximation using a linear combination of \(d_{xz}\) and \(d_{yz}\) orbitals.
After the spread minimization, three MLWFs are obtained, each centered on a distinct site of the kagome lattice and oriented within the central plane of the coordinating octahedra. 
These MLWFs share identical shapes, originating from a linear combination of the \(d_{x^2-y^2}\) orbital (in the local reference frame of this site) and Cl \(p\)-orbitals. 
Details on the orientations of the local reference frames are provided in the SM.
An example MLWF is illustrated in Fig.~\ref{fig:wannierorbitals}\textcolor{blue}{(a),}\textcolor{blue}{(b)} with the orientation of the corresponding octahedron outlined in gray.
$\mathbf{\hat{a}},\mathbf{\hat{b}},\mathbf{\hat{c}}$ span the (global) reference frame of the crystal, while $\mathbf{\hat{z}}$ aligns with the octahedron axis.
The three MLWFs map onto each other under 60$^\circ$ rotation.
These orbitals reflect the $B_{2g}$ elementary band representation of $P6/mmm$, well separated from $B_{3g}$ by the pronounced Jahn-Teller distortion described above~\cite{wu2021nature}.

By extracting bands from this three-orbital tight-binding model and comparing them with DFT band calculations in Fig.~\ref{fig:wannierorbitals}\textcolor{blue}{(c)}, we observe a perfect alignment of the bands, combined with a very small spread.
This allows for a description using MLWF, making the three-orbital description a simple and well-suited model for many-body calculations.
To study possible instabilities of \ce{CsCu3Cl5}, we equip the non-interacting theory with two particle interactions. Following the discussion of the preceeding section and preceeding works on the kagome lattice~\cite{Kiesel2013, Wang2013, Profe2024}, we choose bare interactions consisting of on-site $U$ and NN repulsion $V$,
\begin{equation}
    \hat{H}_I = U \sum_{\mathbf i}{\hat{n}_{\mathbf i\uparrow}\hat{n}_{\mathbf i\downarrow}} 
     + ~ V \sum_{\langle \mathbf i, \mathbf j \rangle, \sigma\sigma'}{\hat{n}_{\mathbf j\sigma}\hat{n}_{\mathbf i\sigma'}} \, ,
\label{eq:interaction}
\end{equation}
where $\hat{n}_{\mathbf j\sigma} = \hat{c}^{\dagger}_{\mathbf j\sigma}\hat{c}_{\mathbf j\sigma}$  is the fermionic number operator on site $\mathbf j$ with spin $\sigma$.
\added{
As discussed above, the reduced SI at the $m$-type vHS - compared to the extensively investigated $p$-type scenario - allows local and long-range interactions to contribute on a more equal footing. 
}
Hence, $V$ cannot be simply neglected as for e.g. done in Ref.~\cite{Yang2024}.
\paragraph{Many-body analysis of magnetic instabilities} 
The small number of bands crossing the Fermi level and the absence of local degrees of freedom in the Wannierized model make it well-suited for exploring the system's ordering tendencies using numerical many-body methods.
We perform functional renormalization group (FRG) calculations, utilizing the truncated unity implementation provided by the divERGe code base~\cite{Profe2024a}.
The FRG provides a well-defined interpolation from the non-interacting model to a low-energy effective theory near the Fermi level by successively integrating out high-energy degrees of freedom~\cite{Metzner2012, Platt_2013}.
During this RG flow, all quantum fluctuations involving states outside this restricted manifold are incorporated in the screening of the effective two-particle interaction.
This is achieved within the FRG by a perturbative expansion of the possible scattering processes, that allows for an unbiased treatment of symmetry breaking transitions in the superconducting, charge, and magnetic channel.
This makes FRG a distinguished method for weak to intermediate coupling strengths, \textit{i.e.} the interaction regime suitable for the exploration of itinerant magnetic orders.
Further details on the FRG calculations can be found in the SM.

\begin{figure}
    \centering
    \includegraphics[width=1.0\linewidth]{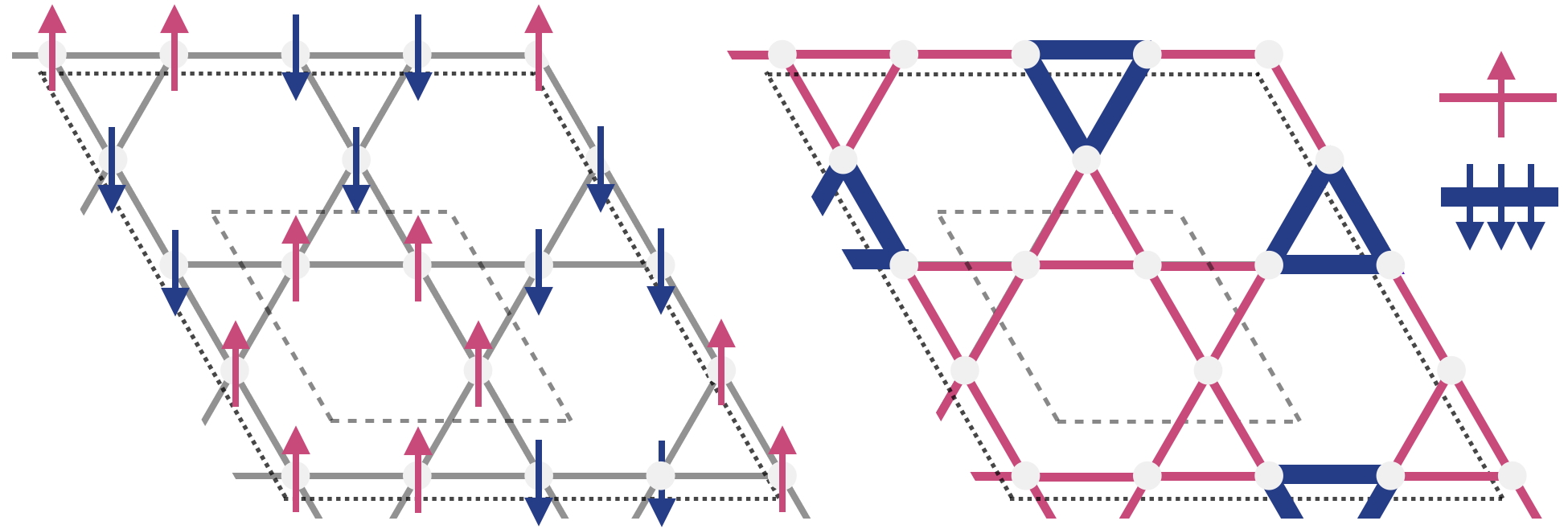}
    \caption{Magnetic ordering vector of the Cu kagome network as obtained from combined FRG + GL analysis at $m$-type vH filling. The order parameter blends an collinear antiferromagnetic arrangement of local magnetic moments given in Eq.~\eqref{eqn:SDW} (left) and NN spin bond terms of Eq.~\eqref{eqn:SBO} (right) \added{reminiscent of a tri-hexagonal pattern. Here, red/blue denote spin up/down on the indicated sites and bonds along the chosen quantization axis. While the relative strength of on-site ($\vec{\Delta}^{SDW}$) and bond magnetization ($\vec{\Delta}^{SBO}$) is dependent on the interaction parameter, their relative sign is fixed by the many-body analysis.To maintain vanishing net magnetization, the spin polarization on the $\downarrow$ bonds surpasses the value on the $\uparrow$ bonds by a factor of 3, as indicated by the line thickness.}}
    \label{fig:mag_order}
    \vspace{-0.3cm}
\end{figure}

Our FRG analysis reveals a $2\times2\times1$ magnetic order depicted in Fig.~\ref{fig:mag_order}, which can be traced to a collaborative effect of two interaction length scales which operate at equal strength for an $m$-type vHS.
The on-site $U$ favors the formation of local spin polarization to avoid double occupancy on the sublattice site -- this leads to long-range antiferromagnetic order at the Fermi surface nesting vector $\mathbf{M}_\gamma$ on the sublattice sites $\gamma$ that do not suffer from SI.
Meanwhile, NN $V$ promotes the coupling of adjacent sites and generically favors bond orders over local particle-hole pairs, as known from the $p$-type vHS scenario~\cite{Kiesel2012}.
This results in a magnetic state $\Vec{\Delta}_\gamma =\Vec{\Delta}^{SDW}_\gamma + \Vec{\Delta}^{SBO}_\gamma$ which combines spin density wave order (SDW)
\begin{equation}
    \Vec{\Delta}^{SDW}_\gamma \propto
    \sum_{\mathbf i} \cos(\mathbf i \cdot\mathbf M_\gamma)
    \langle \hat{c}^\cre_{\mathbf i \gamma \sigma} \vec{\sigma}^{\sigma \sigma^\prime} \hat{c}^\ann_{\mathbf i \gamma \sigma^\prime} \rangle
\label{eqn:SDW}
\end{equation}
with finite relative angular momentum components
\begin{equation}
\begin{split}
    \Vec{\Delta}^{SBO}_\gamma \propto
    \sum_{\langle \mathbf i, \mathbf j\rangle}
    \cos(\mathbf i \cdot\mathbf M_\gamma)
    \Delta_\gamma^{\mathbf i \mathbf j \alpha \beta}
    \langle \hat{c}^\cre_{\mathbf i \alpha \sigma} \vec{\sigma}^{\sigma \sigma^\prime} \hat{c}^\ann_{\mathbf j \beta \sigma^\prime} \rangle \ ,
\end{split}
\label{eqn:SBO}
\end{equation}
that constitute a spin bond order phase (SBO).
Here, $\langle \mathbf i, \mathbf j\rangle$ is the sum over unit cell vecors $\mathbf i,\mathbf j$ such that $\mathbf i, \alpha$ and $\mathbf j, \beta$ describe neighboring sites,
and all doubly occurring indices are summed over.
The real space representations of the SBO order parameter $\Delta_\gamma^{\mathbf i\mathbf j \alpha \beta}$ are given in the SM.

At the $M$-point, the associated little group to the point group $P6/mmm$ is given by $mmm$, which is equivalent to the local symmetry group of the individual kagome sites on the 3f Wyckoff positions.
Consequently, the SDW order parameter, which transforms trivially under all elements of $mmm$, has support only on sublattice $\gamma$.
In order to mix with the $U$ driven on-site magnetization, the SBO phase transforms within the $A_g$ irreducible representation of $mmm$ and thus represents an extended $s$-wave magnetization pattern.
This is a crucial precondition for the magnetic state to lower its free energy via spin fluctuations of both local and non-local moments present in the $m$-type vH scenario.
Contrarily, the $p$-type scenario does not allow for a local moment formation at the Fermi level due to SI~\cite{Kiesel2012}.
While the $p$-type scenario opens up opportunities for bond-magnetization in non-trivial irreducible representations, it also limits the available free-energy gain, since the spin polarization exhibits strong momentum dependence with additional nodes in the quasi-particle spectrum, that cannot contribute from the dominant on-site repulsion $U$.

A Ginzburg-Landau (GL) analysis (cf. SM) shows that near the onset of magnetic order, the favoured state is a collinear order in which SDW and SBO point along the same axis (Fig.~\ref{fig:mag_order}), and modulate with a star-of-David pattern.
The magnetic arrangement breaks $Z_4$ translation and spin rotation symmetries, and transforms within the three-dimensional $F_2^\prime$ irrep of the extended $C_{6v}^{\prime \prime \prime}$ symmetry group of the enlarged $2 \times 2$ unit cell~\cite{Venderbos2016}.
While retaining the original $P6/mmm$ symmetry, the superposition of the three different $A_g$ states leads to a triple enhancement of the hexagonal bonds compared to the star-like bonds to ensure vanishing net magnetization.
This marks the first instance of spontaneous magnetization in kagome metals driven by quantum fluctuations.

\paragraph{Induced spin current order}
This change of paradigm - from the local moment-driven, to itinerant magnetism - comes with the promise of more, yet unprecedented phases. At lower temperature, deeper in the symmetry-broken phase, higher-order terms in GL expansion of the free-energy become important. In this circumstance, the SDW/SBO configuration may cease to be uniaxial, promoting an octahedral spin configuration, which permits a more intriguing coupling of the order parameter to subleading magnetic fluctuations via the spin chirality coupling (cf. SM). For charge-ordered phases, the coupling of real and imaginary bond order is known to promote subsidiary loop current orders~\cite{Guo2024, Neupert2021}; here we draw an interesting parallel to magnetic instabilities. Spin currents are even under time-reversal symmetry (TRS). Hence, a coupling to SBO/SDW order can already appear at first-order in the spin-current order (SCO) parameter via the cubic term
\begin{equation}
\begin{split}
    \mathcal{F}^{(3)} \propto & \sum_{\alpha \beta \gamma} \varepsilon_{\alpha \beta \gamma}
    \,\Vec{\Delta}^{s}_\alpha \cdot (\Vec{\Delta}^{s}_\beta \times \Vec{\Delta}^{SCO}_\gamma)
\end{split}
\label{eqn:sbo_sco}
\end{equation}
where $s=\{SBO,SDW\}$. Minimizing this contribution to the free energy produces a state in which $\vec{\Delta}^{s}_\alpha \perp \vec{\Delta}^{s}_\beta \perp \vec{\Delta}^{SCO}_\gamma$, i.e. the $\vec{\Delta}^{s}$ vectors at the three $M$ points are mutually perpendicular, and locked parallel to the SCO vector, $\vec{\Delta}^{s}_\alpha \parallel \vec{\Delta}^{SCO}_\alpha$. This locking between spin and $M$-point is a kind of ``spontaneous spin-orbit coupling''. Eq. \eqref{eqn:sbo_sco} is linear in $\Vec{\Delta}^{SCO}$, which implies that non collinear $\Vec{\Delta}^{s}$ is expected to immediately induce SCO, and conversely, the presence of SCO can induce canting of the $\Vec{\Delta}^{s}$ state.

This mechanism for the emergence of spin current appears as a direct consequence of cascading phase transitions characteristic of kagome materials~\cite{Zhao2021}. The only instance of spin currents reported to date is in FeGe, in the A-type AFM phase~\cite{Yin2022a, Teng2022, Teng2023}. However, in that system the SCO can be simply viewed as loop current formation of spin-polarized electrons~\cite{Han2024, Teng2024, Zhan2024}. The new mechanism we describe above gives way to uncharted territory of magnetic states due to the unique features of the kagome lattice, combining the effects of geometric frustration and SI in the moderately coupled regime.


\paragraph{Summary}
This study explores the emergence of itinerant magnetism in kagome lattice systems. 
Using \ce{CsCu3Cl5} as theoretical platform, we achieve an electronic structure displaying the three characteristic kagome bands around the Fermi level almost without any interference from additional bands.
A description by means of local reference frames enables the identification of a minimal tight-binding model using three maximally localized Wannier orbitals. While many kagome materials exhibit a more complex description, with several vHS near the Fermi level which may interplay to form novel electronic orders~\cite{Wu2021, Hu2022, Kang2022, Li2024a, Scammell2023, Ingham2024}, our exemplary Cu-based kagome material \ce{CsCu3Cl5} exhibits an isolated $m$-type vHS close to the Fermi level, providing a model realization of the kagome Hubbard model, worth of exploring in future material science experiments.

Our FRG study reveals a unprecedented unconventional $2 \times 2 \times 1$ antiferromagnetic ordering. While kagome magnets have attracted much attention in recent years, the formation of long-range magnetization is usually fostered by dipole interactions of localised magnetic moments rather than an intrinsic electronic mechanism which is at display here~\cite{Norman2016, Yin2022}.
In the presented scenario, quantum fluctuations provide the driving force for the magnetic transition.
This expands the catalogue of magnetic states on the kagome lattice: We predict a spin bond order phase with descendent spin current patterns on the kagome lattice, featuring a finite relative angular momentum of the spin-1 particle-hole pair in analogy to spin-0 charge bond order patterns. 
Our analysis marks the first chapter of emergent magnetic order from itinerant electrons on the kagome lattice, and sets magnetic instabilities on the landscape of symmetry-broken phases in metallic kagome compounds. 


\paragraph{Acknowledgements}
We thank S. Enzner, L.M. Schoop, J.B. Profe, and L. Klebl for valuable discussions and feedback on this work.
A.W., M.D., H.H. and R.T. are supported by the Deutsche Forschungsgemeinschaft (DFG, German Research Foundation) through Project-ID 258499086 - SFB 1170, through the W\"urzburg-Dresden Cluster of Excellence on Complexity and Topology in Quantum Matter - ct.qmat Project-ID 390858490 - EXC 2147, and the research unit QUAST, FOR 5249-449872909 (Project 3).
A.C. acknowledges support from PNRR MUR project PE0000023-NQSTI.
\added{M.D. is grateful for support from a PhD scholarship of the Studienstiftung des Deutschen Volkes.}
A.W., A.C., D.D.S. and R.T. acknowledge the Gauss Centre for Supercomputing e.V. (https://www.gauss-centre.eu) for funding this project by providing computing time on the GCS Supercomputer SuperMUC-NG at Leibniz Supercomputing Centre (https://www.lrz.de) where the DFT and Wannier-based calculations were performed.
M.D., H.H. and R.T. are grateful for HPC resources provided by the Erlangen National High Performance Computing Center (NHR@FAU) of the Friedrich-Alexander-Universit\"at Erlangen-N\"urnberg (FAU), that were used for the FRG calculations. NHR funding is provided by federal and Bavarian state authorities. NHR@FAU hardware is partially funded by the DFG - 440719683.

\paragraph{Author contributions}
R.T. initiated and supervised the project.
R.T., A.W., A.C., and F.v.R. designed the model system. 
A.W., A.C. and D.D.S. conducted the first principles calculations and the related Wannier-based analysis. 
H.H. and M.D. performed the FRG calculations.
M.D. carried out the GL analysis with input from H.D.S. and J.I..
\added{A.W., M.D. and J.I. wrote the manuscript with input from all authors.}

\let\oldaddcontentsline\addcontentsline
\renewcommand{\addcontentsline}[3]{}

\let\addcontentsline\oldaddcontentsline

\supplement{Supplementary Material for \\ Theory of unconventional magnetism in a Cu-based kagome metal}

\tableofcontents

\section{Crystal Structure}
The unit cell of CsCu$_3$Cl$_5$ is defined by the vectors
\begin{equation*}
    \begin{aligned}
        \hat{\textbf{a}}&=(\;\;\,6.025, \; 0.000, \; 0.000) \\
        \hat{\textbf{b}}&=(-3.013, \; 5.218, \; 0.000)\\
        \hat{\textbf{c}}&=(\;\;\,0.000, \; 0.000,\; 7.027).    
    \end{aligned}
\end{equation*}
%
Pristine CsCu$_3$Cl$_5$ is configured in a $P6/mmm$ space group,
with atomic positions specified in fractional coordinates relative to these vectors listed in Tab.\,\ref{pos_pristine}. 

\begin{table}[h]
\centering
\setlength{\tabcolsep}{15pt}
\begin{tabular}{lccc}
\hline
\hline
 \textbf{Atom} & \textbf{a} & \textbf{b} & \textbf{c} \\
\hline
    Cs1   & 0.000 &0.000  &0.000  \\
    Cu1  & 0.500 & 0.500 & 0.500 \\
    Cu2  & 0.500 & 0.000  & 0.500  \\
    Cu3  & 0.000 & 0.500  & 0.500  \\
    Cl1  & 0.000 & 0.000 & 0.500  \\
    Cl2  & 0.667 & 0.333 & 0.736 \\
    Cl3  & 0.333 & 0.667 & 0.264 \\
    Cl4  & 0.333 & 0.667 & 0.736 \\
    Cl5  & 0.667 & 0.333 & 0.264 \\
\hline
\hline
\end{tabular}
\caption{Atomic positions of pristine CsCu$_3$Cl$_5$ in direct coordinates.}
\label{pos_pristine}
\end{table}
\section{Charge Distribution}
To examine the charge distribution within the CsCu$_3$Cl$_5$ compound, we must examine the oxidation states of the involved elements. 
Utilizing the Pauling scale, which spans from 3.98 (fluorine, the most electronegative) to 0.7 (francium, the least electronegative), we find that Cs, with an electronegativity of 0.79, typically adopts an oxidation state of +1 in compounds. 
Conversely, Cl, having an electronegativity of 3.16, predominantly exists in a -1 oxidation state. 
The transition metal copper, with an electronegativity of 1.9, commonly exhibits oxidation states of +2 and, less common, +1 or +3 \cite{Pauling}.
To determine the formal (integer) charges of the atoms in CsCu$_3$Cl$_5$, we distribute (integer) electrons among the unit cell ions to achieve an overall neutral compound. 
The 6$s^1$ electron in cesium is readily relinquished due to its weak binding. Chlorine, requiring only one electron to complete its 3$p^6$ configuration, strongly attracts electrons. 
As a result, we draw four electrons from the three copper atoms. Applying integer charges, two of the Cu atoms are in the less common +1 state, while one is in the +2 state.
Given the $P6/mmm$ symmetry of the compound, the choice of which copper atom assumes the +2 state is equivalent, as they can be transformed into each other through a simple $120^\circ$ rotation. 
With three energetically equivalent possibilities for charge distribution, a Lewis resonance can emerge, allowing electrons to hop between the Cu+1 and Cu+2 sites with equal probability. This resonance creates three Cu+2 sites, with two electrons uniformly hopping between them, resulting in a uniform distribution of -2/3 per copper site. 
Consequently, each Cu has an effective charge of +4/3.
To assess the stability of this charge distribution, an initial magnetic moment imbalance among the Cu atoms is intentionally introduced, followed by relaxation of the electronic degrees of freedom. The outcome consistently reflects the same uniform charge distribution. This suggests that the uniform configuration is not a mere consequence of symmetric initial parameters in the calculations, but rather a stable configuration imposed by the inherent symmetry of the system. 
\\
%
%

\section{Local Frame of Reference}

\begin{figure} [h!]
    \centering
    \includegraphics[width=\linewidth]{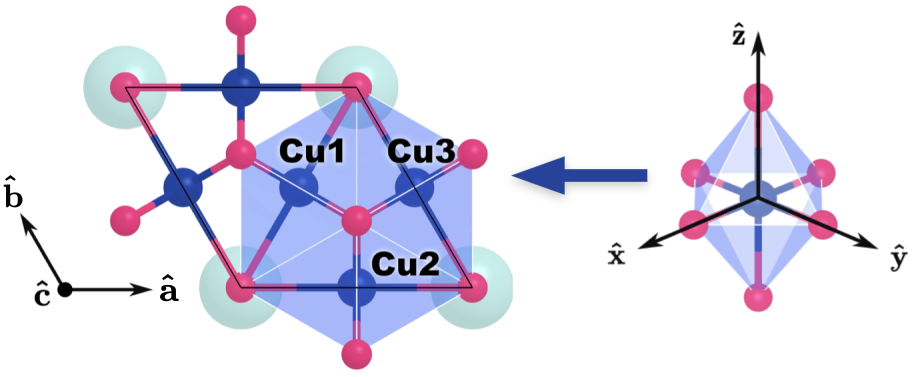}
    \caption{Top view of the kagome plane (left) highlighting the three copper atoms with the local reference frame around each Cu atom (right).}
    \label{fig:octa}
\end{figure}

Fig.~\ref{fig:octa} visualizes the unit cell's three copper atoms (Cu1, Cu2, Cu3) and the orientations of their distorted Cl octahedra within the crystal. 
The elongated side of the Cl octahedron aligns along $\mathbf{\hat{a}+\hat{b}}$ for Cu1, \(\mathbf{\hat{a}}\) for Cu2, and \(\mathbf{\hat{b}}\) for Cu3. 
We chose a local reference frame for each octahedra such that the elongation observed in pristine CsCu$_3$Cl$_5$ is oriented along the $\mathbf{\hat{z}}$-direction, visualized in Fig.~\ref{fig:octa} (right). 
Each octahedron's local reference frame has the Cu atom at the origin, with \(\mathbf{\hat{z}}\) along the elongated tips and \(\mathbf{\hat{x}}\), \(\mathbf{\hat{y}}\) towards the corners.

\section{Structure-Correlation Link in Kagome Materials} 
\added{
To estimate the expected interaction regime for the low energy manifold of \ce{CsCu3Cl5}, we follow the procedure outlined in Ref.~\cite{Schoop2022}:
the degree of electronic correlations in  kagome compounds can be understood based on the relative distances between the transition metal atoms, responsible for the electron hopping, and the surrounding ligands, which influence the strength of the crystal field and ligand hybridization.}
\\
In weakly correlated CsV$_3$Sb$_5$, the V atoms are coordinated by nearly undistorted Sb-octahedra with a bond length ratio between neighboring kagome sites ($d(\text{V-V}) = 2.75 $\;\r{A}) and V-ligand distance ($d(\text{V-Sb}) = 2.75 $\;\r{A}) of 1.00.
This equal distance between V atoms and their surrounding ligands facilitates significant electron hopping between V sites, leading to more delocalized electrons and weaker correlations.
\\
In the strongly correlated Herbertsmithite, each Cu atom is surrounded by a highly distorted octahedron, with four oxygen atoms playing the dominant role in coordination and two Cl atoms contributing less significantly. 
The distance between neighbouring kagome sites is $d(\text{Cu-Cu})=3.53$\;\r{A} while the nearest Cu-ligand distance is $d(\text{Cu-O}) = 1.99$\;\r{A}, giving a bond length ratio of 1.77.
Chlorine atoms also bond to copper, but at a greater distance of $ d(\text{Cu-Cl}) = 2.99$\;\AA.
The close proximity of the oxygen atoms leads to strong Cu-O bonds that confine the electrons, resulting in significant localization.
The large Cu-Cu distance weakens electron hopping between kagome sites, while the strong interaction with oxygen atoms further enhances electron localization, driving Herbertsmithite deep into the Mott-insulating regime with strong electronic correlations.
\\
CsCu$_3$Cl$_5$ falls between these two cases. 
The Cu atoms are coordinated by distorted Cl-octahedra, in the with four equally long bonds out of the kagome plane and two longer bonds in-plane for the pristine case. The Cu-Cu distance is $ d(\text{Cu-Cu}) = 3.01 $\;\r{A}, and the nearest Cu-ligand distance is $ d(\text{Cu-Cl}) = 2.40 $\;\r{A}, giving a bond length ratio of 1.25.
This intermediate coordination environment suggests that CsCu$_3$Cl$_5$ is likely to exhibit moderate electron correlation. 
The Cu-Cu distance is longer than in CsV$_3$Sb$_5$ but shorter than in Herbertsmithite, allowing for some electron delocalization. 
At the same time, the weaker Cu-Cl hybridization, compared to the Cu-O bonds in Herbertsmithite, results in reduced electron-electron interactions. 
This positions CsCu$_3$Cl$_5$ between the weakly correlated metallic state of CsV$_3$Sb$_5$ and the strongly correlated Mott-insulating state of Herbertsmithite, making it a promising candidate for hosting both itinerant and localized magnetic phenomena.
%
\begin{figure}[b]
    \includegraphics[width=\linewidth]{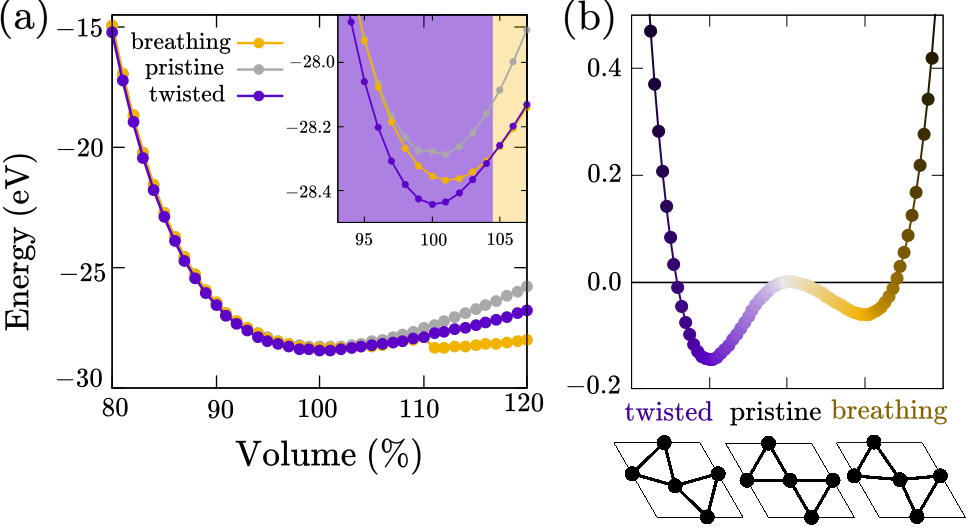}
    \caption{Energy profiles of the pristine, breathing and twisted configurations for different interpolating geometries and unit cell's sizes. (a) Total free energy as a function of an isotropic change of volume, for all three discussed configurations. The twisted configuration is the most favourable one in a wide range of volume variations.
    (b) For a unit cell with constant volume, the energy is plotted across varying degrees of interpolation between the pristine and breathing cases, as well as between the pristine and the twisted ones. The twisted configuration stands out as the most energetically favourable.}
    \label{fig:stability}
\end{figure}
\section{Crystal Distortions and their Stability}
The complex interplay between electronic structure, orbital ordering and structural dynamics becomes noticeable when analyzing the different structural distortions compatible with the phonon modes of the pristine kagome lattice.
This interplay leads indeed to an energy loss known as the Jahn-Teller (JT) effect. 
The sequence of events remains still uncertain, whether the JT distortion precedes orbital ordering or vice versa \cite{Streltsov_2017}.
The focus here lies on the breathing and twisted configurations descending from the pristine structure. Fig.~\ref{fig:stability} illustrates the evolution of the total energy as a function of the geometry, interpolated at constant volume between the pristine and the two distorted cases. The twisted state is the energetically preferred one by $\sim100 $ meV per unit cell with respect to the breathing lattice, and by $\sim150 $ meV per unit cell with respect to the pristine lattice.
Finally, we observe that a consistent conclusion is drawn from the analysis of the total free energy $F$ of the three configurations, as a function of an isotropic volume variation, shown in Fig.~\ref{fig:stability}.
The twisted case remains the most favourable configuration at least in a [-5\%,5\%] range around the equilibrium volume, above which a competition with the other phases can be observed. This fact suggests both hydro-static pressure, as well as uni-axial strain, as valid methods for structural and electronic tuning \cite{PhysRevB.105.165146, PhysRevB.104.205129, lin2024giant, tuniz2024straininduced}, whose detailed treatment is beyond the scope of the present work.
\begin{figure}
    \centering
    \includegraphics[width=0.9\linewidth]{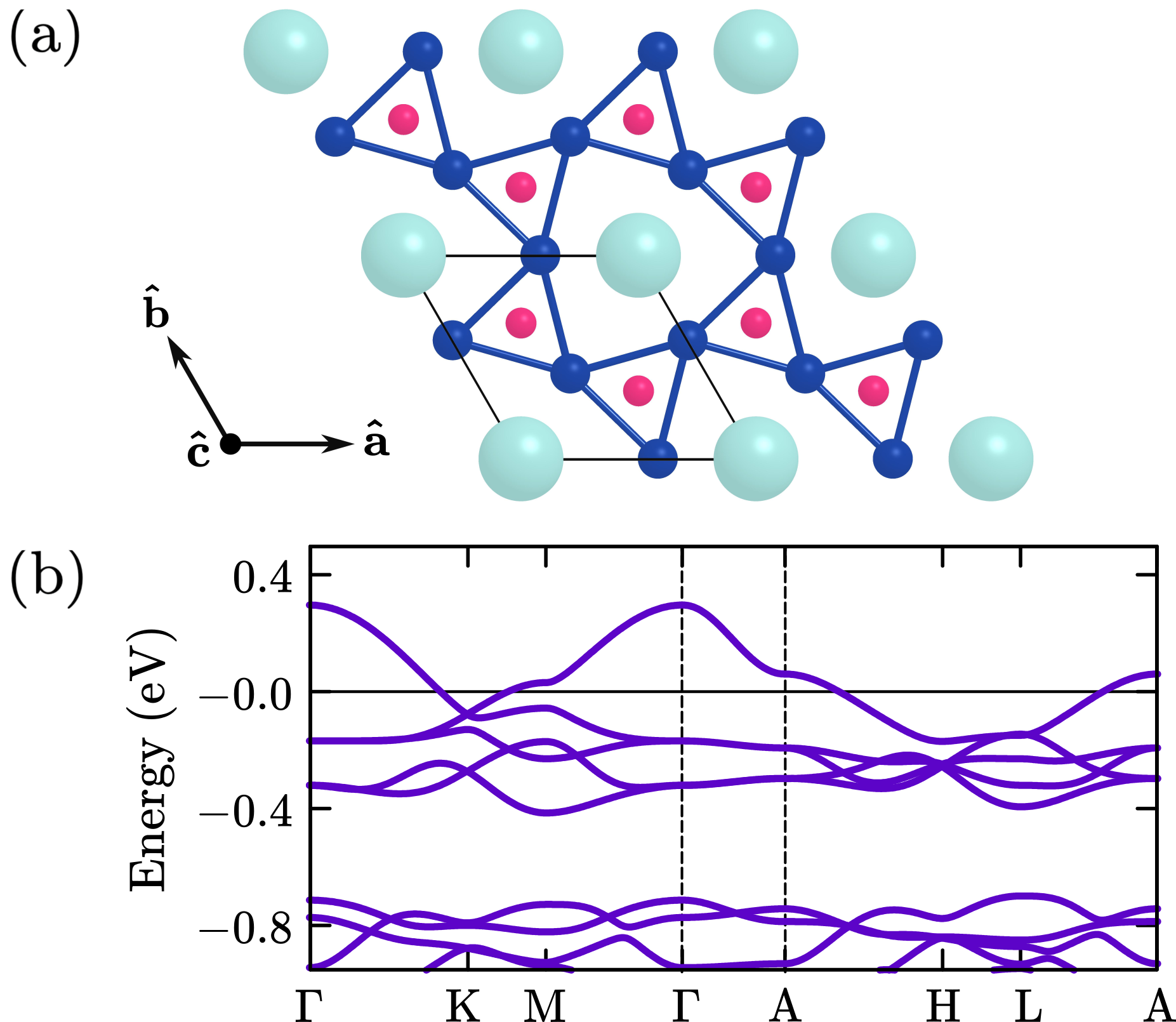}
    \caption{Twisted configuration of CsCu$_3$Cl$_5$. (a) Top view on the crystal with the Cu kagome lattice highlighted by blue lines. Turquoise, blue and pink spheres represent Cs, Cu and Cl atoms. The unit cell is indicated by black lines. (b) Electronic band structure exhibiting a narrower band width and a closer proximity of the $m$-type vHS to the Fermi level, in contrast to the pristine configuration.}
    \label{fig:twistedLatt}
\end{figure}

\subsection{Twisted Configuration}
Within the twisted kagome configuration, taken as an example, the JT distortion leads to a displacement of the Cu atoms as depicted in Fig.~\ref{fig:twistedLatt}\textcolor{blue}{(a)}.
Consequently their orbitals become more spatially separated, leading to a reduction in Coulomb repulsion and hybridization and a corresponding decrease in energy.
The unit cell's top-view reveals that the Cl ions are still positioned at the center of the Cu triangles, albeit in a twisted triangular arrangement that reduces the number of symmetries.
Comparing the band structure of the twisted configuration shown in Fig.~\ref{fig:twistedLatt}\textcolor{blue}{(b)} with the pristine counterpart, the electronic structure of the former demonstrates a reduced dispersion and a slightly lower energy profile. Additionally, the flat band itself exhibits an increased dispersion due to a reduced destructive interference process among the orbitals of Cu atoms.
The Fermi level lies even closer to the (unaffected) $m$-type vHS than in the pristine case.
Given the minimal shift in dispersion, the electronic structure of the twisted configuration remains remarkably close to that of the pristine configuration.
\\
%
\subsection{Breathing Configuration}
\begin{figure}[t]
    \centering
    \includegraphics[width=0.9\linewidth]{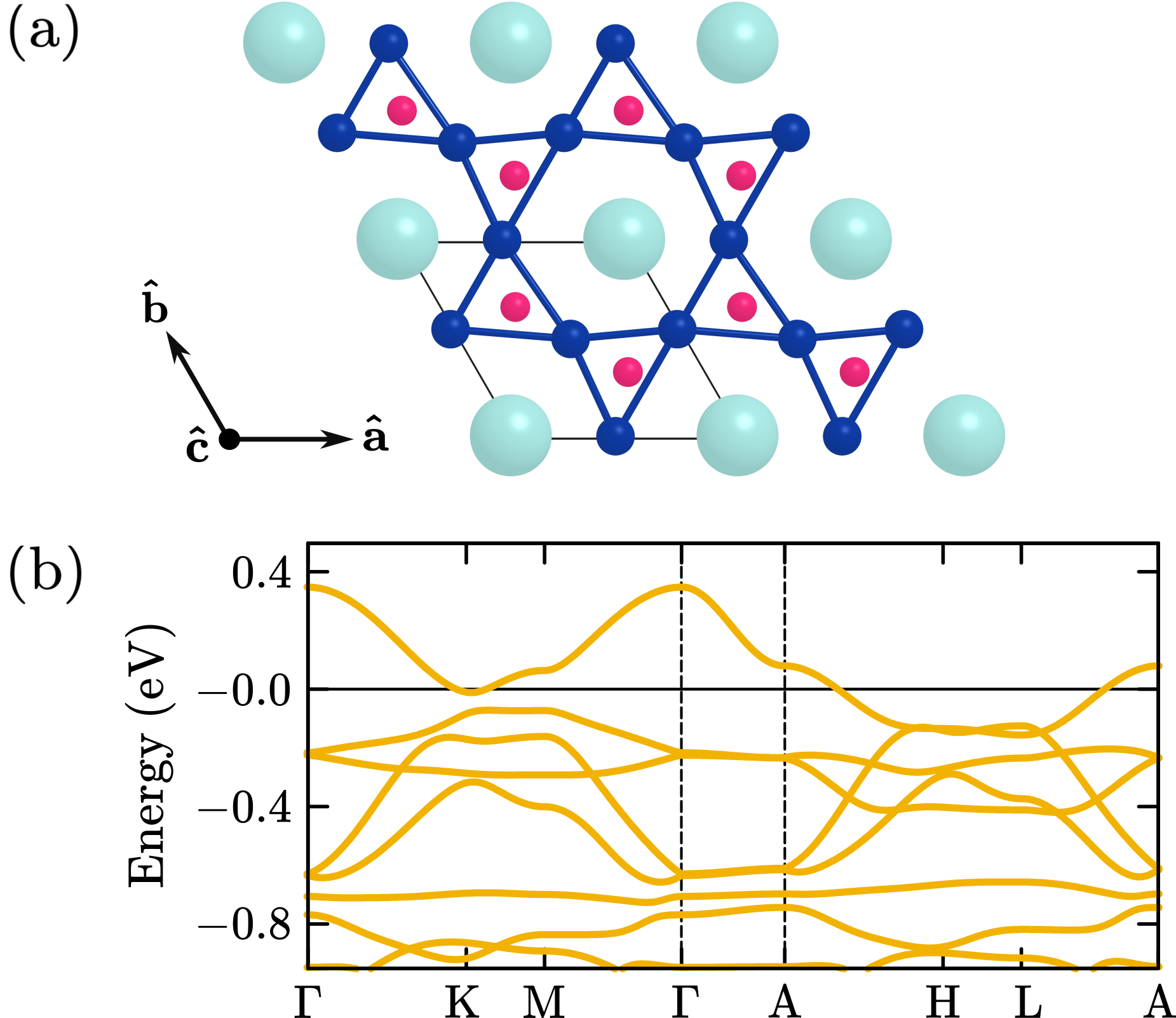}
    \caption{Breathing configuration of CsCu$_3$Cl$_5$. (a) Top view with the kagome lattice highlighted by blue lines. Turquoise, blue and pink spheres represent Cs, Cu and Cl atoms. The unit cell is indicated by black lines. (b) Electronic band structure of breathing CsCu$_3$Cl$_5$. As the mirror symmetry is broken, the Dirac cone gaps out; the $m$-type vHS stays close to the Fermi level.}
    \label{fig:breathingLatt}
\end{figure}
Another structural distortion compatible with the system's phonon modes is the breathing configuration, depicted in Fig.~\ref{fig:breathingLatt}\textcolor{blue}{(a)}.
The energy of this breathing kagome lattice is lower than the pristine lattice, yet higher than the twisted counterpart. 
The central observation is a deviation from the previous triangular arrangement, with Cl ions no longer positioned in the center of the Cu triangles.
Half of the triangles now possess two long sides and one short side, while the other half exhibit two short sides and one long side. 
Despite the distortion exhibiting a lower deviation compared to the twisted kagome lattice, it results in an additional reduction in symmetry. As the mirror symmetry is broken, the Dirac cones are no longer preserved and gap out whereas the $m$-type vHS still remains in the vicinity of the Fermi level.
%

\subsection{Phonon Analysis}
\added{
A conclusive assessment of compound stability and synthetization of \ce{CsCu3Cl5} can not be drawn from a purely \textit{ab initio} perspective and consequently lies beyond the scope of our work and methodology.
It is well known that DFT can produce unreliable predictions concerning a material's stability due to approximations involved in the derivation of the phonon spectra. For instance, \ce{KTaO3} is a material where DFT incorrectly predicts instability due to imaginary phonon modes. However, when anharmonic effects are included, the phonon spectrum becomes stable \cite{KTaO3}.
Additionally, even in the presence of imaginary phonon modes the system might be still stable at ambient conditions and eventually exhibit a phonon driven symmetry breaking only at low temperatures. This is a common feature in comparable 135 kagome compounds like \ce{$A$V3Sb5}~\cite{Enzner2025p}. In this case, electronic correlations might overtake the phononic contribution and lead to an ordered state irrespective of a subsequent phonon anomaly. 
\\
To give a first estimate on the experimental feasability of \ce{CsCu3Cl5}, we perform DFT phonon calculations at the $\Gamma$-point within the harmonic approximation (cf. \textit{Computational Methods} for further details on the calculation). The pristine configuration exhibits imaginary phonon modes, confirming the metastable state. In contrast, the twisted configuration yields 24 stable optical phonons modes with frequencies ranging from 0.68 to 7.14 THz.
Although restricted only to $q=\Gamma$, this phonon analysis provides a first indication of a possible stability of the twisted configuration.}

\subsection{Wannier Model of twisted \protect\ensuremath{\mathbf{CsCu_3Cl_5}}}

\added{
Since the twisted configuration is energetically favorable, we additionally derive the Wannier model for this structure. The band structure of the twisted kagome compound closely resembles that of the pristine case, allowing for the construction of a qualitatively equivalent Wannier model using the same orbitals and local axis orientation (see \textit{Local frame of reference} above). In the twisted structure, however, Cu and Cl atoms no longer lie in the same plane. As a result, the Wannier orbital lobes are also displaced from the plane and exhibit a slight tilt of approximately 12$^\circ$. Despite this geometric distortion, the orbital remains centered on the Cu site and retains a similar shape, originating from a linear combination of the Cu $d_{x^2-y^2}$ orbital, the neighboring Cl $p$ orbitals, and a small contribution from the nearest-neighbor Cu $d_{x^2-y^2}$ orbitals. Rotational symmetry is now limited to 120$^\circ$ rotations implying the reduced $P\bar{6}2m$ space group, while retaining the same unit cell, with fractional atomic coordinates listed in Tab.\,\ref{pos_twisted}.
The band structure obtained from this three-orbital tight-binding model shows excellent agreement with the DFT results, as illustrated in Fig.\,\ref{fig:wannierorbitals_twisted}\textcolor{blue}{(c)}, with minimal spread in the Wannier functions. This validates the use of MLWFs and confirms that the three-orbital model offers a compact and reliable basis for many-body calculations, even in the twisted configuration.
Using these MLWFs as the target subspace for cRPA calculations yields a screened on-site Coulomb interaction of $U_{eff} = 3.3\,$eV.
This slightly reduced value compared to the pristine case is consistent with the band structure, where two bands located below the kagome manifold lie closer in energy and contribute additional screening.
Combined with a narrower bandwidth, this leads to a ratio $U/t \approx 5.3$, indicating a modest increase in electronic correlations.
}

\begin{table}[h]
\centering
\setlength{\tabcolsep}{15pt}
\begin{tabular}{lccc}
\hline
\hline
 \textbf{Atom} & \textbf{a} & \textbf{b} & \textbf{c} \\
\hline
    Cs1   & 0.000 &0.000  &0.000  \\
    Cu1  & 0.418 & 0.418 & 0.500 \\
    Cu2  & 0.582 & 0.000  & 0.500  \\
    Cu3  & 0.000 & 0.582  & 0.500  \\
    Cl1  & 0.000 & 0.000 & 0.500  \\
    Cl2  & 0.667 & 0.333 & 0.732 \\
    Cl3  & 0.333 & 0.667 & 0.268 \\
    Cl4  & 0.333 & 0.667 & 0.732 \\
    Cl5  & 0.667 & 0.333 & 0.268 \\
\hline
\hline
\end{tabular}
\caption{Atomic positions of twisted CsCu$_3$Cl$_5$ in direct coordinates.}
\label{pos_twisted}
\end{table}

\section{Computational Methods}
\subsection{Density Functional Theory}
Electronic structure calculations were performed with the Vienna Ab Initio Simulation Package (VASP)~\cite{PhysRevB.59.1758, PhysRevB.54.11169}, using the projector augmented wave (PAW) method. Generalized Gradient Approximation (GGA), within the
Perdew-Burke-Ernzerhof (PBE) method, has been used to handle exchange and correlation effects~\cite{PhysRevLett.77.3865}. Volume and structural relaxations have been computed with a plane-wave cutoff of 500 eV. The relaxation of the electronic degrees of
freedom was considered converged when the energy difference between two steps was equal or smaller than $1.0e^{-8}\,$eV. The ionic relaxation loop was considered converged when the norms of the forces acting on all atoms were equal or smaller than $1.0 e^{-6}\,$eV/\AA. 
The number of $k$-points was set to $18\times18\times12$, both for relaxation and self-consistent loop. Partial occupancies have been determined according to a Gaussian smearing, with a width of $0.05\,$eV.  Band structures results have been visualized using the VASPKIT postprocessing tool~\cite{WANG2021108033}, while VESTA~\cite{VESTA} has been used to visualize crystal structures and charge density isosurfaces. Wannier models have been constructed by using the WANNIER90 package~\cite{MOSTOFI2008685}.
\added{The data of all calculations can be accessed on the Zenodo platform~\cite{nomad}.}

\begin{figure}
    \centering
    \includegraphics[width=\linewidth]{w90_twisted_2.jpg}
    \caption{(a) Top and (b) side views of the Wannier orbital centered on the Cu site in the twisted structure. The site-local reference axis $\mathbf{\hat{z}}$ points toward the center of the hexagonal plaquettes. The purple and yellow lobes indicate the phase of the MLWF. (c) Band structure comparison between DFT (gray) and the three-orbital Wannier tight-binding model (yellow).}
    \label{fig:wannierorbitals_twisted}
    \vspace{-0.3cm}
\end{figure}

\subsection{Constrained Random Phase Approximation}
\added{
We obtained interaction parameters from first-principles electronic structure calculations, employing the constrained random phase approximation (cRPA) as implemented in VASP \cite{kaltak2015, PhysRevB.54.11169}, based on WANNIER90 \cite{MOSTOFI2008685} and the VASP2WANNIER90 interface.
The calculations were performed using a $9\times 9 \times 6$ $k$-point mesh, a plane-wave energy cutoff of $600\,$eV, a $400\,$eV cutoff for the response functions and approximately 200 unoccupied bands. Consistent with the underlying DFT calculations, the PBE exchange-correlation functional was used throughout, a Gaussian smearing of $0.05\,$eV was applied, and total energies were converged to within $10^{-8}\,$eV.\\
In cRPA, the interaction tensor inside the correlated manifold is obtained by considering the screening effect of all other bands on the Coulomb repulsion: These rest bands provide a background whose polarizability is evaluated by calculating the bare particle-hole susceptibility within linear response theory. This polarization bubble is then inserted in a Dyson equation for the dielectric screening of the Coulomb repulsion leading to renormalized interaction parameter in the target space. \\
In addition to the missing contribution from diagrammatic screening processes outside the RPA ladder series, the incomplete disentanglement between the low-energy subspace (target space) and the higher-energy bands poses additional difficulties for reliable cRPA calculations: The bands generated from the MLWFs, which are used to define the target space, overlap with other lower-lying bands, thereby undermining the assumption that high-energy states can be cleanly integrated out. This leads in particular to a high sensitivity to the calculation parameters impeding quantitative estimates of the interaction strength. For a detailed discussion of the drawbacks of the cRPA approximation we refer to Ref.~\cite{Honerkamp2012e, Profe2025e, Kaltak2025c}.
Hence, we employ the obtained cRPA parameter as a qualitative estimate for the degree of correlations in \ce{CsCu3Cl5} compared to other kagome compounds as discussed in the main paper. Remarkably, this classification is consistent with the intuitive bond length analysis presented above reassuring the initial assessment of \ce{CsCu3Cl5} as an intermediately correlated kagome material. Notably, the obtained $U/V$ ratio is consistent with the universal scaling behavior of long-range screened Coulomb interaction obtained in independent cRPA studies on V-based kagome compounds~\cite{DiSante2023}.}
\begin{figure}[t]
    \centering
    \includegraphics[width=0.9\linewidth]{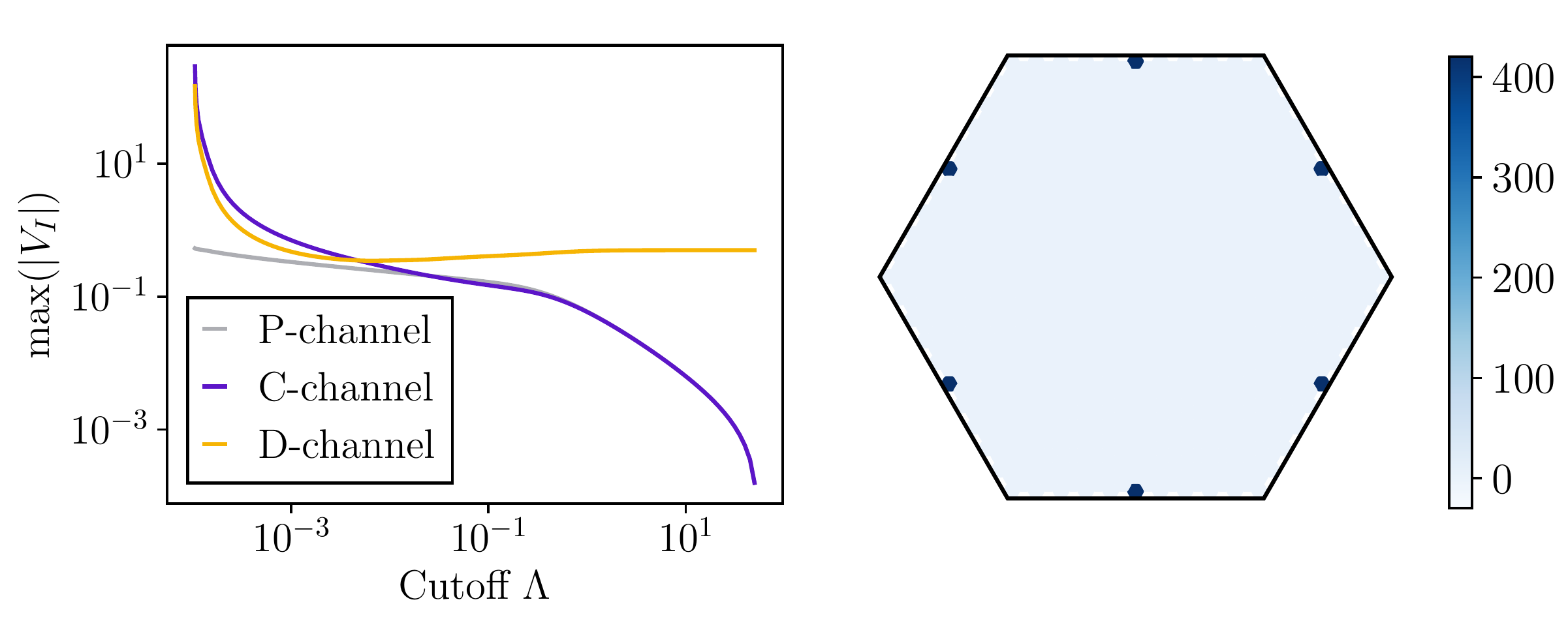}
    \caption{\added{Representative FRG analysis for the pristine model with bare interaction vertex defined in Eq.~(1) of the main paper with  $U = 0.5\,$eV and $V = 0.25\,$eV. (left) Flow of the maximal vertex element in the three channels. The divergent C channel indicates a magnetic instability. (right) Eigenvalue spectrum of the C channel vertex at the end of the flow. The system exhibits degenerate instabilities at each of the three $M$ points, that are symmetry related by a $C_3$ rotation. The three corresponding eigenvectors are depicted in Fig.~\ref{fig:3Q}.}}
    \label{fig:frg_explained}
\end{figure}

\subsection{Functional Renormalization Group}

\begin{figure}[t]
    \centering
    \includegraphics[width=0.9\linewidth]{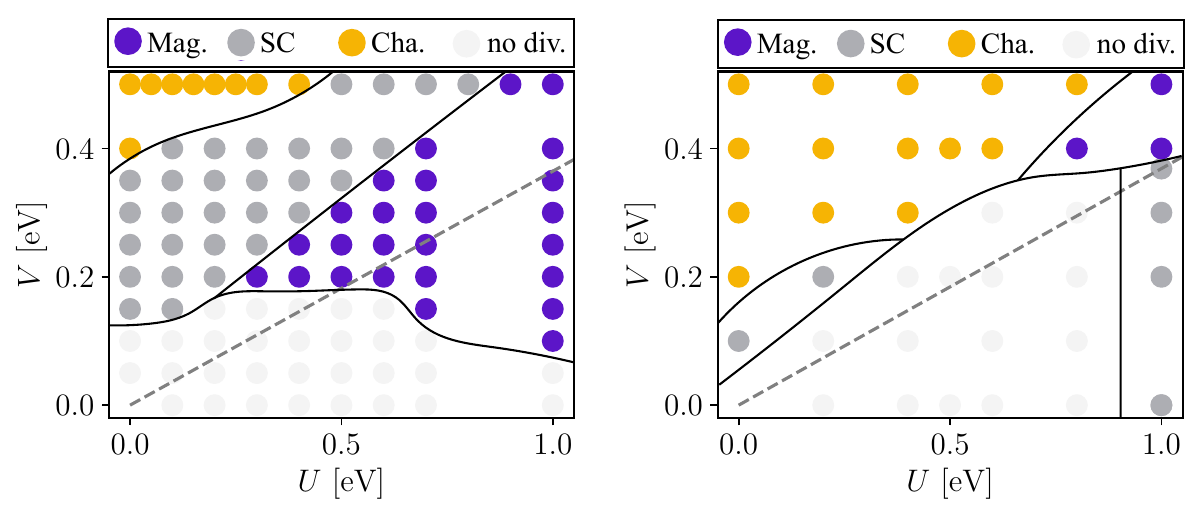}
    \caption{\added{FRG phase diagram for \ce{CsCu3Cl5} in the pristine (left) and twisted (right) configuration as a function of interaction values $U$ and $V$ defined in Eq.~(1) in the main paper. Different colors indicate magnetic (mag.), charge (Cha.) and superconucting (SC) instabilities. Light grey marks points without a divergent susceptibility, \textit{i.e.} no symmetry breaking within the resolution of the calculation. Solid black lines denote the approximate phase boundaries, while dashed black lines indicate the $U/V$ ratio obtained from cRPA.}}
    \label{fig:phase_diag}
\end{figure}

\added{Kagome materials exhibit a variety of competing electronic instabilities, which can intertwine in complex ways. 
To identify possible ordering tendencies without bias, a method capable of capturing all channels on equal footing is required.
For this purpose, we employ the functional renormalization group (FRG)~\cite{Metzner2012,Platt_2013}. \\
The FRG solves the many electron problem by parametrizing the flow of the effective action by an artificial cutoff $\Lambda$ from a solvable bare theory ($\Lambda \rightarrow \infty$) to the full interacting system ($ \Lambda = 0$). This is achieved by dressing the bare propagators with a $\Lambda$ dependent regulator, that selectively suppresses electronic modes in the partition sum.
In general, this flow is described by an infinite set of integro-differential equations containing all kinds of higher order screening processes. To arrive at a computationally feasible problem, the hierarchy of diagrams is truncated at the one loop level, \textit{i.e.} only screening processes involving the effective two particle interaction are kept in the equations. While this restricts the FRG to the weak to intermediate coupling regime, where the 1-loop approximation is valid, it allows to write the FRG flow equations in a closed form and numerically solve them. \\
In order to do so, we use the truncated unity (TU) parametrization of FRG to separate the flow of the full vertex $\mathcal V$ into the three diagrammatic channels according to the momentum transfer $\mathbf q$ of the fermionic loop integrals: the particle-particle (P), direct particle-hole (D) and crossed particle-hole (C) channel~\cite{Lichtenstein2017h, Beyer2022r, Profe2022t}.
The flow is stopped upon encountering a divergence in one of the channels, that indicates a superconducting (P), charge (D) or magnetic (C) instability. The predicted phase transition is determined by solving the linearised gap equation in the divergent channel, that amounts to an eigenvalue equation: The eigenvector corresponding to the leading eigenvalue gives the order parameter in the symmetry broken phase.
\begin{equation}
    \lambda(\mathbf q) \Delta_{f}(\mathbf q) = \sum_{f'}
    \mathcal V_{ff'}^\text{P/C/D}(\mathbf q) \, \Delta_{f'}(\mathbf q) \ .
\end{equation}
Here, we suppressed all other potential indices. 
In TUFRG, this eigenvector is given in terms of the ordering vector $\mathbf q$ equal to the center of mass momentum  of the fermionic bilinear and the real space relative coordinate on the crystal lattice $f$.
An archetypical flow and the resulting eigenvalue spectrum $\lambda(\mathbf q)$ in the BZ is depicted in Fig.~\ref{fig:frg_explained}.}
\\
The FRG calculations were performed with the TUFRG backend of the divERGe library~\cite{Profe2024a}. 
We employed a $30 \times 30$ mesh for the bosonic momenta of the vertices, with an additional refinement of $61 \times 61$ for the integration of the loop. The form-factor cutoff distance is chosen as 1.99 in units of the lattice vectors. 
We check for convergence by calculating selected points in parameter space with increased number of momentum points and form-factor cutoff ($42 \times 42$, refinement: $81 \times 81$, formfactor cutoff: $3.32$). We utilized the Euler 34 integrator of the divERGe library.
\\
Taking into account the importance of long-range interactions on the kagome, we employ on-site and nearest neighbour density-density interactions via the interaction Hamiltonian Eq.\,(3) in the main text.
\added{As discussed above, the absolute interaction values of the cRPA are not trustworthy. Consequently, we perform a phase scan in $U$ and $V$ to determine the system's symmetry breaking propensities for a wide range of interaction scenarios presented in Fig.~\ref{fig:phase_diag}. For the pristine case, the abundant order in the reasonable interaction regime $U > V$ is given the spin bond order discussed in the main text. In the FRG, this surfaces as divergent vertex at $\mathbf q = M_i \ \forall i = 1,2,3$ (compare Fig.~\ref{fig:frg_explained}). \\
For the twisted configuration, the broken $C_6$ symmetry significantly reduces the Fermi surface nesting away from the vHs points. This leads to a suppression of emergent orders in the $U < V$ regime and favor weak coupling superconductivity at sufficiently strong $U$. However, still the SBO phase persists in the stronger correlated part of the phase diagram around the $U/V$ ratio predicted by cRPA. Hence, we conclude that critical fluctuations in \ce{CsCu3Cl5} are still of SBO type and the lattice distortion only marginally effects the material's symmetry breaking behavior. \\
The results shown in the main part of this work were obtained  with representative interaction values $U = 0.5\,$eV and $V = 0.25\,$eV close to the $U/V$ ratio from cRPA.}

\subsection{Ginzburg-Landau Analysis of the 3Q Order}
\begin{figure}
    \includegraphics[width=1\linewidth]{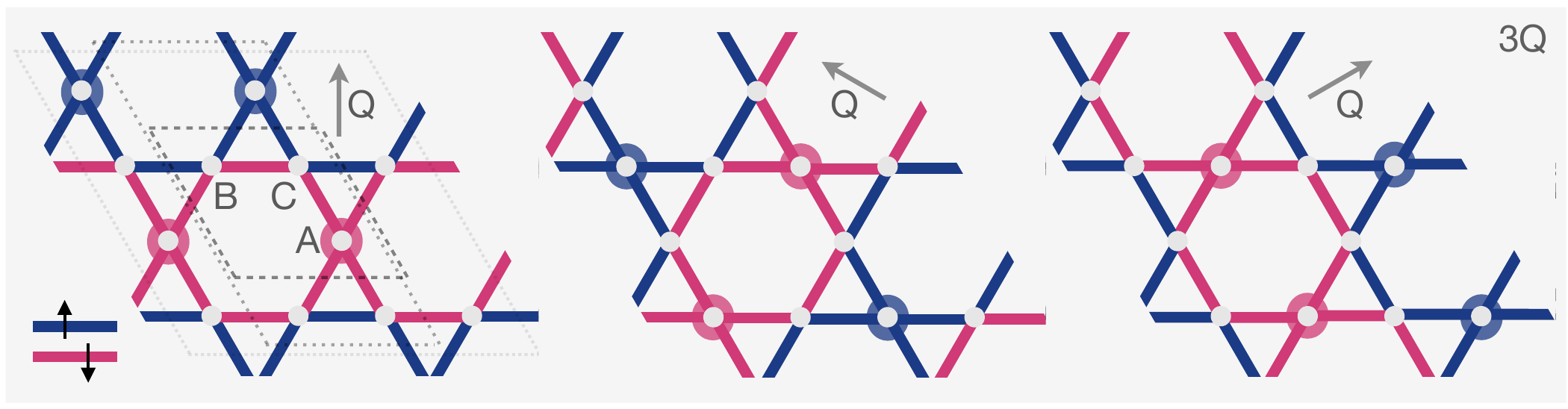}
    \caption{Three symmetry equivalent magnetic orders with largest susceptibility in the FRG flow \added{given by the eigenvectors corresponding to the leading eigenvalues of Fig.~\ref{fig:frg_explained}}. Colored sites (bonds) \added{indicate} a finite local magnetisation on the associated site (bond).
    \added{Dotted (dashed) polygons indicate the extended (original) unit cell corresponding to the translation symmetry breaking indicated by $Q$. All NN bonds share the same absolute value of the magnetization with sign structure indicated by different colors. More distant bonds likewise display a finite magnetization but with absolute values of 2 orders of magnitude smaller in size and are hence not shown. The relative strength of on-site and bond magnetization is highly dependent on the phase space point in Fig.~\ref{fig:phase_diag} and hence only schematically drawn here.}}
    \label{fig:3Q}
\end{figure}

The FRG calculations exhibit a divergent susceptibility simultaneously at the three inequivalent $M$- points in the hexagonal Brillouin zone indicating a phase transition with the given wave vector \added{(cf. Fig.~\ref{fig:frg_explained})}. Solving the linearised gap equation at the phase transition for the three different ordering vectors gives the pattern displayed in Fig.~\ref{fig:3Q}, where we only depict the most dominant nearest neighbor bond magnetisation and neglect longer range bonds for clarity. To determine the relative strength of the degenerate ordering vectors, we consider coupling between the different ordering vectors by means of Ginzburg-Landau (GL) theory at the Fermi level. In contrast to the previously studied 3$Q$ charge orders on the kagome lattice (see \textit{e.g.} Ref.~\cite{Neupert2021}), the relative orientation of the magnetic ordering vector per $Q$ adds three continuous degrees of freedom to the relative strength and sign of the three independent order parameters subject to the GL analysis.
To obtain the free energy functional after the FRG flow and avoid double counting, we project the interacting theory at the FRG cutoff scale onto the three van Hove momenta, which carry the dominant spectral weight on the Fermi surface~\cite{Park2021,Scammell2023}. The prevalent contribution to this Hamiltonian is given by the most divergent part of the FRG vertex, that we expand up to second order in the fluctuations around the static order parameter using a Hubbard-Stratonovich transformation. The resulting action for the low energy effective theory on the Fermi level reads
\begin{equation}
\begin{split}
    S = -T \sum_{\omega_n} & \sum_{\alpha s} c^\cre_{\alpha s}(\omega_n) G_\alpha (\omega_n) c^\ann_{\alpha s}(\omega_n) \\ 
    + & \sum_{\alpha \beta \gamma} \epsilon_{\alpha\beta\gamma} \vec{\Delta}_{\gamma} \vec{S}_{\alpha \beta}(\omega_n)
		+ g^{-1} \sum_{\gamma} \vec{\Delta}_{\gamma}^2 \, ,
\end{split}
\end{equation}
with the non-interacting Green's function $G_{\alpha}(\omega_{n})=1/(i\omega_n-\varepsilon_{\alpha})$ and the spin operator $\vec{S}_{\alpha \beta}(\omega_n) = \sum_{s s'} c^\cre_{\alpha s}(\omega_n) \vec{\sigma}^{s s'} c^\ann_{\beta s}(\omega_n)$, where $\vec{\sigma}$ are Pauli matrices acting on spin space, and the sum runs over all fermionic Matsubara frequencies $\omega_n$.
Due to the mixed orbital content of the vHS, there is no one-to-one correspondence between the vH index $\alpha$ and the sublattice degree of the original TB model opposed to the $p$-type scenario~\cite{Park2021}. However, the scattering in the magnetic channel is purely off-diagonal in this new basis analogous to the $p$-type scenario, since the different instabilities $\vec{\Delta}_{\gamma} = \vec{\Delta}(M_\gamma)$ for $\gamma \in \{1, 2, 3\}$  involve a momentum transfer between the different vHS points at $M_\alpha, M_\beta$ with $\alpha \neq \beta \neq \gamma$.
By carrying out the Gaussian integrals in the remaining Grassman fields, one obtains the free energy
\begin{equation}
    F = g^{-1} \sum_\gamma \vec{\Delta}_\gamma^2 -\operatorname{Tr} \ln \left(-\mathcal{G}^{-1}\right) \, 
\end{equation}
where we define the fully interacting Green's function $\mathcal{G}^{-1} = G^{-1} + \Sigma^{-1}$ in the tensor product Hilbert space of spin and patch degrees of freedom, \textit{i.e.} $G = 1_{2 \times 2} \otimes \text{diag}(G_1, G_2, G_3)$ and
\begin{equation}
    \Sigma = \left(\begin{array}{ccc}
			0 & \vec{\Delta}_3 & -\vec{\Delta}_2 \\
			-\vec{\Delta}_3 & 0 & \vec{\Delta}_1 \\
			\vec{\Delta}_2 & -\vec{\Delta}_1 & 0
		\end{array}\right) \, .
\end{equation}
Each $\vec{\Delta}_\gamma$ is hence a $2 \times 2$ matrix in spin space, whose Pauli vector is kept as tuning parameter.
Expanding the free energy in terms of the order parameter fields up to forth order we obtain the GL functional
\begin{equation}
	\begin{aligned}
		F = & g^{-1} \sum_\gamma \vec{\Delta}_\gamma^2 + \frac{1}{2} \operatorname{Tr}\left(G \Sigma\right)^2
          + \frac{1}{4} \operatorname{Tr}\left(G \Sigma\right)^4 + \mathcal O(\Sigma^6)\, ,
	\end{aligned}
\end{equation}
that is valid in vicinity of the phase transition.
The quadratic order 
\begin{equation}
	\begin{aligned}
		F^{(2)}= \sum_\gamma \biggl(\frac{1}{g} - \chi^{ph}(T)\biggr) \vec{\Delta}_\gamma^2 \propto T - T_c \, ,
	\end{aligned}
\end{equation}
is given by the non-interacting particle-hole susceptibility $\chi^{ph}(T) = \operatorname{Tr}(G_1G_2)$ and scales linear around the transition temperature $T_c$.
Since the free energy inherits time-reversal and global SO(3) symmetry of the spin quantisation axis from the kinetic Hamiltonian Eq.~(1) in the main text, the cubic term is absent and a coupling of the different order parameter $\vec{\Delta}_\gamma$ first appears at quartic order
\begin{equation}
\begin{aligned}
    F^{(4)} =& Z_1 \big( \sum_\gamma \vec{\Delta}_\gamma^2 \big)^2 \\
    +& 2 (Z_2 - Z_1 - Z_3) \left( \vec{\Delta}_1^2 \vec{\Delta}_2^2 + \vec{\Delta}_2^2 \vec{\Delta}_3^2 + \vec{\Delta}_3^2 \vec{\Delta}_1^2 \right) \\
    +& 4 Z_3 \left( (\vec{\Delta}_1 \cdot \vec{\Delta}_2)^2 + (\vec{\Delta}_2 \cdot \vec{\Delta}_3)^2 + (\vec{\Delta}_3 \cdot \vec{\Delta}_1)^2\right) \, .
\end{aligned}
\label{eq:f4}
\end{equation}
The expansion coefficients can be calculated from the kinetic theory on the Fermi level and are given by
\begin{equation}
\begin{aligned}
    Z_1 &= \operatorname{Tr}(G_1^2G_2^2), \\
    Z_2 &= \operatorname{Tr}(G_1^2G_2G_3), \\
    Z_3 &= \operatorname{Tr}(G_1G_2G_3G_0) \, .
\end{aligned}
\end{equation}
These prefactors can be calculated analytically by expanding the TB bandstructure around the saddle-points; the resulting action is identical to that of Ref.~\cite{Nandkishore2012}.
While $Z_2 - Z_1 - Z_3 < 0$ results in an equal strength of the three magnetic ordering vectors, $Z_3 < 0$ aligns the magnetisation vectors of the different $\vec{\Delta}_\gamma$. 
This uniaxial spin arrangement of the 3$Q$ state persists within the symmetry broken phase as evidenced by the sixth order term $F^{(6)} = - 4 Z_4 (\vec{\Delta}_1 \cdot \vec{\Delta}_2 \times \vec{\Delta}_3 )^2$ with $Z_4 < 0$~\cite{Nandkishore2012}.
The final result of the combined FRG and GL analysis is presented in Fig.~3 in the main text and resembles other itinerant SDW patterns on the kagome lattice Hubbard model discussed in the literature~\cite{Park2021, Kiesel2013, Wang2013, Xu2023, Profe2024a}.

The real space order parameter of the parent spin bond order phase determined by our FRG analysis breaks $Z_4$ translation symmetry and time reversal symmetry $\mathcal{T}$.
Since the global spin quantisation axis remains undefined, the magnetic states can be classified by the irreducible representations (irreps) of the magnetic space group $\mathcal M = C_{6v}^{\prime \prime \prime} \otimes \mathcal{T}$, that is a direct product of enlarge space group in the $2 \times 2$ unit cell and TRS symmetry.
The $\vec{\Delta}_\gamma^{SDW/SBO}$ obtained by the presented GL analysis transforms within the $F_2^\prime$ irrep, where the prime indicates an odd transformation behaviour under the TRS $\mathcal{T}$. This results from an $F_1$ structure for the spatial part of the order parameter combined with an odd transformation behaviour of the spin degree of freedom under all mirror operations (corresponding to an $A_2$ irrep for the spin part). 
Despite Eq.~\eqref{eq:f4} does not determine the spin polarisation value along the unidirectional quantisation axis, all possible parallel and antiparallel magnetic arrangements of the different $\vec{\Delta}_\gamma$ result in the same physical state, potentially translated by one original kagome lattice vector, that constitute the three basis states of the $F_2^\prime$ irrep.

As a direct consequence, the uniaxial spin alignment of the obtained 3$Q$ order differs drastically from studies of the kagome Heisenberg model, that suggests an octahedral order as preferred superposition of the different $\vec{\Delta}(M_\gamma)$ ordering vectors~\cite{Messio2011}. The reason for this apparent discrepancy roots in the dissimilar nature of the coupling between the 3 ordering vectors: In the kagome Heisenberg model, the different orders interact via the bare short-range first and second nearest-neighbor spin-exchange couplings $J_1$ and $J_2$. In the itinerant picture, however, these couplings are generated dynamically and are driven by bandstructure effects, namely the pronounced nesting of the $M$- point vHS.

\section{Emergence of Octahedral Order inside the Collinear Magnetic Phase}
Within the symmetry broken phase, the spin exchange coupling is expected to become larger as the GL expansion breaks well below the N\'eel temperature $T_N$ and eventually promote a perpendicular spin orientation of the different $\vec{\Delta}_\gamma$ in accordance with self-consistent mean field calculations on the honeycomb lattice~\cite{Nandkishore2012}.
This tendency can be directly inferred from the highest order GL term considered above: Exploiting the universal relation $|\vec{\Delta}_\gamma|_{T=0} \propto T_N$, we can reevaluate the prefactors $Z_i$ of the GL functional at low temperatures in the limit $T_N \lesssim t$ with $t$ is the kinetic energy scale of the system to obtain an intuition of the leading coupling terms between the magnetic ordering vectors. We recognize, that $Z_4$ changes sign already at small values of $T_N$, thus favoring a pairwise orthogonal orientation of the magnetic vectors $\vec{\Delta}_\alpha \perp \vec{\Delta}_\beta \perp \vec{\Delta}_\gamma \perp \vec{\Delta}_\alpha$~\cite{Nandkishore2012}.
We thus recover the octahedral spin order deep inside the magnetic phase, that is separated from the uniaxial $F_2^\prime$ order by a first order transition. As discussed in the main text, the non uniaxial spin order couples linearly to spin loop current order, immediately inducing a coexisting spin current order.

\begin{table}[h]
\centering
\begin{tabular}{lccc}
\hline
\hline
\textbf{Property} & \textbf{CsCu$_3$Cl$_5$} & \textbf{CsCu$_3$Cl$_4$Br} & \textbf{CsCu$_3$Cl$_4$I} \\
\hline
$|\hat{c}|$ (\AA)       & 7.027 & 7.179 & 7.494 \\
$E_{\text{vHS},m}$ (eV)       & 0.06  & 0.03  & -0.04 \\
$\Delta E_{\text{vHS},m}$ (eV) & 0.26  & 0.22  & 0.15  \\
\hline
\hline
\end{tabular}
\caption{Comparison of structural and electronic properties for CsCu$_3$Cl$_5$, CsCu$_3$Cl$_4$Br, and CsCu$_3$Cl$_4$I. $|\hat{c}|$ represents the length of the unit vector perpendicular to the kagome plane, while $E_{\text{vHS},m}$ indicates the energy of the $m$-type vHS relative to the Fermi level. $\Delta E_{\text{vHS},m}$ denotes its shift in energy from the $M$-point to the $L$-point in k-space.}
\label{table:comparison}
\end{table}
\begin{figure}[t]
    \centering
    \includegraphics[width=0.9\linewidth]{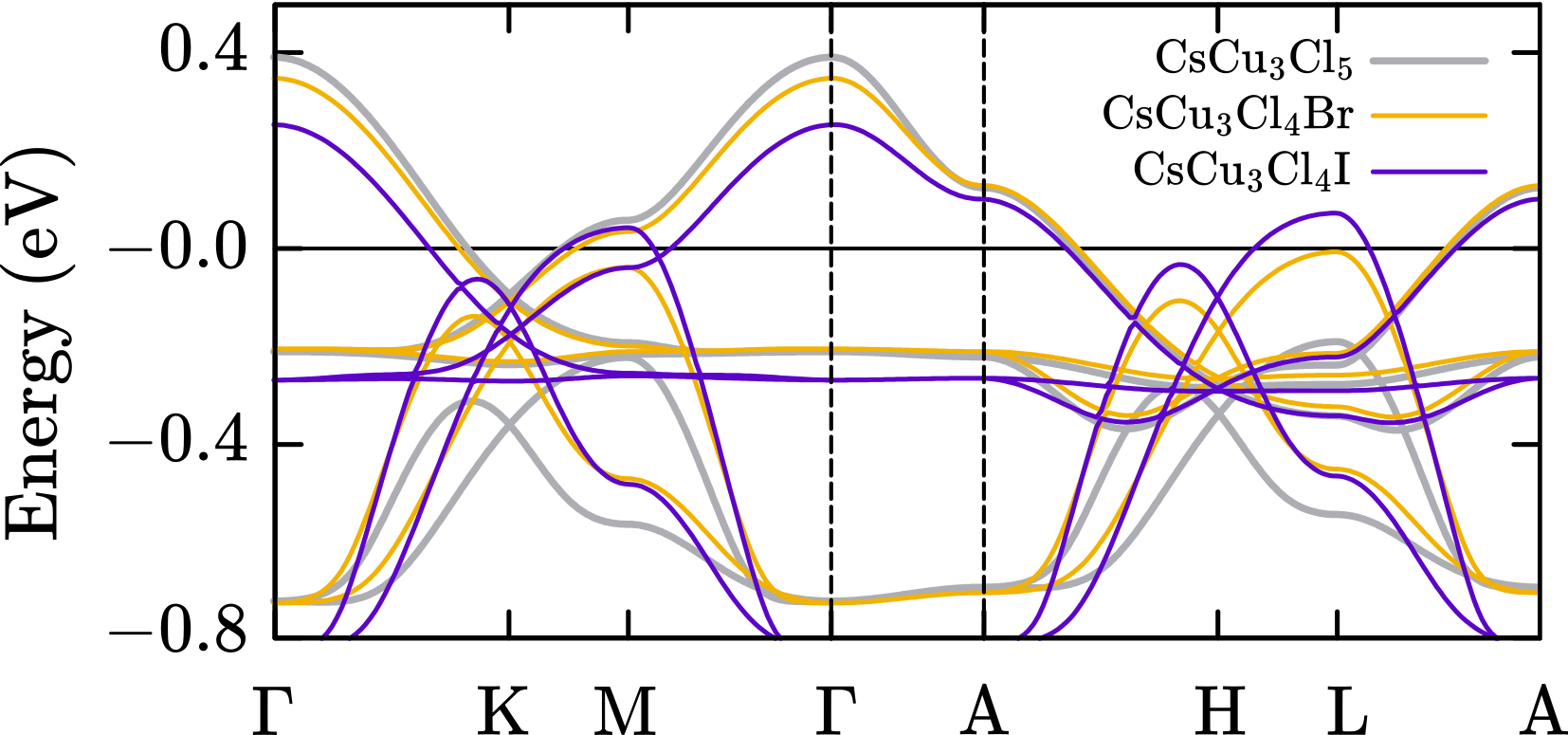}
    \caption{Comparison of band structures \added{of different Cu based kagome compounds} illustrating the effects of increasing substituting atom size on the kagome bands. As the substituting atom size increases, the kagome bands narrow and shift downward in energy, while the other two bands broaden and shift upward.}
    \label{fig:subs_vgl}
\end{figure}
\section{The Role of the \protect\ensuremath{\mathbf{k_z}}-Dispersion}
Due to the layered structure of the compound, we have conducted FRG calculations in the $k_z = 0$ plane, thus omitting the dispersion in the third direction in accordance with other theoretical studies, \textit{e.g.} on $A$V$_3$Sb$_5$~\cite{Wu2021}.
CsCu$_3$Cl$_5$ exhibits some hybridization along the \( k_z \) direction, introducing out-of-plane dispersion in the bandstructure that softens the logarithmic divergence in the density of states at the 2D van Hove points. 
\\
A future goal is to enhance the two-dimensional characteristics of the layered material by either effectively screening interlayer hopping or by increasing the interlayer spacing between the planes along the z-axis.
The latter can be approached in two ways. First, the thickness of the triangular layer can be increased by substituting Cs with a molecule that has a +1 oxidation state, or by adding a neutral molecule to the layer. To preserve the unique kagome physics, it is essential that any added structures maintain kagome symmetry and are carefully examined to ensure that the $m$-type vHS remain close to the Fermi level.
The second approach involves increasing the spacing between the kagome Cu$_3$Cl layer and the chlorine layers that sandwich it. This can be achieved by substituting the Cl atom at the Wyckoff position 1b with larger elements from the same group, such as Br or I.
The $k_z$ dispersion can be measured by the change in energy of the vHS between the $M$-point ($k_z=0$) and the $L$-point ($k_z=1/2$).
\\
Fig.~\ref{fig:subs_vgl} compares the band structures of CsCu$_3$Cl$_5$, CsCu$_3$Cl$_4$Br, and CsCu$_3$Cl$_4$I. As the size of the substituting atom increases, the kagome bands narrow and shift downward in energy, while the other two bands widen and shift upward.
To quantify the effects of substitution, Tab.~\ref{table:comparison} lists the length of the \(\hat{c}\) unit vector perpendicular to the plane, the position of the $m$-type vHS in the $k_z$=0 plane relative to the Fermi energy, and the change in energy of this vHS from $k_z$=0 to $k_z$=1/2 for the different substitutions.
Comparing these values shows that as the substituting atom increases in size, the \(\hat{c}\) out-of-plane unit vector indeed grows, decreasing interlayer hopping and reducing $k_z$-dispersion. For example, substituting with Iod increases the unit cell size along the \(\hat{z}\)-direction from 7.027\,\AA\ to 7.494\,\AA\ and reduces the $k_z$ dispersion of the $m$-type vHS from 0.26$\,$eV to 0.15\,eV. The vHS shifts from 0.06 eV above the Fermi level to 0.04 eV below it.
As observed for CsCu$_3$Cl$_5$, orbital projections show that the three characteristic kagome bands originate from Cu \(d_{xz}\) + \(d_{yz}\) + Cl \(p\) orbitals. The Iod \(p\)-orbitals only contribute to the two lower bands in Fig.~\ref{fig:subs_vgl}.
Projections onto the single Cu atoms confirm the presence of sublattice interference.


\begin{thebibliography}{86}%
\makeatletter
\providecommand \@ifxundefined [1]{%
 \@ifx{#1\undefined}
}%
\providecommand \@ifnum [1]{%
 \ifnum #1\expandafter \@firstoftwo
 \else \expandafter \@secondoftwo
 \fi
}%
\providecommand \@ifx [1]{%
 \ifx #1\expandafter \@firstoftwo
 \else \expandafter \@secondoftwo
 \fi
}%
\providecommand \natexlab [1]{#1}%
\providecommand \enquote  [1]{``#1''}%
\providecommand \bibnamefont  [1]{#1}%
\providecommand \bibfnamefont [1]{#1}%
\providecommand \citenamefont [1]{#1}%
\providecommand \href@noop [0]{\@secondoftwo}%
\providecommand \href [0]{\begingroup \@sanitize@url \@href}%
\providecommand \@href[1]{\@@startlink{#1}\@@href}%
\providecommand \@@href[1]{\endgroup#1\@@endlink}%
\providecommand \@sanitize@url [0]{\catcode `\\12\catcode `\$12\catcode
  `\&12\catcode `\#12\catcode `\^12\catcode `\_12\catcode `\%12\relax}%
\providecommand \@@startlink[1]{}%
\providecommand \@@endlink[0]{}%
\providecommand \url  [0]{\begingroup\@sanitize@url \@url }%
\providecommand \@url [1]{\endgroup\@href {#1}{\urlprefix }}%
\providecommand \urlprefix  [0]{URL }%
\providecommand \Eprint [0]{\href }%
\providecommand \doibase [0]{http://dx.doi.org/}%
\providecommand \selectlanguage [0]{\@gobble}%
\providecommand \bibinfo  [0]{\@secondoftwo}%
\providecommand \bibfield  [0]{\@secondoftwo}%
\providecommand \translation [1]{[#1]}%
\providecommand \BibitemOpen [0]{}%
\providecommand \bibitemStop [0]{}%
\providecommand \bibitemNoStop [0]{.\EOS\space}%
\providecommand \EOS [0]{\spacefactor3000\relax}%
\providecommand \BibitemShut  [1]{\csname bibitem#1\endcsname}%
\let\auto@bib@innerbib\@empty
\bibitem [{\citenamefont {Jiang}\ \emph {et~al.}(2021)\citenamefont {Jiang},
  \citenamefont {Yin}, \citenamefont {Denner}, \citenamefont {Shumiya},
  \citenamefont {Ortiz}, \citenamefont {Xu}, \citenamefont {Guguchia},
  \citenamefont {He}, \citenamefont {Hossain}, \citenamefont {Liu},
  \citenamefont {Ruff}, \citenamefont {Kautzsch}, \citenamefont {Zhang},
  \citenamefont {Chang}, \citenamefont {Belopolski}, \citenamefont {Zhang},
  \citenamefont {Cochran}, \citenamefont {Multer}, \citenamefont {Litskevich},
  \citenamefont {Cheng}, \citenamefont {Yang}, \citenamefont {Wang},
  \citenamefont {Thomale}, \citenamefont {Neupert}, \citenamefont {Wilson},\
  and\ \citenamefont {Hasan}}]{Jiang2021}%
  \BibitemOpen
  \bibfield  {author} {\bibinfo {author} {\bibfnamefont {Yu-Xiao}\ \bibnamefont
  {Jiang}}, \bibinfo {author} {\bibfnamefont {Jia-Xin}\ \bibnamefont {Yin}},
  \bibinfo {author} {\bibfnamefont {M.~Michael}\ \bibnamefont {Denner}},
  \bibinfo {author} {\bibfnamefont {Nana}\ \bibnamefont {Shumiya}}, \bibinfo
  {author} {\bibfnamefont {Brenden~R.}\ \bibnamefont {Ortiz}}, \bibinfo
  {author} {\bibfnamefont {Gang}\ \bibnamefont {Xu}}, \bibinfo {author}
  {\bibfnamefont {Zurab}\ \bibnamefont {Guguchia}}, \bibinfo {author}
  {\bibfnamefont {Junyi}\ \bibnamefont {He}}, \bibinfo {author} {\bibfnamefont
  {Md~Shafayat}\ \bibnamefont {Hossain}}, \bibinfo {author} {\bibfnamefont
  {Xiaoxiong}\ \bibnamefont {Liu}}, \bibinfo {author} {\bibfnamefont {Jacob}\
  \bibnamefont {Ruff}}, \bibinfo {author} {\bibfnamefont {Linus}\ \bibnamefont
  {Kautzsch}}, \bibinfo {author} {\bibfnamefont {Songtian~S.}\ \bibnamefont
  {Zhang}}, \bibinfo {author} {\bibfnamefont {Guoqing}\ \bibnamefont {Chang}},
  \bibinfo {author} {\bibfnamefont {Ilya}\ \bibnamefont {Belopolski}}, \bibinfo
  {author} {\bibfnamefont {Qi}~\bibnamefont {Zhang}}, \bibinfo {author}
  {\bibfnamefont {Tyler~A.}\ \bibnamefont {Cochran}}, \bibinfo {author}
  {\bibfnamefont {Daniel}\ \bibnamefont {Multer}}, \bibinfo {author}
  {\bibfnamefont {Maksim}\ \bibnamefont {Litskevich}}, \bibinfo {author}
  {\bibfnamefont {Zi-Jia}\ \bibnamefont {Cheng}}, \bibinfo {author}
  {\bibfnamefont {Xian~P.}\ \bibnamefont {Yang}}, \bibinfo {author}
  {\bibfnamefont {Ziqiang}\ \bibnamefont {Wang}}, \bibinfo {author}
  {\bibfnamefont {Ronny}\ \bibnamefont {Thomale}}, \bibinfo {author}
  {\bibfnamefont {Titus}\ \bibnamefont {Neupert}}, \bibinfo {author}
  {\bibfnamefont {Stephen~D.}\ \bibnamefont {Wilson}}, \ and\ \bibinfo {author}
  {\bibfnamefont {M.~Zahid}\ \bibnamefont {Hasan}},\ }\bibfield  {title}
  {\enquote {\bibinfo {title} {Unconventional chiral charge order in kagome
  superconductor {KV}$_3${Sb}$_5$},}\ }\href {\doibase
  10.1038/s41563-021-01034-y} {\bibfield  {journal} {\bibinfo  {journal} {Nat.
  Mater.}\ }\textbf {\bibinfo {volume} {20}},\ \bibinfo {pages} {1353--1357}
  (\bibinfo {year} {2021})}\BibitemShut {NoStop}%
\bibitem [{\citenamefont {Mielke}\ \emph {et~al.}(2022)\citenamefont {Mielke},
  \citenamefont {Das}, \citenamefont {Yin}, \citenamefont {Liu}, \citenamefont
  {Gupta}, \citenamefont {Jiang}, \citenamefont {Medarde}, \citenamefont {Wu},
  \citenamefont {Lei}, \citenamefont {Chang}, \citenamefont {Dai},
  \citenamefont {Si}, \citenamefont {Miao}, \citenamefont {Thomale},
  \citenamefont {Neupert}, \citenamefont {Shi}, \citenamefont {Khasanov},
  \citenamefont {Hasan}, \citenamefont {Luetkens},\ and\ \citenamefont
  {Guguchia}}]{Mielke2022}%
  \BibitemOpen
  \bibfield  {author} {\bibinfo {author} {\bibfnamefont {C.}~\bibnamefont
  {Mielke}}, \bibinfo {author} {\bibfnamefont {D.}~\bibnamefont {Das}},
  \bibinfo {author} {\bibfnamefont {J.-X.}\ \bibnamefont {Yin}}, \bibinfo
  {author} {\bibfnamefont {H.}~\bibnamefont {Liu}}, \bibinfo {author}
  {\bibfnamefont {R.}~\bibnamefont {Gupta}}, \bibinfo {author} {\bibfnamefont
  {Y.-X.}\ \bibnamefont {Jiang}}, \bibinfo {author} {\bibfnamefont
  {M.}~\bibnamefont {Medarde}}, \bibinfo {author} {\bibfnamefont
  {X.}~\bibnamefont {Wu}}, \bibinfo {author} {\bibfnamefont {H.~C.}\
  \bibnamefont {Lei}}, \bibinfo {author} {\bibfnamefont {J.}~\bibnamefont
  {Chang}}, \bibinfo {author} {\bibfnamefont {Pengcheng}\ \bibnamefont {Dai}},
  \bibinfo {author} {\bibfnamefont {Q.}~\bibnamefont {Si}}, \bibinfo {author}
  {\bibfnamefont {H.}~\bibnamefont {Miao}}, \bibinfo {author} {\bibfnamefont
  {R.}~\bibnamefont {Thomale}}, \bibinfo {author} {\bibfnamefont
  {T.}~\bibnamefont {Neupert}}, \bibinfo {author} {\bibfnamefont
  {Y.}~\bibnamefont {Shi}}, \bibinfo {author} {\bibfnamefont {R.}~\bibnamefont
  {Khasanov}}, \bibinfo {author} {\bibfnamefont {M.~Z.}\ \bibnamefont {Hasan}},
  \bibinfo {author} {\bibfnamefont {H.}~\bibnamefont {Luetkens}}, \ and\
  \bibinfo {author} {\bibfnamefont {Z.}~\bibnamefont {Guguchia}},\ }\bibfield
  {title} {\enquote {\bibinfo {title} {Time-reversal symmetry-breaking charge
  order in a kagome superconductor},}\ }\href {\doibase
  10.1038/s41586-021-04327-z} {\bibfield  {journal} {\bibinfo  {journal}
  {Nature}\ }\textbf {\bibinfo {volume} {602}},\ \bibinfo {pages} {245--250}
  (\bibinfo {year} {2022})}\BibitemShut {NoStop}%
\bibitem [{\citenamefont {Li}\ \emph {et~al.}(2023)\citenamefont {Li},
  \citenamefont {Cheng}, \citenamefont {Ortiz}, \citenamefont {Tan},
  \citenamefont {Werhahn}, \citenamefont {Zeng}, \citenamefont {Johrendt},
  \citenamefont {Yan}, \citenamefont {Wang}, \citenamefont {Wilson} \emph
  {et~al.}}]{li2023electronic}%
  \BibitemOpen
  \bibfield  {author} {\bibinfo {author} {\bibfnamefont {Hong}\ \bibnamefont
  {Li}}, \bibinfo {author} {\bibfnamefont {Siyu}\ \bibnamefont {Cheng}},
  \bibinfo {author} {\bibfnamefont {Brenden~R}\ \bibnamefont {Ortiz}}, \bibinfo
  {author} {\bibfnamefont {Hengxin}\ \bibnamefont {Tan}}, \bibinfo {author}
  {\bibfnamefont {Dominik}\ \bibnamefont {Werhahn}}, \bibinfo {author}
  {\bibfnamefont {Keyu}\ \bibnamefont {Zeng}}, \bibinfo {author} {\bibfnamefont
  {Dirk}\ \bibnamefont {Johrendt}}, \bibinfo {author} {\bibfnamefont {Binghai}\
  \bibnamefont {Yan}}, \bibinfo {author} {\bibfnamefont {Ziqiang}\ \bibnamefont
  {Wang}}, \bibinfo {author} {\bibfnamefont {Stephen~D}\ \bibnamefont
  {Wilson}},  \emph {et~al.},\ }\bibfield  {title} {\enquote {\bibinfo {title}
  {Electronic nematicity without charge density waves in titanium-based kagome
  metal},}\ }\href@noop {} {\bibfield  {journal} {\bibinfo  {journal} {Nature
  Physics}\ }\textbf {\bibinfo {volume} {19}},\ \bibinfo {pages} {1591--1598}
  (\bibinfo {year} {2023})}\BibitemShut {NoStop}%
\bibitem [{\citenamefont {Jiang}\ \emph {et~al.}(2023)\citenamefont {Jiang},
  \citenamefont {Liu}, \citenamefont {Ma}, \citenamefont {Xia}, \citenamefont
  {Liu}, \citenamefont {Liu}, \citenamefont {Cho}, \citenamefont {Yang},
  \citenamefont {Ding}, \citenamefont {Liu} \emph {et~al.}}]{jiang2023flat}%
  \BibitemOpen
  \bibfield  {author} {\bibinfo {author} {\bibfnamefont {Zhicheng}\
  \bibnamefont {Jiang}}, \bibinfo {author} {\bibfnamefont {Zhengtai}\
  \bibnamefont {Liu}}, \bibinfo {author} {\bibfnamefont {Haiyang}\ \bibnamefont
  {Ma}}, \bibinfo {author} {\bibfnamefont {Wei}\ \bibnamefont {Xia}}, \bibinfo
  {author} {\bibfnamefont {Zhonghao}\ \bibnamefont {Liu}}, \bibinfo {author}
  {\bibfnamefont {Jishan}\ \bibnamefont {Liu}}, \bibinfo {author}
  {\bibfnamefont {Soohyun}\ \bibnamefont {Cho}}, \bibinfo {author}
  {\bibfnamefont {Yichen}\ \bibnamefont {Yang}}, \bibinfo {author}
  {\bibfnamefont {Jianyang}\ \bibnamefont {Ding}}, \bibinfo {author}
  {\bibfnamefont {Jiayu}\ \bibnamefont {Liu}},  \emph {et~al.},\ }\bibfield
  {title} {\enquote {\bibinfo {title} {Flat bands, non-trivial band topology
  and rotation symmetry breaking in layered kagome-lattice \ce{RbTi3Bi5}},}\
  }\href@noop {} {\bibfield  {journal} {\bibinfo  {journal} {Nature
  Communications}\ }\textbf {\bibinfo {volume} {14}},\ \bibinfo {pages} {4892}
  (\bibinfo {year} {2023})}\BibitemShut {NoStop}%
\bibitem [{\citenamefont {Nag}\ \emph {et~al.}(2024)\citenamefont {Nag},
  \citenamefont {Batabyal}, \citenamefont {Ingham}, \citenamefont {Morali},
  \citenamefont {Tan}, \citenamefont {Koo}, \citenamefont {Consiglio},
  \citenamefont {Liu}, \citenamefont {Avraham}, \citenamefont {Queiroz} \emph
  {et~al.}}]{nag2024pomeranchuk}%
  \BibitemOpen
  \bibfield  {author} {\bibinfo {author} {\bibfnamefont {Pranab~Kumar}\
  \bibnamefont {Nag}}, \bibinfo {author} {\bibfnamefont {Rajib}\ \bibnamefont
  {Batabyal}}, \bibinfo {author} {\bibfnamefont {Julian}\ \bibnamefont
  {Ingham}}, \bibinfo {author} {\bibfnamefont {Noam}\ \bibnamefont {Morali}},
  \bibinfo {author} {\bibfnamefont {Hengxin}\ \bibnamefont {Tan}}, \bibinfo
  {author} {\bibfnamefont {Jahyun}\ \bibnamefont {Koo}}, \bibinfo {author}
  {\bibfnamefont {Armando}\ \bibnamefont {Consiglio}}, \bibinfo {author}
  {\bibfnamefont {Enke}\ \bibnamefont {Liu}}, \bibinfo {author} {\bibfnamefont
  {Nurit}\ \bibnamefont {Avraham}}, \bibinfo {author} {\bibfnamefont {Raquel}\
  \bibnamefont {Queiroz}},  \emph {et~al.},\ }\bibfield  {title} {\enquote
  {\bibinfo {title} {Pomeranchuk instability induced by an emergent
  higher-order van hove singularity on the distorted kagome surface of co $ \_3
  $ sn $ \_2 $ s $ \_2$},}\ }\href {https://arxiv.org/abs/2410.01994}
  {\bibfield  {journal} {\bibinfo  {journal} {arXiv preprint arXiv:2410.01994}\
  } (\bibinfo {year} {2024})}\BibitemShut {NoStop}%
\bibitem [{\citenamefont {Jiang}\ \emph {et~al.}(2024)\citenamefont {Jiang},
  \citenamefont {Shao}, \citenamefont {Xia}, \citenamefont {Denner},
  \citenamefont {Ingham}, \citenamefont {Hossain}, \citenamefont {Qiu},
  \citenamefont {Zheng}, \citenamefont {Chen}, \citenamefont {Cheng} \emph
  {et~al.}}]{jiang2024van}%
  \BibitemOpen
  \bibfield  {author} {\bibinfo {author} {\bibfnamefont {Yu-Xiao}\ \bibnamefont
  {Jiang}}, \bibinfo {author} {\bibfnamefont {Sen}\ \bibnamefont {Shao}},
  \bibinfo {author} {\bibfnamefont {Wei}\ \bibnamefont {Xia}}, \bibinfo
  {author} {\bibfnamefont {M~Michael}\ \bibnamefont {Denner}}, \bibinfo
  {author} {\bibfnamefont {Julian}\ \bibnamefont {Ingham}}, \bibinfo {author}
  {\bibfnamefont {Md~Shafayat}\ \bibnamefont {Hossain}}, \bibinfo {author}
  {\bibfnamefont {Qingzheng}\ \bibnamefont {Qiu}}, \bibinfo {author}
  {\bibfnamefont {Xiquan}\ \bibnamefont {Zheng}}, \bibinfo {author}
  {\bibfnamefont {Hongyu}\ \bibnamefont {Chen}}, \bibinfo {author}
  {\bibfnamefont {Zi-Jia}\ \bibnamefont {Cheng}},  \emph {et~al.},\ }\bibfield
  {title} {\enquote {\bibinfo {title} {Van {Hove} annihilation and nematic
  instability on a kagome lattice},}\ }\href
  {https://doi.org/10.1038/s41563-024-01914-z} {\bibfield  {journal} {\bibinfo
  {journal} {Nat. Mater.}\ }\textbf {\bibinfo {volume} {23}},\ \bibinfo {pages}
  {1214--1221} (\bibinfo {year} {2024})}\BibitemShut {NoStop}%
\bibitem [{\citenamefont {Zhong}\ \emph {et~al.}(2023)\citenamefont {Zhong},
  \citenamefont {Li}, \citenamefont {Liu}, \citenamefont {Dong}, \citenamefont
  {Aido}, \citenamefont {Arai}, \citenamefont {Li}, \citenamefont {Zhang},
  \citenamefont {Shi}, \citenamefont {Wang}, \citenamefont {Shin},
  \citenamefont {Lee}, \citenamefont {Miao}, \citenamefont {Kondo},\ and\
  \citenamefont {Okazaki}}]{Zhong2023}%
  \BibitemOpen
  \bibfield  {author} {\bibinfo {author} {\bibfnamefont {Yigui}\ \bibnamefont
  {Zhong}}, \bibinfo {author} {\bibfnamefont {Shaozhi}\ \bibnamefont {Li}},
  \bibinfo {author} {\bibfnamefont {Hongxiong}\ \bibnamefont {Liu}}, \bibinfo
  {author} {\bibfnamefont {Yuyang}\ \bibnamefont {Dong}}, \bibinfo {author}
  {\bibfnamefont {Kohei}\ \bibnamefont {Aido}}, \bibinfo {author}
  {\bibfnamefont {Yosuke}\ \bibnamefont {Arai}}, \bibinfo {author}
  {\bibfnamefont {Haoxiang}\ \bibnamefont {Li}}, \bibinfo {author}
  {\bibfnamefont {Weilu}\ \bibnamefont {Zhang}}, \bibinfo {author}
  {\bibfnamefont {Youguo}\ \bibnamefont {Shi}}, \bibinfo {author}
  {\bibfnamefont {Ziqiang}\ \bibnamefont {Wang}}, \bibinfo {author}
  {\bibfnamefont {Shik}\ \bibnamefont {Shin}}, \bibinfo {author} {\bibfnamefont
  {H.~N.}\ \bibnamefont {Lee}}, \bibinfo {author} {\bibfnamefont
  {H.}~\bibnamefont {Miao}}, \bibinfo {author} {\bibfnamefont {Takeshi}\
  \bibnamefont {Kondo}}, \ and\ \bibinfo {author} {\bibfnamefont {Kozo}\
  \bibnamefont {Okazaki}},\ }\bibfield  {title} {\enquote {\bibinfo {title}
  {Testing electron--phonon coupling for the superconductivity in kagome metal
  {CsV}$_3${Sb}$_5$},}\ }\href {\doibase 10.1038/s41467-023-37605-7} {\bibfield
   {journal} {\bibinfo  {journal} {Nat. Commun.}\ }\textbf {\bibinfo {volume}
  {14}},\ \bibinfo {pages} {1945} (\bibinfo {year} {2023})}\BibitemShut
  {NoStop}%
\bibitem [{\citenamefont {Xie}\ \emph {et~al.}(2022)\citenamefont {Xie},
  \citenamefont {Li}, \citenamefont {Bourges}, \citenamefont {Ivanov},
  \citenamefont {Ye}, \citenamefont {Yin}, \citenamefont {Hasan}, \citenamefont
  {Luo}, \citenamefont {Yao}, \citenamefont {Wang}, \citenamefont {Xu},\ and\
  \citenamefont {Dai}}]{Xie2022}%
  \BibitemOpen
  \bibfield  {author} {\bibinfo {author} {\bibfnamefont {Yaofeng}\ \bibnamefont
  {Xie}}, \bibinfo {author} {\bibfnamefont {Yongkai}\ \bibnamefont {Li}},
  \bibinfo {author} {\bibfnamefont {Philippe}\ \bibnamefont {Bourges}},
  \bibinfo {author} {\bibfnamefont {Alexandre}\ \bibnamefont {Ivanov}},
  \bibinfo {author} {\bibfnamefont {Zijin}\ \bibnamefont {Ye}}, \bibinfo
  {author} {\bibfnamefont {Jia-Xin}\ \bibnamefont {Yin}}, \bibinfo {author}
  {\bibfnamefont {M.~Zahid}\ \bibnamefont {Hasan}}, \bibinfo {author}
  {\bibfnamefont {Aiyun}\ \bibnamefont {Luo}}, \bibinfo {author} {\bibfnamefont
  {Yugui}\ \bibnamefont {Yao}}, \bibinfo {author} {\bibfnamefont {Zhiwei}\
  \bibnamefont {Wang}}, \bibinfo {author} {\bibfnamefont {Gang}\ \bibnamefont
  {Xu}}, \ and\ \bibinfo {author} {\bibfnamefont {Pengcheng}\ \bibnamefont
  {Dai}},\ }\bibfield  {title} {\enquote {\bibinfo {title} {Electron-phonon
  coupling in the charge density wave state of {CsV}$_3${Sb}$_5$},}\ }\href
  {\doibase 10.1103/PhysRevB.105.L140501} {\bibfield  {journal} {\bibinfo
  {journal} {Phys. Rev. B}\ }\textbf {\bibinfo {volume} {105}},\ \bibinfo
  {pages} {L140501} (\bibinfo {year} {2022})}\BibitemShut {NoStop}%
\bibitem [{\citenamefont {Kiesel}\ and\ \citenamefont
  {Thomale}(2012)}]{Kiesel2012}%
  \BibitemOpen
  \bibfield  {author} {\bibinfo {author} {\bibfnamefont {Maximilian~L}\
  \bibnamefont {Kiesel}}\ and\ \bibinfo {author} {\bibfnamefont {Ronny}\
  \bibnamefont {Thomale}},\ }\bibfield  {title} {\enquote {\bibinfo {title}
  {Sublattice interference in the kagome hubbard model},}\ }\href
  {https://link.aps.org/doi/10.1103/PhysRevB.86.121105} {\bibfield  {journal}
  {\bibinfo  {journal} {Phys. Rev. B}\ }\textbf {\bibinfo {volume} {86}},\
  \bibinfo {pages} {121105} (\bibinfo {year} {2012})}\BibitemShut {NoStop}%
\bibitem [{\citenamefont {Hu}\ \emph {et~al.}(2022)\citenamefont {Hu},
  \citenamefont {Wu}, \citenamefont {Ortiz}, \citenamefont {Ju}, \citenamefont
  {Han}, \citenamefont {Ma}, \citenamefont {Plumb}, \citenamefont {Radovic},
  \citenamefont {Thomale}, \citenamefont {Wilson}, \citenamefont {Schnyder},\
  and\ \citenamefont {Shi}}]{Hu2022}%
  \BibitemOpen
  \bibfield  {author} {\bibinfo {author} {\bibfnamefont {Yong}\ \bibnamefont
  {Hu}}, \bibinfo {author} {\bibfnamefont {Xianxin}\ \bibnamefont {Wu}},
  \bibinfo {author} {\bibfnamefont {Brenden~R.}\ \bibnamefont {Ortiz}},
  \bibinfo {author} {\bibfnamefont {Sailong}\ \bibnamefont {Ju}}, \bibinfo
  {author} {\bibfnamefont {Xinloong}\ \bibnamefont {Han}}, \bibinfo {author}
  {\bibfnamefont {Junzhang}\ \bibnamefont {Ma}}, \bibinfo {author}
  {\bibfnamefont {Nicholas~C.}\ \bibnamefont {Plumb}}, \bibinfo {author}
  {\bibfnamefont {Milan}\ \bibnamefont {Radovic}}, \bibinfo {author}
  {\bibfnamefont {Ronny}\ \bibnamefont {Thomale}}, \bibinfo {author}
  {\bibfnamefont {Stephen~D.}\ \bibnamefont {Wilson}}, \bibinfo {author}
  {\bibfnamefont {Andreas~P.}\ \bibnamefont {Schnyder}}, \ and\ \bibinfo
  {author} {\bibfnamefont {Ming}\ \bibnamefont {Shi}},\ }\bibfield  {title}
  {\enquote {\bibinfo {title} {Rich nature of van {H}ove singularities in
  kagome superconductor {CsV$_3$Sb$_5$}},}\ }\href {\doibase
  10.1038/s41467-022-29828-x} {\bibfield  {journal} {\bibinfo  {journal} {Nat.
  Commun.}\ }\textbf {\bibinfo {volume} {13}},\ \bibinfo {pages} {2220}
  (\bibinfo {year} {2022})}\BibitemShut {NoStop}%
\bibitem [{\citenamefont {Kang}\ \emph {et~al.}(2022)\citenamefont {Kang},
  \citenamefont {Fang}, \citenamefont {Kim}, \citenamefont {Ortiz},
  \citenamefont {Ryu}, \citenamefont {Kim}, \citenamefont {Yoo}, \citenamefont
  {Sangiovanni}, \citenamefont {Di~Sante}, \citenamefont {Park}, \citenamefont
  {Jozwiak}, \citenamefont {Bostwick}, \citenamefont {Rotenberg}, \citenamefont
  {Kaxiras}, \citenamefont {Wilson}, \citenamefont {Park},\ and\ \citenamefont
  {Comin}}]{Kang2022}%
  \BibitemOpen
  \bibfield  {author} {\bibinfo {author} {\bibfnamefont {Mingu}\ \bibnamefont
  {Kang}}, \bibinfo {author} {\bibfnamefont {Shiang}\ \bibnamefont {Fang}},
  \bibinfo {author} {\bibfnamefont {Jeong-Kyu}\ \bibnamefont {Kim}}, \bibinfo
  {author} {\bibfnamefont {Brenden~R.}\ \bibnamefont {Ortiz}}, \bibinfo
  {author} {\bibfnamefont {Sae~Hee}\ \bibnamefont {Ryu}}, \bibinfo {author}
  {\bibfnamefont {Jimin}\ \bibnamefont {Kim}}, \bibinfo {author} {\bibfnamefont
  {Jonggyu}\ \bibnamefont {Yoo}}, \bibinfo {author} {\bibfnamefont {Giorgio}\
  \bibnamefont {Sangiovanni}}, \bibinfo {author} {\bibfnamefont {Domenico}\
  \bibnamefont {Di~Sante}}, \bibinfo {author} {\bibfnamefont {Byeong-Gyu}\
  \bibnamefont {Park}}, \bibinfo {author} {\bibfnamefont {Chris}\ \bibnamefont
  {Jozwiak}}, \bibinfo {author} {\bibfnamefont {Aaron}\ \bibnamefont
  {Bostwick}}, \bibinfo {author} {\bibfnamefont {Eli}\ \bibnamefont
  {Rotenberg}}, \bibinfo {author} {\bibfnamefont {Efthimios}\ \bibnamefont
  {Kaxiras}}, \bibinfo {author} {\bibfnamefont {Stephen~D.}\ \bibnamefont
  {Wilson}}, \bibinfo {author} {\bibfnamefont {Jae-Hoon}\ \bibnamefont {Park}},
  \ and\ \bibinfo {author} {\bibfnamefont {Riccardo}\ \bibnamefont {Comin}},\
  }\bibfield  {title} {\enquote {\bibinfo {title} {Twofold van {Hove}
  singularity and origin of charge order in topological kagome superconductor
  {CsV}$_3${Sb}$_5$},}\ }\href {\doibase 10.1038/s41567-021-01451-5} {\bibfield
   {journal} {\bibinfo  {journal} {Nat. Phys.}\ }\textbf {\bibinfo {volume}
  {18}},\ \bibinfo {pages} {301--308} (\bibinfo {year} {2022})}\BibitemShut
  {NoStop}%
\bibitem [{\citenamefont {Kiesel}\ \emph {et~al.}(2013)\citenamefont {Kiesel},
  \citenamefont {Platt},\ and\ \citenamefont {Thomale}}]{Kiesel2013}%
  \BibitemOpen
  \bibfield  {author} {\bibinfo {author} {\bibfnamefont {Maximilian~L}\
  \bibnamefont {Kiesel}}, \bibinfo {author} {\bibfnamefont {Christian}\
  \bibnamefont {Platt}}, \ and\ \bibinfo {author} {\bibfnamefont {Ronny}\
  \bibnamefont {Thomale}},\ }\bibfield  {title} {\enquote {\bibinfo {title}
  {Unconventional fermi surface instabilities in the kagome hubbard model},}\
  }\href {https://link.aps.org/doi/10.1103/PhysRevLett.110.126405} {\bibfield
  {journal} {\bibinfo  {journal} {Phys. Rev. Lett.}\ }\textbf {\bibinfo
  {volume} {110}},\ \bibinfo {pages} {126405} (\bibinfo {year}
  {2013})}\BibitemShut {NoStop}%
\bibitem [{\citenamefont {Wang}\ \emph {et~al.}(2013)\citenamefont {Wang},
  \citenamefont {Li}, \citenamefont {Xiang},\ and\ \citenamefont
  {Wang}}]{Wang2013}%
  \BibitemOpen
  \bibfield  {author} {\bibinfo {author} {\bibfnamefont {Wan-Sheng}\
  \bibnamefont {Wang}}, \bibinfo {author} {\bibfnamefont {Zheng-Zhao}\
  \bibnamefont {Li}}, \bibinfo {author} {\bibfnamefont {Yuan-Yuan}\
  \bibnamefont {Xiang}}, \ and\ \bibinfo {author} {\bibfnamefont {Qiang-Hua}\
  \bibnamefont {Wang}},\ }\bibfield  {title} {\enquote {\bibinfo {title}
  {Competing electronic orders on kagome lattices at van {Hove} filling},}\
  }\href {\doibase 10.1103/PhysRevB.87.115135} {\bibfield  {journal} {\bibinfo
  {journal} {Phys. Rev. B}\ }\textbf {\bibinfo {volume} {87}},\ \bibinfo
  {pages} {115135} (\bibinfo {year} {2013})}\BibitemShut {NoStop}%
\bibitem [{\citenamefont {Zhan}\ \emph {et~al.}(2024)\citenamefont {Zhan},
  \citenamefont {Hohmann}, \citenamefont {D\"urrnagel}, \citenamefont {Fu},
  \citenamefont {Zhou}, \citenamefont {Wang}, \citenamefont {Thomale},
  \citenamefont {Wu},\ and\ \citenamefont {Hu}}]{Zhan2024}%
  \BibitemOpen
  \bibfield  {author} {\bibinfo {author} {\bibfnamefont {Jun}\ \bibnamefont
  {Zhan}}, \bibinfo {author} {\bibfnamefont {Hendrik}\ \bibnamefont {Hohmann}},
  \bibinfo {author} {\bibfnamefont {Matteo}\ \bibnamefont {D\"urrnagel}},
  \bibinfo {author} {\bibfnamefont {Ruiqing}\ \bibnamefont {Fu}}, \bibinfo
  {author} {\bibfnamefont {Sen}\ \bibnamefont {Zhou}}, \bibinfo {author}
  {\bibfnamefont {Ziqiang}\ \bibnamefont {Wang}}, \bibinfo {author}
  {\bibfnamefont {Ronny}\ \bibnamefont {Thomale}}, \bibinfo {author}
  {\bibfnamefont {Xianxin}\ \bibnamefont {Wu}}, \ and\ \bibinfo {author}
  {\bibfnamefont {Jiangping}\ \bibnamefont {Hu}},\ }\href@noop {} {\enquote
  {\bibinfo {title} {Loop current order on the kagome lattice},}\ } (\bibinfo
  {year} {2024}),\ \bibinfo {note} {submitted}\BibitemShut {NoStop}%
\bibitem [{\citenamefont {Bert}\ \emph {et~al.}(2009)\citenamefont {Bert},
  \citenamefont {Olariu}, \citenamefont {Zorko}, \citenamefont {Mendels},
  \citenamefont {Trombe}, \citenamefont {Duc}, \citenamefont {de~Vries},
  \citenamefont {Harrison}, \citenamefont {Hillier}, \citenamefont {Lord},
  \citenamefont {Amato},\ and\ \citenamefont {Baines}}]{Bert2009}%
  \BibitemOpen
  \bibfield  {author} {\bibinfo {author} {\bibfnamefont {F}~\bibnamefont
  {Bert}}, \bibinfo {author} {\bibfnamefont {A}~\bibnamefont {Olariu}},
  \bibinfo {author} {\bibfnamefont {A}~\bibnamefont {Zorko}}, \bibinfo {author}
  {\bibfnamefont {P}~\bibnamefont {Mendels}}, \bibinfo {author} {\bibfnamefont
  {J~C}\ \bibnamefont {Trombe}}, \bibinfo {author} {\bibfnamefont
  {F}~\bibnamefont {Duc}}, \bibinfo {author} {\bibfnamefont {M~A}\ \bibnamefont
  {de~Vries}}, \bibinfo {author} {\bibfnamefont {A}~\bibnamefont {Harrison}},
  \bibinfo {author} {\bibfnamefont {A~D}\ \bibnamefont {Hillier}}, \bibinfo
  {author} {\bibfnamefont {J}~\bibnamefont {Lord}}, \bibinfo {author}
  {\bibfnamefont {A}~\bibnamefont {Amato}}, \ and\ \bibinfo {author}
  {\bibfnamefont {C}~\bibnamefont {Baines}},\ }\bibfield  {title} {\enquote
  {\bibinfo {title} {Frustrated magnetism in the quantum kagome herbertsmithite
  \ce{ZnCu$_3$(OH)$_6$Cl$_2$} antiferromagnet},}\ }\href {\doibase
  10.1088/1742-6596/145/1/012004} {\bibfield  {journal} {\bibinfo  {journal}
  {Journal of Physics: Conference Series}\ }\textbf {\bibinfo {volume} {145}},\
  \bibinfo {pages} {012004} (\bibinfo {year} {2009})}\BibitemShut {NoStop}%
\bibitem [{\citenamefont {Norman}(2016)}]{Norman2016}%
  \BibitemOpen
  \bibfield  {author} {\bibinfo {author} {\bibfnamefont {M.~R.}\ \bibnamefont
  {Norman}},\ }\bibfield  {title} {\enquote {\bibinfo {title} {Colloquium:
  Herbertsmithite and the search for the quantum spin liquid},}\ }\href
  {\doibase 10.1103/RevModPhys.88.041002} {\bibfield  {journal} {\bibinfo
  {journal} {Rev. Mod. Phys.}\ }\textbf {\bibinfo {volume} {88}},\ \bibinfo
  {pages} {041002} (\bibinfo {year} {2016})}\BibitemShut {NoStop}%
\bibitem [{\citenamefont {McGinnety}(1972)}]{mcginnety1972cesium}%
  \BibitemOpen
  \bibfield  {author} {\bibinfo {author} {\bibfnamefont {John~A}\ \bibnamefont
  {McGinnety}},\ }\bibfield  {title} {\enquote {\bibinfo {title} {Cesium
  tetrachlorocuprate. structure, crystal forces, and charge distribution},}\
  }\href@noop {} {\bibfield  {journal} {\bibinfo  {journal} {Journal of the
  American Chemical Society}\ }\textbf {\bibinfo {volume} {94}},\ \bibinfo
  {pages} {8406--8413} (\bibinfo {year} {1972})}\BibitemShut {NoStop}%
\bibitem [{\citenamefont {Schlueter}\ \emph {et~al.}(1966)\citenamefont
  {Schlueter}, \citenamefont {Jacobson},\ and\ \citenamefont
  {Rundle}}]{schlueter1966redetermination}%
  \BibitemOpen
  \bibfield  {author} {\bibinfo {author} {\bibfnamefont {Albert~W}\
  \bibnamefont {Schlueter}}, \bibinfo {author} {\bibfnamefont {Robert~A}\
  \bibnamefont {Jacobson}}, \ and\ \bibinfo {author} {\bibfnamefont {Robert~E}\
  \bibnamefont {Rundle}},\ }\bibfield  {title} {\enquote {\bibinfo {title} {A
  redetermination of the crystal structure of \ce{CsCuCl3}},}\ }\href@noop {}
  {\bibfield  {journal} {\bibinfo  {journal} {Inorganic Chemistry}\ }\textbf
  {\bibinfo {volume} {5}},\ \bibinfo {pages} {277--280} (\bibinfo {year}
  {1966})}\BibitemShut {NoStop}%
\bibitem [{\citenamefont {Hull}\ and\ \citenamefont
  {Berastegui}(2004)}]{hull2004crystal}%
  \BibitemOpen
  \bibfield  {author} {\bibinfo {author} {\bibfnamefont {S.}~\bibnamefont
  {Hull}}\ and\ \bibinfo {author} {\bibfnamefont {P.}~\bibnamefont
  {Berastegui}},\ }\bibfield  {title} {\enquote {\bibinfo {title} {{Crystal
  structures and ionic conductivities of ternary derivatives of the silver and
  copper monohalides-II: ordered phases within the (AgX)$_x$--(MX)$_{1-x}$ and
  (CuX)$_x$--(MX)$_{1-x}$ (M = K, Rb and Cs; X = Cl, Br and I) systems}},}\
  }\href@noop {} {\bibfield  {journal} {\bibinfo  {journal} {Journal of Solid
  State Chemistry}\ }\textbf {\bibinfo {volume} {177}},\ \bibinfo {pages}
  {3156--3173} (\bibinfo {year} {2004})}\BibitemShut {NoStop}%
\bibitem [{\citenamefont {Shannon}(1976)}]{shannon1976revised}%
  \BibitemOpen
  \bibfield  {author} {\bibinfo {author} {\bibfnamefont {Robert~D}\
  \bibnamefont {Shannon}},\ }\bibfield  {title} {\enquote {\bibinfo {title}
  {Revised effective ionic radii and systematic studies of interatomic
  distances in halides and chalcogenides},}\ }\href@noop {} {\bibfield
  {journal} {\bibinfo  {journal} {Foundations of Crystallography}\ }\textbf
  {\bibinfo {volume} {32}},\ \bibinfo {pages} {751--767} (\bibinfo {year}
  {1976})}\BibitemShut {NoStop}%
\bibitem [{\citenamefont {Veidis}\ \emph {et~al.}(1969)\citenamefont {Veidis},
  \citenamefont {Schreiber}, \citenamefont {Gough},\ and\ \citenamefont
  {Palenik}}]{veidis1969jahn}%
  \BibitemOpen
  \bibfield  {author} {\bibinfo {author} {\bibfnamefont {MV}~\bibnamefont
  {Veidis}}, \bibinfo {author} {\bibfnamefont {GH}~\bibnamefont {Schreiber}},
  \bibinfo {author} {\bibfnamefont {TE}~\bibnamefont {Gough}}, \ and\ \bibinfo
  {author} {\bibfnamefont {Gus~J}\ \bibnamefont {Palenik}},\ }\bibfield
  {title} {\enquote {\bibinfo {title} {Jahn-teller distortions in octahedral
  copper {(II)} complexes},}\ }\href@noop {} {\bibfield  {journal} {\bibinfo
  {journal} {Journal of the American Chemical Society}\ }\textbf {\bibinfo
  {volume} {91}},\ \bibinfo {pages} {1859--1860} (\bibinfo {year}
  {1969})}\BibitemShut {NoStop}%
\bibitem [{\citenamefont {Werhahn}\ \emph {et~al.}(2022)\citenamefont
  {Werhahn}, \citenamefont {Ortiz}, \citenamefont {Hay}, \citenamefont
  {Wilson}, \citenamefont {Seshadri},\ and\ \citenamefont
  {Johrendt}}]{Werhahn2022}%
  \BibitemOpen
  \bibfield  {author} {\bibinfo {author} {\bibfnamefont {Dominik}\ \bibnamefont
  {Werhahn}}, \bibinfo {author} {\bibfnamefont {Brenden~R.}\ \bibnamefont
  {Ortiz}}, \bibinfo {author} {\bibfnamefont {Aurland~K.}\ \bibnamefont {Hay}},
  \bibinfo {author} {\bibfnamefont {Stephen~D.}\ \bibnamefont {Wilson}},
  \bibinfo {author} {\bibfnamefont {Ram}\ \bibnamefont {Seshadri}}, \ and\
  \bibinfo {author} {\bibfnamefont {Dirk}\ \bibnamefont {Johrendt}},\
  }\bibfield  {title} {\enquote {\bibinfo {title} {The kagom\'e metals
  \ce{RbTi3Bi5} and \ce{CsTi3Bi5}},}\ }\href {\doibase
  doi:10.1515/znb-2022-0125} {\bibfield  {journal} {\bibinfo  {journal}
  {Zeitschrift f\"ur Naturforschung B}\ }\textbf {\bibinfo {volume} {77}},\
  \bibinfo {pages} {757--764} (\bibinfo {year} {2022})}\BibitemShut {NoStop}%
\bibitem [{\citenamefont {Yang}\ \emph {et~al.}(2023)\citenamefont {Yang},
  \citenamefont {Yi}, \citenamefont {Zhao}, \citenamefont {Xie}, \citenamefont
  {Miao}, \citenamefont {Luo}, \citenamefont {Chen}, \citenamefont {Liang},
  \citenamefont {Zhu}, \citenamefont {Ye}, \citenamefont {You}, \citenamefont
  {Gu}, \citenamefont {Zhang}, \citenamefont {Zhang}, \citenamefont {Yang},
  \citenamefont {Wang}, \citenamefont {Peng}, \citenamefont {Mao},
  \citenamefont {Liu}, \citenamefont {Xu}, \citenamefont {Chen}, \citenamefont
  {Yang}, \citenamefont {Su}, \citenamefont {Gao}, \citenamefont {Zhao},\ and\
  \citenamefont {Zhou}}]{Yang2023}%
  \BibitemOpen
  \bibfield  {author} {\bibinfo {author} {\bibfnamefont {Jiangang}\
  \bibnamefont {Yang}}, \bibinfo {author} {\bibfnamefont {Xinwei}\ \bibnamefont
  {Yi}}, \bibinfo {author} {\bibfnamefont {Zhen}\ \bibnamefont {Zhao}},
  \bibinfo {author} {\bibfnamefont {Yuyang}\ \bibnamefont {Xie}}, \bibinfo
  {author} {\bibfnamefont {Taimin}\ \bibnamefont {Miao}}, \bibinfo {author}
  {\bibfnamefont {Hailan}\ \bibnamefont {Luo}}, \bibinfo {author}
  {\bibfnamefont {Hao}\ \bibnamefont {Chen}}, \bibinfo {author} {\bibfnamefont
  {Bo}~\bibnamefont {Liang}}, \bibinfo {author} {\bibfnamefont {Wenpei}\
  \bibnamefont {Zhu}}, \bibinfo {author} {\bibfnamefont {Yuhan}\ \bibnamefont
  {Ye}}, \bibinfo {author} {\bibfnamefont {Jing-Yang}\ \bibnamefont {You}},
  \bibinfo {author} {\bibfnamefont {Bo}~\bibnamefont {Gu}}, \bibinfo {author}
  {\bibfnamefont {Shenjin}\ \bibnamefont {Zhang}}, \bibinfo {author}
  {\bibfnamefont {Fengfeng}\ \bibnamefont {Zhang}}, \bibinfo {author}
  {\bibfnamefont {Feng}\ \bibnamefont {Yang}}, \bibinfo {author} {\bibfnamefont
  {Zhimin}\ \bibnamefont {Wang}}, \bibinfo {author} {\bibfnamefont {Qinjun}\
  \bibnamefont {Peng}}, \bibinfo {author} {\bibfnamefont {Hanqing}\
  \bibnamefont {Mao}}, \bibinfo {author} {\bibfnamefont {Guodong}\ \bibnamefont
  {Liu}}, \bibinfo {author} {\bibfnamefont {Zuyan}\ \bibnamefont {Xu}},
  \bibinfo {author} {\bibfnamefont {Hui}\ \bibnamefont {Chen}}, \bibinfo
  {author} {\bibfnamefont {Haitao}\ \bibnamefont {Yang}}, \bibinfo {author}
  {\bibfnamefont {Gang}\ \bibnamefont {Su}}, \bibinfo {author} {\bibfnamefont
  {Hongjun}\ \bibnamefont {Gao}}, \bibinfo {author} {\bibfnamefont {Lin}\
  \bibnamefont {Zhao}}, \ and\ \bibinfo {author} {\bibfnamefont {X.~J.}\
  \bibnamefont {Zhou}},\ }\bibfield  {title} {\enquote {\bibinfo {title}
  {Observation of flat band, {Dirac} nodal lines and topological surface states
  in kagome superconductor {CsTi}$_3${Bi}$_5$},}\ }\href {\doibase
  10.1038/s41467-023-39620-0} {\bibfield  {journal} {\bibinfo  {journal} {Nat.
  Commun.}\ }\textbf {\bibinfo {volume} {14}},\ \bibinfo {pages} {4089}
  (\bibinfo {year} {2023})}\BibitemShut {NoStop}%
\bibitem [{\citenamefont {Bigi}\ \emph {et~al.}(2024)\citenamefont {Bigi},
  \citenamefont {D\"urrnagel}, \citenamefont {Klebl}, \citenamefont
  {Consiglio}, \citenamefont {Pokharel}, \citenamefont {Bertran}, \citenamefont
  {F\'evre}, \citenamefont {Jaouen}, \citenamefont {Tchouekem}, \citenamefont
  {Turban}, \citenamefont {Vita}, \citenamefont {Miwa}, \citenamefont {Wells},
  \citenamefont {Oh}, \citenamefont {Comin}, \citenamefont {Thomale},
  \citenamefont {Zeljkovic}, \citenamefont {Ortiz}, \citenamefont {Wilson},
  \citenamefont {Sangiovanni}, \citenamefont {Mazzola},\ and\ \citenamefont
  {Sante}}]{Bigi2024p}%
  \BibitemOpen
  \bibfield  {author} {\bibinfo {author} {\bibfnamefont {Chiara}\ \bibnamefont
  {Bigi}}, \bibinfo {author} {\bibfnamefont {Matteo}\ \bibnamefont
  {D\"urrnagel}}, \bibinfo {author} {\bibfnamefont {Lennart}\ \bibnamefont
  {Klebl}}, \bibinfo {author} {\bibfnamefont {Armando}\ \bibnamefont
  {Consiglio}}, \bibinfo {author} {\bibfnamefont {Ganesh}\ \bibnamefont
  {Pokharel}}, \bibinfo {author} {\bibfnamefont {Francois}\ \bibnamefont
  {Bertran}}, \bibinfo {author} {\bibfnamefont {Patrick~Le}\ \bibnamefont
  {F\'evre}}, \bibinfo {author} {\bibfnamefont {Thomas}\ \bibnamefont
  {Jaouen}}, \bibinfo {author} {\bibfnamefont {Hulerich~C.}\ \bibnamefont
  {Tchouekem}}, \bibinfo {author} {\bibfnamefont {Pascal}\ \bibnamefont
  {Turban}}, \bibinfo {author} {\bibfnamefont {Alessandro~De}\ \bibnamefont
  {Vita}}, \bibinfo {author} {\bibfnamefont {Jill~A.}\ \bibnamefont {Miwa}},
  \bibinfo {author} {\bibfnamefont {Justin~W.}\ \bibnamefont {Wells}}, \bibinfo
  {author} {\bibfnamefont {Dongjin}\ \bibnamefont {Oh}}, \bibinfo {author}
  {\bibfnamefont {Riccardo}\ \bibnamefont {Comin}}, \bibinfo {author}
  {\bibfnamefont {Ronny}\ \bibnamefont {Thomale}}, \bibinfo {author}
  {\bibfnamefont {Ilija}\ \bibnamefont {Zeljkovic}}, \bibinfo {author}
  {\bibfnamefont {Brenden~R.}\ \bibnamefont {Ortiz}}, \bibinfo {author}
  {\bibfnamefont {Stephen~D.}\ \bibnamefont {Wilson}}, \bibinfo {author}
  {\bibfnamefont {Giorgio}\ \bibnamefont {Sangiovanni}}, \bibinfo {author}
  {\bibfnamefont {Federico}\ \bibnamefont {Mazzola}}, \ and\ \bibinfo {author}
  {\bibfnamefont {Domenico~Di}\ \bibnamefont {Sante}},\ }\href
  {https://arxiv.org/abs/2410.22929} {\enquote {\bibinfo {title} {Pomeranchuk
  instability from electronic correlations in csti$_3$bi$_5$ kagome metal},}\ }
  (\bibinfo {year} {2024}),\ \Eprint {http://arxiv.org/abs/2410.22929}
  {arXiv:2410.22929 [cond-mat.str-el]} \BibitemShut {NoStop}%
\bibitem [{\citenamefont {Ortiz}\ \emph {et~al.}(2021)\citenamefont {Ortiz},
  \citenamefont {Teicher}, \citenamefont {Kautzsch}, \citenamefont {Sarte},
  \citenamefont {Ratcliff}, \citenamefont {Harter}, \citenamefont {Ruff},
  \citenamefont {Seshadri},\ and\ \citenamefont {Wilson}}]{Ortiz2021}%
  \BibitemOpen
  \bibfield  {author} {\bibinfo {author} {\bibfnamefont {Brenden~R.}\
  \bibnamefont {Ortiz}}, \bibinfo {author} {\bibfnamefont {Samuel M.~L.}\
  \bibnamefont {Teicher}}, \bibinfo {author} {\bibfnamefont {Linus}\
  \bibnamefont {Kautzsch}}, \bibinfo {author} {\bibfnamefont {Paul~M.}\
  \bibnamefont {Sarte}}, \bibinfo {author} {\bibfnamefont {Noah}\ \bibnamefont
  {Ratcliff}}, \bibinfo {author} {\bibfnamefont {John}\ \bibnamefont {Harter}},
  \bibinfo {author} {\bibfnamefont {Jacob P.~C.}\ \bibnamefont {Ruff}},
  \bibinfo {author} {\bibfnamefont {Ram}\ \bibnamefont {Seshadri}}, \ and\
  \bibinfo {author} {\bibfnamefont {Stephen~D.}\ \bibnamefont {Wilson}},\
  }\bibfield  {title} {\enquote {\bibinfo {title} {Fermi surface mapping and
  the nature of charge-density-wave order in the kagome superconductor
  ${\mathrm{csv}}_{3}{\mathrm{sb}}_{5}$},}\ }\href {\doibase
  10.1103/PhysRevX.11.041030} {\bibfield  {journal} {\bibinfo  {journal} {Phys.
  Rev. X}\ }\textbf {\bibinfo {volume} {11}},\ \bibinfo {pages} {041030}
  (\bibinfo {year} {2021})}\BibitemShut {NoStop}%
\bibitem [{\citenamefont {He}\ \emph {et~al.}(2024)\citenamefont {He},
  \citenamefont {Peis}, \citenamefont {Cuddy}, \citenamefont {Zhao},
  \citenamefont {Li}, \citenamefont {Zhang}, \citenamefont {Stumberger},
  \citenamefont {Moritz}, \citenamefont {Yang}, \citenamefont {Gao},
  \citenamefont {Devereaux},\ and\ \citenamefont {Hackl}}]{He2024}%
  \BibitemOpen
  \bibfield  {author} {\bibinfo {author} {\bibfnamefont {Ge}~\bibnamefont
  {He}}, \bibinfo {author} {\bibfnamefont {Leander}\ \bibnamefont {Peis}},
  \bibinfo {author} {\bibfnamefont {Emma~Frances}\ \bibnamefont {Cuddy}},
  \bibinfo {author} {\bibfnamefont {Zhen}\ \bibnamefont {Zhao}}, \bibinfo
  {author} {\bibfnamefont {Dong}\ \bibnamefont {Li}}, \bibinfo {author}
  {\bibfnamefont {Yuhang}\ \bibnamefont {Zhang}}, \bibinfo {author}
  {\bibfnamefont {Romona}\ \bibnamefont {Stumberger}}, \bibinfo {author}
  {\bibfnamefont {Brian}\ \bibnamefont {Moritz}}, \bibinfo {author}
  {\bibfnamefont {Haitao}\ \bibnamefont {Yang}}, \bibinfo {author}
  {\bibfnamefont {Hongjun}\ \bibnamefont {Gao}}, \bibinfo {author}
  {\bibfnamefont {Thomas~Peter}\ \bibnamefont {Devereaux}}, \ and\ \bibinfo
  {author} {\bibfnamefont {Rudi}\ \bibnamefont {Hackl}},\ }\bibfield  {title}
  {\enquote {\bibinfo {title} {Anharmonic strong-coupling effects at the origin
  of the charge density wave in {CsV}$_3${Sb}$_5$},}\ }\href {\doibase
  10.1038/s41467-024-45865-0} {\bibfield  {journal} {\bibinfo  {journal} {Nat.
  Commun.}\ }\textbf {\bibinfo {volume} {15}},\ \bibinfo {pages} {1895}
  (\bibinfo {year} {2024})}\BibitemShut {NoStop}%
\bibitem [{\citenamefont {Aryasetiawan}\ \emph {et~al.}(2004)\citenamefont
  {Aryasetiawan}, \citenamefont {Imada}, \citenamefont {Georges}, \citenamefont
  {Kotliar}, \citenamefont {Biermann},\ and\ \citenamefont
  {Lichtenstein}}]{PhysRevB.70.195104}%
  \BibitemOpen
  \bibfield  {author} {\bibinfo {author} {\bibfnamefont {F.}~\bibnamefont
  {Aryasetiawan}}, \bibinfo {author} {\bibfnamefont {M.}~\bibnamefont {Imada}},
  \bibinfo {author} {\bibfnamefont {A.}~\bibnamefont {Georges}}, \bibinfo
  {author} {\bibfnamefont {G.}~\bibnamefont {Kotliar}}, \bibinfo {author}
  {\bibfnamefont {S.}~\bibnamefont {Biermann}}, \ and\ \bibinfo {author}
  {\bibfnamefont {A.~I.}\ \bibnamefont {Lichtenstein}},\ }\bibfield  {title}
  {\enquote {\bibinfo {title} {Frequency-dependent local interactions and
  low-energy effective models from electronic structure calculations},}\ }\href
  {\doibase 10.1103/PhysRevB.70.195104} {\bibfield  {journal} {\bibinfo
  {journal} {Phys. Rev. B}\ }\textbf {\bibinfo {volume} {70}},\ \bibinfo
  {pages} {195104} (\bibinfo {year} {2004})}\BibitemShut {NoStop}%
\bibitem [{\citenamefont {Miyake}\ \emph {et~al.}(2009)\citenamefont {Miyake},
  \citenamefont {Aryasetiawan},\ and\ \citenamefont
  {Imada}}]{PhysRevB.80.155134}%
  \BibitemOpen
  \bibfield  {author} {\bibinfo {author} {\bibfnamefont {Takashi}\ \bibnamefont
  {Miyake}}, \bibinfo {author} {\bibfnamefont {Ferdi}\ \bibnamefont
  {Aryasetiawan}}, \ and\ \bibinfo {author} {\bibfnamefont {Masatoshi}\
  \bibnamefont {Imada}},\ }\bibfield  {title} {\enquote {\bibinfo {title} {Ab
  initio procedure for constructing effective models of correlated materials
  with entangled band structure},}\ }\href {\doibase
  10.1103/PhysRevB.80.155134} {\bibfield  {journal} {\bibinfo  {journal} {Phys.
  Rev. B}\ }\textbf {\bibinfo {volume} {80}},\ \bibinfo {pages} {155134}
  (\bibinfo {year} {2009})}\BibitemShut {NoStop}%
\bibitem [{\citenamefont {Vaugier}\ \emph {et~al.}(2012)\citenamefont
  {Vaugier}, \citenamefont {Jiang},\ and\ \citenamefont
  {Biermann}}]{PhysRevB.86.165105}%
  \BibitemOpen
  \bibfield  {author} {\bibinfo {author} {\bibfnamefont {Lo\"{\i}g}\
  \bibnamefont {Vaugier}}, \bibinfo {author} {\bibfnamefont {Hong}\
  \bibnamefont {Jiang}}, \ and\ \bibinfo {author} {\bibfnamefont {Silke}\
  \bibnamefont {Biermann}},\ }\bibfield  {title} {\enquote {\bibinfo {title}
  {Hubbard $u$ and hund exchange $j$ in transition metal oxides: Screening
  versus localization trends from constrained random phase approximation},}\
  }\href {\doibase 10.1103/PhysRevB.86.165105} {\bibfield  {journal} {\bibinfo
  {journal} {Phys. Rev. B}\ }\textbf {\bibinfo {volume} {86}},\ \bibinfo
  {pages} {165105} (\bibinfo {year} {2012})}\BibitemShut {NoStop}%
\bibitem [{\citenamefont {Di~Sante}\ \emph {et~al.}(2023)\citenamefont
  {Di~Sante}, \citenamefont {Kim}, \citenamefont {Hanke}, \citenamefont
  {Wehling}, \citenamefont {Franchini}, \citenamefont {Thomale},\ and\
  \citenamefont {Sangiovanni}}]{DiSante2023}%
  \BibitemOpen
  \bibfield  {author} {\bibinfo {author} {\bibfnamefont {Domenico}\
  \bibnamefont {Di~Sante}}, \bibinfo {author} {\bibfnamefont {Bongjae}\
  \bibnamefont {Kim}}, \bibinfo {author} {\bibfnamefont {Werner}\ \bibnamefont
  {Hanke}}, \bibinfo {author} {\bibfnamefont {Tim}\ \bibnamefont {Wehling}},
  \bibinfo {author} {\bibfnamefont {Cesare}\ \bibnamefont {Franchini}},
  \bibinfo {author} {\bibfnamefont {Ronny}\ \bibnamefont {Thomale}}, \ and\
  \bibinfo {author} {\bibfnamefont {Giorgio}\ \bibnamefont {Sangiovanni}},\
  }\bibfield  {title} {\enquote {\bibinfo {title} {Electronic correlations and
  universal long-range scaling in kagome metals},}\ }\href {\doibase
  10.1103/PhysRevResearch.5.L012008} {\bibfield  {journal} {\bibinfo  {journal}
  {Phys. Rev. Res.}\ }\textbf {\bibinfo {volume} {5}},\ \bibinfo {pages}
  {L012008} (\bibinfo {year} {2023})}\BibitemShut {NoStop}%
\bibitem [{\citenamefont {Mazin}\ \emph {et~al.}(2014)\citenamefont {Mazin},
  \citenamefont {Jeschke}, \citenamefont {Lechermann}, \citenamefont {Lee},
  \citenamefont {Fink}, \citenamefont {Thomale},\ and\ \citenamefont
  {Valent{\'\i}}}]{Mazin2014}%
  \BibitemOpen
  \bibfield  {author} {\bibinfo {author} {\bibfnamefont {I.~I.}\ \bibnamefont
  {Mazin}}, \bibinfo {author} {\bibfnamefont {Harald~O.}\ \bibnamefont
  {Jeschke}}, \bibinfo {author} {\bibfnamefont {Frank}\ \bibnamefont
  {Lechermann}}, \bibinfo {author} {\bibfnamefont {Hunpyo}\ \bibnamefont
  {Lee}}, \bibinfo {author} {\bibfnamefont {Mario}\ \bibnamefont {Fink}},
  \bibinfo {author} {\bibfnamefont {Ronny}\ \bibnamefont {Thomale}}, \ and\
  \bibinfo {author} {\bibfnamefont {Roser}\ \bibnamefont {Valent{\'\i}}},\
  }\bibfield  {title} {\enquote {\bibinfo {title} {Theoretical prediction of a
  strongly correlated {D}irac metal},}\ }\href {\doibase 10.1038/ncomms5261}
  {\bibfield  {journal} {\bibinfo  {journal} {Nat. Commun.}\ }\textbf {\bibinfo
  {volume} {5}},\ \bibinfo {pages} {4261} (\bibinfo {year} {2014})}\BibitemShut
  {NoStop}%
\bibitem [{\citenamefont {Jovanovic}\ and\ \citenamefont
  {Schoop}(2022)}]{Schoop2022}%
  \BibitemOpen
  \bibfield  {author} {\bibinfo {author} {\bibfnamefont {Milena}\ \bibnamefont
  {Jovanovic}}\ and\ \bibinfo {author} {\bibfnamefont {Leslie~M}\ \bibnamefont
  {Schoop}},\ }\bibfield  {title} {\enquote {\bibinfo {title} {Simple chemical
  rules for predicting band structures of kagome materials},}\ }\href@noop {}
  {\bibfield  {journal} {\bibinfo  {journal} {Journal of the American Chemical
  Society}\ }\textbf {\bibinfo {volume} {144}},\ \bibinfo {pages}
  {10978--10991} (\bibinfo {year} {2022})}\BibitemShut {NoStop}%
\bibitem [{\citenamefont {Ortiz}\ \emph {et~al.}(2020)\citenamefont {Ortiz},
  \citenamefont {Teicher}, \citenamefont {Hu}, \citenamefont {Zuo},
  \citenamefont {Sarte}, \citenamefont {Schueller}, \citenamefont {Abeykoon},
  \citenamefont {Krogstad}, \citenamefont {Rosenkranz}, \citenamefont {Osborn},
  \citenamefont {Seshadri}, \citenamefont {Balents}, \citenamefont {He},\ and\
  \citenamefont {Wilson}}]{OrtizWilson}%
  \BibitemOpen
  \bibfield  {author} {\bibinfo {author} {\bibfnamefont {Brenden~R.}\
  \bibnamefont {Ortiz}}, \bibinfo {author} {\bibfnamefont {Samuel M.~L.}\
  \bibnamefont {Teicher}}, \bibinfo {author} {\bibfnamefont {Yong}\
  \bibnamefont {Hu}}, \bibinfo {author} {\bibfnamefont {Julia~L.}\ \bibnamefont
  {Zuo}}, \bibinfo {author} {\bibfnamefont {Paul~M.}\ \bibnamefont {Sarte}},
  \bibinfo {author} {\bibfnamefont {Emily~C.}\ \bibnamefont {Schueller}},
  \bibinfo {author} {\bibfnamefont {A.~M.~Milinda}\ \bibnamefont {Abeykoon}},
  \bibinfo {author} {\bibfnamefont {Matthew~J.}\ \bibnamefont {Krogstad}},
  \bibinfo {author} {\bibfnamefont {Stephan}\ \bibnamefont {Rosenkranz}},
  \bibinfo {author} {\bibfnamefont {Raymond}\ \bibnamefont {Osborn}}, \bibinfo
  {author} {\bibfnamefont {Ram}\ \bibnamefont {Seshadri}}, \bibinfo {author}
  {\bibfnamefont {Leon}\ \bibnamefont {Balents}}, \bibinfo {author}
  {\bibfnamefont {Junfeng}\ \bibnamefont {He}}, \ and\ \bibinfo {author}
  {\bibfnamefont {Stephen~D.}\ \bibnamefont {Wilson}},\ }\bibfield  {title}
  {\enquote {\bibinfo {title} {Cs{V$_3$}{Sb$_5$}: A {${\mathbb{Z}}_{2}$}
  topological kagome metal with a superconducting ground state},}\ }\href
  {\doibase 10.1103/PhysRevLett.125.247002} {\bibfield  {journal} {\bibinfo
  {journal} {Phys. Rev. Lett.}\ }\textbf {\bibinfo {volume} {125}},\ \bibinfo
  {pages} {247002} (\bibinfo {year} {2020})}\BibitemShut {NoStop}%
\bibitem [{\citenamefont {Liu}\ \emph {et~al.}(2024)\citenamefont {Liu},
  \citenamefont {Liu}, \citenamefont {Bao}, \citenamefont {Yang}, \citenamefont
  {Ji}, \citenamefont {Wu}, \citenamefont {Shen}, \citenamefont {Luo},
  \citenamefont {Yang}, \citenamefont {Liu}, \citenamefont {Xu}, \citenamefont
  {Yang}, \citenamefont {Chai}, \citenamefont {Lu}, \citenamefont {Liu},
  \citenamefont {Wang}, \citenamefont {Jiang}, \citenamefont {Tao},
  \citenamefont {Ren}, \citenamefont {Xu}, \citenamefont {Cao}, \citenamefont
  {Xu}, \citenamefont {Zhou}, \citenamefont {Cheng},\ and\ \citenamefont
  {Cao}}]{Liu2024a}%
  \BibitemOpen
  \bibfield  {author} {\bibinfo {author} {\bibfnamefont {Yi}~\bibnamefont
  {Liu}}, \bibinfo {author} {\bibfnamefont {Zi-Yi}\ \bibnamefont {Liu}},
  \bibinfo {author} {\bibfnamefont {Jin-Ke}\ \bibnamefont {Bao}}, \bibinfo
  {author} {\bibfnamefont {Peng-Tao}\ \bibnamefont {Yang}}, \bibinfo {author}
  {\bibfnamefont {Liang-Wen}\ \bibnamefont {Ji}}, \bibinfo {author}
  {\bibfnamefont {Si-Qi}\ \bibnamefont {Wu}}, \bibinfo {author} {\bibfnamefont
  {Qin-Xin}\ \bibnamefont {Shen}}, \bibinfo {author} {\bibfnamefont {Jun}\
  \bibnamefont {Luo}}, \bibinfo {author} {\bibfnamefont {Jie}\ \bibnamefont
  {Yang}}, \bibinfo {author} {\bibfnamefont {Ji-Yong}\ \bibnamefont {Liu}},
  \bibinfo {author} {\bibfnamefont {Chen-Chao}\ \bibnamefont {Xu}}, \bibinfo
  {author} {\bibfnamefont {Wu-Zhang}\ \bibnamefont {Yang}}, \bibinfo {author}
  {\bibfnamefont {Wan-Li}\ \bibnamefont {Chai}}, \bibinfo {author}
  {\bibfnamefont {Jia-Yi}\ \bibnamefont {Lu}}, \bibinfo {author} {\bibfnamefont
  {Chang-Chao}\ \bibnamefont {Liu}}, \bibinfo {author} {\bibfnamefont {Bo-Sen}\
  \bibnamefont {Wang}}, \bibinfo {author} {\bibfnamefont {Hao}\ \bibnamefont
  {Jiang}}, \bibinfo {author} {\bibfnamefont {Qian}\ \bibnamefont {Tao}},
  \bibinfo {author} {\bibfnamefont {Zhi}\ \bibnamefont {Ren}}, \bibinfo
  {author} {\bibfnamefont {Xiao-Feng}\ \bibnamefont {Xu}}, \bibinfo {author}
  {\bibfnamefont {Chao}\ \bibnamefont {Cao}}, \bibinfo {author} {\bibfnamefont
  {Zhu-An}\ \bibnamefont {Xu}}, \bibinfo {author} {\bibfnamefont {Rui}\
  \bibnamefont {Zhou}}, \bibinfo {author} {\bibfnamefont {Jin-Guang}\
  \bibnamefont {Cheng}}, \ and\ \bibinfo {author} {\bibfnamefont {Guang-Han}\
  \bibnamefont {Cao}},\ }\bibfield  {title} {\enquote {\bibinfo {title}
  {Superconductivity under pressure in a chromium-based kagome metal},}\ }\href
  {\doibase 10.1038/s41586-024-07761-x} {\bibfield  {journal} {\bibinfo
  {journal} {Nature}\ }\textbf {\bibinfo {volume} {632}},\ \bibinfo {pages}
  {1032--1037} (\bibinfo {year} {2024})}\BibitemShut {NoStop}%
\bibitem [{\citenamefont {{Sangiovanni}}(2024)}]{Sangiovanni2024}%
  \BibitemOpen
  \bibfield  {author} {\bibinfo {author} {\bibfnamefont {Giorgio}\ \bibnamefont
  {{Sangiovanni}}},\ }\bibfield  {title} {\enquote {\bibinfo {title}
  {{Superconductor surprises with strongly interacting electrons}},}\ }\href
  {\doibase 10.1038/d41586-024-02559-3} {\bibfield  {journal} {\bibinfo
  {journal} {Nature}\ }\textbf {\bibinfo {volume} {632}},\ \bibinfo {pages}
  {988--989} (\bibinfo {year} {2024})}\BibitemShut {NoStop}%
\bibitem [{\citenamefont {Li}\ \emph {et~al.}(2025)\citenamefont {Li},
  \citenamefont {Liu}, \citenamefont {Du}, \citenamefont {Wu}, \citenamefont
  {Zhao}, \citenamefont {Zhai}, \citenamefont {Hu}, \citenamefont {Zhang},
  \citenamefont {Chen}, \citenamefont {Liu}, \citenamefont {Yang},
  \citenamefont {Peng}, \citenamefont {Hashimoto}, \citenamefont {Lu},
  \citenamefont {Liu}, \citenamefont {Wang}, \citenamefont {Chen},
  \citenamefont {Cao},\ and\ \citenamefont {Yang}}]{Li2025e}%
  \BibitemOpen
  \bibfield  {author} {\bibinfo {author} {\bibfnamefont {Yidian}\ \bibnamefont
  {Li}}, \bibinfo {author} {\bibfnamefont {Yi}~\bibnamefont {Liu}}, \bibinfo
  {author} {\bibfnamefont {Xian}\ \bibnamefont {Du}}, \bibinfo {author}
  {\bibfnamefont {Siqi}\ \bibnamefont {Wu}}, \bibinfo {author} {\bibfnamefont
  {Wenxuan}\ \bibnamefont {Zhao}}, \bibinfo {author} {\bibfnamefont {Kaiyi}\
  \bibnamefont {Zhai}}, \bibinfo {author} {\bibfnamefont {Yinqi}\ \bibnamefont
  {Hu}}, \bibinfo {author} {\bibfnamefont {Senyao}\ \bibnamefont {Zhang}},
  \bibinfo {author} {\bibfnamefont {Houke}\ \bibnamefont {Chen}}, \bibinfo
  {author} {\bibfnamefont {Jieyi}\ \bibnamefont {Liu}}, \bibinfo {author}
  {\bibfnamefont {Yiheng}\ \bibnamefont {Yang}}, \bibinfo {author}
  {\bibfnamefont {Cheng}\ \bibnamefont {Peng}}, \bibinfo {author}
  {\bibfnamefont {Makoto}\ \bibnamefont {Hashimoto}}, \bibinfo {author}
  {\bibfnamefont {Donghui}\ \bibnamefont {Lu}}, \bibinfo {author}
  {\bibfnamefont {Zhongkai}\ \bibnamefont {Liu}}, \bibinfo {author}
  {\bibfnamefont {Yilin}\ \bibnamefont {Wang}}, \bibinfo {author}
  {\bibfnamefont {Yulin}\ \bibnamefont {Chen}}, \bibinfo {author}
  {\bibfnamefont {Guanghan}\ \bibnamefont {Cao}}, \ and\ \bibinfo {author}
  {\bibfnamefont {Lexian}\ \bibnamefont {Yang}},\ }\bibfield  {title} {\enquote
  {\bibinfo {title} {Electron correlation and incipient flat bands in the
  kagome superconductor cscr3sb5},}\ }\href {\doibase
  10.1038/s41467-025-58487-x} {\bibfield  {journal} {\bibinfo  {journal}
  {Nature Communications}\ }\textbf {\bibinfo {volume} {16}},\ \bibinfo {pages}
  {3229} (\bibinfo {year} {2025})}\BibitemShut {NoStop}%
\bibitem [{\citenamefont {Wu}\ \emph {et~al.}(2025)\citenamefont {Wu},
  \citenamefont {Xu}, \citenamefont {Wang}, \citenamefont {Lin}, \citenamefont
  {Cao},\ and\ \citenamefont {Cao}}]{Wu2025f}%
  \BibitemOpen
  \bibfield  {author} {\bibinfo {author} {\bibfnamefont {Siqi}\ \bibnamefont
  {Wu}}, \bibinfo {author} {\bibfnamefont {Chenchao}\ \bibnamefont {Xu}},
  \bibinfo {author} {\bibfnamefont {Xiaoqun}\ \bibnamefont {Wang}}, \bibinfo
  {author} {\bibfnamefont {Hai-Qing}\ \bibnamefont {Lin}}, \bibinfo {author}
  {\bibfnamefont {Chao}\ \bibnamefont {Cao}}, \ and\ \bibinfo {author}
  {\bibfnamefont {Guang-Han}\ \bibnamefont {Cao}},\ }\bibfield  {title}
  {\enquote {\bibinfo {title} {Flat-band enhanced antiferromagnetic
  fluctuations and superconductivity in pressurized cscr3sb5},}\ }\href
  {\doibase 10.1038/s41467-025-56582-7} {\bibfield  {journal} {\bibinfo
  {journal} {Nature Communications}\ }\textbf {\bibinfo {volume} {16}},\
  \bibinfo {pages} {1375} (\bibinfo {year} {2025})}\BibitemShut {NoStop}%
\bibitem [{\citenamefont {Xie}\ \emph {et~al.}(2025)\citenamefont {Xie},
  \citenamefont {Fang}, \citenamefont {Li}, \citenamefont {Huang},
  \citenamefont {Chen}, \citenamefont {Setty}, \citenamefont {Sur},
  \citenamefont {Yakobson}, \citenamefont {Valent\'{\i}},\ and\ \citenamefont
  {Si}}]{Xie2025e}%
  \BibitemOpen
  \bibfield  {author} {\bibinfo {author} {\bibfnamefont {Fang}\ \bibnamefont
  {Xie}}, \bibinfo {author} {\bibfnamefont {Yuan}\ \bibnamefont {Fang}},
  \bibinfo {author} {\bibfnamefont {Ying}\ \bibnamefont {Li}}, \bibinfo
  {author} {\bibfnamefont {Yuefei}\ \bibnamefont {Huang}}, \bibinfo {author}
  {\bibfnamefont {Lei}\ \bibnamefont {Chen}}, \bibinfo {author} {\bibfnamefont
  {Chandan}\ \bibnamefont {Setty}}, \bibinfo {author} {\bibfnamefont {Shouvik}\
  \bibnamefont {Sur}}, \bibinfo {author} {\bibfnamefont {Boris}\ \bibnamefont
  {Yakobson}}, \bibinfo {author} {\bibfnamefont {Roser}\ \bibnamefont
  {Valent\'{\i}}}, \ and\ \bibinfo {author} {\bibfnamefont {Qimiao}\
  \bibnamefont {Si}},\ }\bibfield  {title} {\enquote {\bibinfo {title}
  {Electron correlations in the kagome flat band metal
  ${\mathrm{cscr}}_{3}{\mathrm{sb}}_{5}$},}\ }\href {\doibase
  10.1103/PhysRevResearch.7.L022061} {\bibfield  {journal} {\bibinfo  {journal}
  {Phys. Rev. Res.}\ }\textbf {\bibinfo {volume} {7}},\ \bibinfo {pages}
  {L022061} (\bibinfo {year} {2025})}\BibitemShut {NoStop}%
\bibitem [{\citenamefont {Wang}(2025)}]{Wang2025h}%
  \BibitemOpen
  \bibfield  {author} {\bibinfo {author} {\bibfnamefont {Yilin}\ \bibnamefont
  {Wang}},\ }\bibfield  {title} {\enquote {\bibinfo {title} {Heavy fermions in
  frustrated hund's metal with portions of incipient flat bands},}\ }\href
  {\doibase 10.1103/PhysRevB.111.035127} {\bibfield  {journal} {\bibinfo
  {journal} {Phys. Rev. B}\ }\textbf {\bibinfo {volume} {111}},\ \bibinfo
  {pages} {035127} (\bibinfo {year} {2025})}\BibitemShut {NoStop}%
\bibitem [{\citenamefont {Denner}\ \emph {et~al.}(2021)\citenamefont {Denner},
  \citenamefont {Thomale},\ and\ \citenamefont {Neupert}}]{Neupert2021}%
  \BibitemOpen
  \bibfield  {author} {\bibinfo {author} {\bibfnamefont {M.~Michael}\
  \bibnamefont {Denner}}, \bibinfo {author} {\bibfnamefont {Ronny}\
  \bibnamefont {Thomale}}, \ and\ \bibinfo {author} {\bibfnamefont {Titus}\
  \bibnamefont {Neupert}},\ }\bibfield  {title} {\enquote {\bibinfo {title}
  {Analysis of charge order in the kagome metal
  {$A{\mathrm{V}}_{3}{\mathrm{Sb}}_{5}$
  ($A=\mathrm{K},\mathrm{Rb},\mathrm{Cs}$)}},}\ }\href {\doibase
  10.1103/PhysRevLett.127.217601} {\bibfield  {journal} {\bibinfo  {journal}
  {Phys. Rev. Lett.}\ }\textbf {\bibinfo {volume} {127}},\ \bibinfo {pages}
  {217601} (\bibinfo {year} {2021})}\BibitemShut {NoStop}%
\bibitem [{\citenamefont {Profe}\ \emph
  {et~al.}(2024{\natexlab{a}})\citenamefont {Profe}, \citenamefont {Klebl},
  \citenamefont {Grandi}, \citenamefont {Hohmann}, \citenamefont
  {D{\"u}rrnagel}, \citenamefont {Schwemmer}, \citenamefont {Thomale},\ and\
  \citenamefont {Kennes}}]{Profe2024}%
  \BibitemOpen
  \bibfield  {author} {\bibinfo {author} {\bibfnamefont {Jonas~B.}\
  \bibnamefont {Profe}}, \bibinfo {author} {\bibfnamefont {Lennart}\
  \bibnamefont {Klebl}}, \bibinfo {author} {\bibfnamefont {Francesco}\
  \bibnamefont {Grandi}}, \bibinfo {author} {\bibfnamefont {Hendrik}\
  \bibnamefont {Hohmann}}, \bibinfo {author} {\bibfnamefont {Matteo}\
  \bibnamefont {D{\"u}rrnagel}}, \bibinfo {author} {\bibfnamefont {Tilman}\
  \bibnamefont {Schwemmer}}, \bibinfo {author} {\bibfnamefont {Ronny}\
  \bibnamefont {Thomale}}, \ and\ \bibinfo {author} {\bibfnamefont {Dante~M.}\
  \bibnamefont {Kennes}},\ }\href@noop {} {\enquote {\bibinfo {title} {The
  kagome hubbard model from a functional renormalization group perspective},}\
  } (\bibinfo {year} {2024}{\natexlab{a}}),\ \Eprint
  {http://arxiv.org/abs/2402.11916} {arXiv:2402.11916 [cond-mat.str-el]}
  \BibitemShut {NoStop}%
\bibitem [{\citenamefont {Wang}\ \emph {et~al.}(2023)\citenamefont {Wang},
  \citenamefont {Wu}, \citenamefont {McCandless}, \citenamefont {Chan},\ and\
  \citenamefont {Ali}}]{wang2023quantum}%
  \BibitemOpen
  \bibfield  {author} {\bibinfo {author} {\bibfnamefont {Yaojia}\ \bibnamefont
  {Wang}}, \bibinfo {author} {\bibfnamefont {Heng}\ \bibnamefont {Wu}},
  \bibinfo {author} {\bibfnamefont {Gregory~T}\ \bibnamefont {McCandless}},
  \bibinfo {author} {\bibfnamefont {Julia~Y}\ \bibnamefont {Chan}}, \ and\
  \bibinfo {author} {\bibfnamefont {Mazhar~N}\ \bibnamefont {Ali}},\ }\bibfield
   {title} {\enquote {\bibinfo {title} {Quantum states and intertwining phases
  in kagome materials},}\ }\href
  {https://www.nature.com/articles/s41586-023-06363-3} {\bibfield  {journal}
  {\bibinfo  {journal} {Nature Reviews Physics}\ }\textbf {\bibinfo {volume}
  {5}},\ \bibinfo {pages} {635--658} (\bibinfo {year} {2023})}\BibitemShut
  {NoStop}%
\bibitem [{\citenamefont {Enzner}\ \emph {et~al.}(2025)\citenamefont {Enzner},
  \citenamefont {Berges}, \citenamefont {Schobert}, \citenamefont {Oh},
  \citenamefont {Kang}, \citenamefont {Comin}, \citenamefont {Thomale},
  \citenamefont {Wehling}, \citenamefont {Sante},\ and\ \citenamefont
  {Sangiovanni}}]{Enzner2025p}%
  \BibitemOpen
  \bibfield  {author} {\bibinfo {author} {\bibfnamefont {Stefan}\ \bibnamefont
  {Enzner}}, \bibinfo {author} {\bibfnamefont {Jan}\ \bibnamefont {Berges}},
  \bibinfo {author} {\bibfnamefont {Arne}\ \bibnamefont {Schobert}}, \bibinfo
  {author} {\bibfnamefont {Dongjin}\ \bibnamefont {Oh}}, \bibinfo {author}
  {\bibfnamefont {Mingu}\ \bibnamefont {Kang}}, \bibinfo {author}
  {\bibfnamefont {Riccardo}\ \bibnamefont {Comin}}, \bibinfo {author}
  {\bibfnamefont {Ronny}\ \bibnamefont {Thomale}}, \bibinfo {author}
  {\bibfnamefont {Tim~O.}\ \bibnamefont {Wehling}}, \bibinfo {author}
  {\bibfnamefont {Domenico~Di}\ \bibnamefont {Sante}}, \ and\ \bibinfo {author}
  {\bibfnamefont {Giorgio}\ \bibnamefont {Sangiovanni}},\ }\href
  {https://arxiv.org/abs/2504.07883} {\enquote {\bibinfo {title} {Phonon
  fluctuation diagnostics: Origin of charge order in av$_3$sb$_5$ kagome
  metals},}\ } (\bibinfo {year} {2025}),\ \Eprint
  {http://arxiv.org/abs/2504.07883} {arXiv:2504.07883 [cond-mat.str-el]}
  \BibitemShut {NoStop}%
\bibitem [{\citenamefont {Wu}\ \emph {et~al.}(2021{\natexlab{a}})\citenamefont
  {Wu}, \citenamefont {Schwemmer}, \citenamefont {M{\"u}ller}, \citenamefont
  {Consiglio}, \citenamefont {Sangiovanni}, \citenamefont {Di~Sante},
  \citenamefont {Iqbal}, \citenamefont {Hanke}, \citenamefont {Schnyder},
  \citenamefont {Denner} \emph {et~al.}}]{wu2021nature}%
  \BibitemOpen
  \bibfield  {author} {\bibinfo {author} {\bibfnamefont {Xianxin}\ \bibnamefont
  {Wu}}, \bibinfo {author} {\bibfnamefont {Tilman}\ \bibnamefont {Schwemmer}},
  \bibinfo {author} {\bibfnamefont {Tobias}\ \bibnamefont {M{\"u}ller}},
  \bibinfo {author} {\bibfnamefont {Armando}\ \bibnamefont {Consiglio}},
  \bibinfo {author} {\bibfnamefont {Giorgio}\ \bibnamefont {Sangiovanni}},
  \bibinfo {author} {\bibfnamefont {Domenico}\ \bibnamefont {Di~Sante}},
  \bibinfo {author} {\bibfnamefont {Yasir}\ \bibnamefont {Iqbal}}, \bibinfo
  {author} {\bibfnamefont {Werner}\ \bibnamefont {Hanke}}, \bibinfo {author}
  {\bibfnamefont {Andreas~P}\ \bibnamefont {Schnyder}}, \bibinfo {author}
  {\bibfnamefont {M~Michael}\ \bibnamefont {Denner}},  \emph {et~al.},\ }\href
  {https://journals.aps.org/prl/abstract/10.1103/PhysRevLett.127.177001}
  {\bibfield  {journal} {\bibinfo  {journal} {Phys. Rev. Lett.}\ }\textbf
  {\bibinfo {volume} {127}},\ \bibinfo {pages} {177001} (\bibinfo {year}
  {2021}{\natexlab{a}})}\BibitemShut {NoStop}%
\bibitem [{\citenamefont {Yang}\ \emph {et~al.}(2024)\citenamefont {Yang},
  \citenamefont {Yao}, \citenamefont {Wang},\ and\ \citenamefont
  {Wang}}]{Yang2024}%
  \BibitemOpen
  \bibfield  {author} {\bibinfo {author} {\bibfnamefont {Qing-Geng}\
  \bibnamefont {Yang}}, \bibinfo {author} {\bibfnamefont {Meng}\ \bibnamefont
  {Yao}}, \bibinfo {author} {\bibfnamefont {Da}~\bibnamefont {Wang}}, \ and\
  \bibinfo {author} {\bibfnamefont {Qiang-Hua}\ \bibnamefont {Wang}},\
  }\bibfield  {title} {\enquote {\bibinfo {title} {Charge bond order and
  $s$-wave superconductivity in the kagome lattice with electron-phonon
  coupling and electron-electron interaction},}\ }\href {\doibase
  10.1103/PhysRevB.109.075130} {\bibfield  {journal} {\bibinfo  {journal}
  {Phys. Rev. B}\ }\textbf {\bibinfo {volume} {109}},\ \bibinfo {pages}
  {075130} (\bibinfo {year} {2024})}\BibitemShut {NoStop}%
\bibitem [{\citenamefont {Profe}\ \emph
  {et~al.}(2024{\natexlab{b}})\citenamefont {Profe}, \citenamefont {Kennes},\
  and\ \citenamefont {Klebl}}]{Profe2024a}%
  \BibitemOpen
  \bibfield  {author} {\bibinfo {author} {\bibfnamefont {Jonas~B.}\
  \bibnamefont {Profe}}, \bibinfo {author} {\bibfnamefont {Dante~M.}\
  \bibnamefont {Kennes}}, \ and\ \bibinfo {author} {\bibfnamefont {Lennart}\
  \bibnamefont {Klebl}},\ }\bibfield  {title} {\enquote {\bibinfo {title}
  {{divERGe implements various Exact Renormalization Group examples}},}\ }\href
  {\doibase 10.21468/SciPostPhysCodeb.26} {\bibfield  {journal} {\bibinfo
  {journal} {SciPost Phys. Codebases}\ ,\ \bibinfo {pages} {26}} (\bibinfo
  {year} {2024}{\natexlab{b}})}\BibitemShut {NoStop}%
\bibitem [{\citenamefont {Metzner}\ \emph {et~al.}(2012)\citenamefont
  {Metzner}, \citenamefont {Salmhofer}, \citenamefont {Honerkamp},
  \citenamefont {Meden},\ and\ \citenamefont {Sch\"onhammer}}]{Metzner2012}%
  \BibitemOpen
  \bibfield  {author} {\bibinfo {author} {\bibfnamefont {Walter}\ \bibnamefont
  {Metzner}}, \bibinfo {author} {\bibfnamefont {Manfred}\ \bibnamefont
  {Salmhofer}}, \bibinfo {author} {\bibfnamefont {Carsten}\ \bibnamefont
  {Honerkamp}}, \bibinfo {author} {\bibfnamefont {Volker}\ \bibnamefont
  {Meden}}, \ and\ \bibinfo {author} {\bibfnamefont {Kurt}\ \bibnamefont
  {Sch\"onhammer}},\ }\bibfield  {title} {\enquote {\bibinfo {title}
  {Functional renormalization group approach to correlated fermion systems},}\
  }\href {\doibase 10.1103/RevModPhys.84.299} {\bibfield  {journal} {\bibinfo
  {journal} {Rev. Mod. Phys.}\ }\textbf {\bibinfo {volume} {84}},\ \bibinfo
  {pages} {299--352} (\bibinfo {year} {2012})}\BibitemShut {NoStop}%
\bibitem [{\citenamefont {Platt}\ \emph {et~al.}(2013)\citenamefont {Platt},
  \citenamefont {Hanke},\ and\ \citenamefont {Thomale}}]{Platt_2013}%
  \BibitemOpen
  \bibfield  {author} {\bibinfo {author} {\bibfnamefont {C.}~\bibnamefont
  {Platt}}, \bibinfo {author} {\bibfnamefont {W.}~\bibnamefont {Hanke}}, \ and\
  \bibinfo {author} {\bibfnamefont {R.}~\bibnamefont {Thomale}},\ }\bibfield
  {title} {\enquote {\bibinfo {title} {Functional renormalization group for
  multi-orbital fermi surface instabilities},}\ }\href {\doibase
  10.1080/00018732.2013.862020} {\bibfield  {journal} {\bibinfo  {journal}
  {Advances in Physics}\ }\textbf {\bibinfo {volume} {62}},\ \bibinfo {pages}
  {453--562} (\bibinfo {year} {2013})}\BibitemShut {NoStop}%
\bibitem [{\citenamefont {Venderbos}(2016)}]{Venderbos2016}%
  \BibitemOpen
  \bibfield  {author} {\bibinfo {author} {\bibfnamefont {J.~W.~F.}\
  \bibnamefont {Venderbos}},\ }\bibfield  {title} {\enquote {\bibinfo {title}
  {Symmetry analysis of translational symmetry broken density waves:
  Application to hexagonal lattices in two dimensions},}\ }\href {\doibase
  10.1103/PhysRevB.93.115107} {\bibfield  {journal} {\bibinfo  {journal} {Phys.
  Rev. B}\ }\textbf {\bibinfo {volume} {93}},\ \bibinfo {pages} {115107}
  (\bibinfo {year} {2016})}\BibitemShut {NoStop}%
\bibitem [{\citenamefont {Guo}\ \emph {et~al.}(2024)\citenamefont {Guo},
  \citenamefont {Wagner}, \citenamefont {Putzke}, \citenamefont {Chen},
  \citenamefont {Wang}, \citenamefont {Zhang}, \citenamefont {Gutierrez-Amigo},
  \citenamefont {Errea}, \citenamefont {Vergniory}, \citenamefont {Felser},
  \citenamefont {Fischer}, \citenamefont {Neupert},\ and\ \citenamefont
  {Moll}}]{Guo2024}%
  \BibitemOpen
  \bibfield  {author} {\bibinfo {author} {\bibfnamefont {Chunyu}\ \bibnamefont
  {Guo}}, \bibinfo {author} {\bibfnamefont {Glenn}\ \bibnamefont {Wagner}},
  \bibinfo {author} {\bibfnamefont {Carsten}\ \bibnamefont {Putzke}}, \bibinfo
  {author} {\bibfnamefont {Dong}\ \bibnamefont {Chen}}, \bibinfo {author}
  {\bibfnamefont {Kaize}\ \bibnamefont {Wang}}, \bibinfo {author}
  {\bibfnamefont {Ling}\ \bibnamefont {Zhang}}, \bibinfo {author}
  {\bibfnamefont {Martin}\ \bibnamefont {Gutierrez-Amigo}}, \bibinfo {author}
  {\bibfnamefont {Ion}\ \bibnamefont {Errea}}, \bibinfo {author} {\bibfnamefont
  {Maia~G.}\ \bibnamefont {Vergniory}}, \bibinfo {author} {\bibfnamefont
  {Claudia}\ \bibnamefont {Felser}}, \bibinfo {author} {\bibfnamefont
  {Mark~H.}\ \bibnamefont {Fischer}}, \bibinfo {author} {\bibfnamefont {Titus}\
  \bibnamefont {Neupert}}, \ and\ \bibinfo {author} {\bibfnamefont {Philip
  J.~W.}\ \bibnamefont {Moll}},\ }\bibfield  {title} {\enquote {\bibinfo
  {title} {Correlated order at the tipping point in the kagome metal
  {CsV}$_3${Sb}$_5$},}\ }\href {\doibase 10.1038/s41567-023-02374-z} {\bibfield
   {journal} {\bibinfo  {journal} {Nat. Phys.}\ }\textbf {\bibinfo {volume}
  {20}},\ \bibinfo {pages} {579--584} (\bibinfo {year} {2024})}\BibitemShut
  {NoStop}%
\bibitem [{\citenamefont {Zhao}\ \emph {et~al.}(2021)\citenamefont {Zhao},
  \citenamefont {Li}, \citenamefont {Ortiz}, \citenamefont {Teicher},
  \citenamefont {Park}, \citenamefont {Ye}, \citenamefont {Wang}, \citenamefont
  {Balents}, \citenamefont {Wilson},\ and\ \citenamefont
  {Zeljkovic}}]{Zhao2021}%
  \BibitemOpen
  \bibfield  {author} {\bibinfo {author} {\bibfnamefont {He}~\bibnamefont
  {Zhao}}, \bibinfo {author} {\bibfnamefont {Hong}\ \bibnamefont {Li}},
  \bibinfo {author} {\bibfnamefont {Brenden~R.}\ \bibnamefont {Ortiz}},
  \bibinfo {author} {\bibfnamefont {Samuel M.~L.}\ \bibnamefont {Teicher}},
  \bibinfo {author} {\bibfnamefont {Takamori}\ \bibnamefont {Park}}, \bibinfo
  {author} {\bibfnamefont {Mengxing}\ \bibnamefont {Ye}}, \bibinfo {author}
  {\bibfnamefont {Ziqiang}\ \bibnamefont {Wang}}, \bibinfo {author}
  {\bibfnamefont {Leon}\ \bibnamefont {Balents}}, \bibinfo {author}
  {\bibfnamefont {Stephen~D.}\ \bibnamefont {Wilson}}, \ and\ \bibinfo {author}
  {\bibfnamefont {Ilija}\ \bibnamefont {Zeljkovic}},\ }\bibfield  {title}
  {\enquote {\bibinfo {title} {Cascade of correlated electron states in the
  kagome superconductor {CsV}$_3${Sb}$_5$},}\ }\href {\doibase
  10.1038/s41586-021-03946-w} {\bibfield  {journal} {\bibinfo  {journal}
  {Nature}\ }\textbf {\bibinfo {volume} {599}},\ \bibinfo {pages} {216--221}
  (\bibinfo {year} {2021})}\BibitemShut {NoStop}%
\bibitem [{\citenamefont {Yin}\ \emph {et~al.}(2022{\natexlab{a}})\citenamefont
  {Yin}, \citenamefont {Jiang}, \citenamefont {Teng}, \citenamefont {Hossain},
  \citenamefont {Mardanya}, \citenamefont {Chang}, \citenamefont {Ye},
  \citenamefont {Xu}, \citenamefont {Denner}, \citenamefont {Neupert},
  \citenamefont {Lienhard}, \citenamefont {Deng}, \citenamefont {Setty},
  \citenamefont {Si}, \citenamefont {Chang}, \citenamefont {Guguchia},
  \citenamefont {Gao}, \citenamefont {Shumiya}, \citenamefont {Zhang},
  \citenamefont {Cochran}, \citenamefont {Multer}, \citenamefont {Yi},
  \citenamefont {Dai},\ and\ \citenamefont {Hasan}}]{Yin2022a}%
  \BibitemOpen
  \bibfield  {author} {\bibinfo {author} {\bibfnamefont {Jia-Xin}\ \bibnamefont
  {Yin}}, \bibinfo {author} {\bibfnamefont {Yu-Xiao}\ \bibnamefont {Jiang}},
  \bibinfo {author} {\bibfnamefont {Xiaokun}\ \bibnamefont {Teng}}, \bibinfo
  {author} {\bibfnamefont {Md.~Shafayat}\ \bibnamefont {Hossain}}, \bibinfo
  {author} {\bibfnamefont {Sougata}\ \bibnamefont {Mardanya}}, \bibinfo
  {author} {\bibfnamefont {Tay-Rong}\ \bibnamefont {Chang}}, \bibinfo {author}
  {\bibfnamefont {Zijin}\ \bibnamefont {Ye}}, \bibinfo {author} {\bibfnamefont
  {Gang}\ \bibnamefont {Xu}}, \bibinfo {author} {\bibfnamefont {M.~Michael}\
  \bibnamefont {Denner}}, \bibinfo {author} {\bibfnamefont {Titus}\
  \bibnamefont {Neupert}}, \bibinfo {author} {\bibfnamefont {Benjamin}\
  \bibnamefont {Lienhard}}, \bibinfo {author} {\bibfnamefont {Han-Bin}\
  \bibnamefont {Deng}}, \bibinfo {author} {\bibfnamefont {Chandan}\
  \bibnamefont {Setty}}, \bibinfo {author} {\bibfnamefont {Qimiao}\
  \bibnamefont {Si}}, \bibinfo {author} {\bibfnamefont {Guoqing}\ \bibnamefont
  {Chang}}, \bibinfo {author} {\bibfnamefont {Zurab}\ \bibnamefont {Guguchia}},
  \bibinfo {author} {\bibfnamefont {Bin}\ \bibnamefont {Gao}}, \bibinfo
  {author} {\bibfnamefont {Nana}\ \bibnamefont {Shumiya}}, \bibinfo {author}
  {\bibfnamefont {Qi}~\bibnamefont {Zhang}}, \bibinfo {author} {\bibfnamefont
  {Tyler~A.}\ \bibnamefont {Cochran}}, \bibinfo {author} {\bibfnamefont
  {Daniel}\ \bibnamefont {Multer}}, \bibinfo {author} {\bibfnamefont {Ming}\
  \bibnamefont {Yi}}, \bibinfo {author} {\bibfnamefont {Pengcheng}\
  \bibnamefont {Dai}}, \ and\ \bibinfo {author} {\bibfnamefont {M.~Zahid}\
  \bibnamefont {Hasan}},\ }\bibfield  {title} {\enquote {\bibinfo {title}
  {Discovery of charge order and corresponding edge state in kagome magnet
  \ce{FeGe}},}\ }\href {\doibase 10.1103/PhysRevLett.129.166401} {\bibfield
  {journal} {\bibinfo  {journal} {Phys. Rev. Lett.}\ }\textbf {\bibinfo
  {volume} {129}},\ \bibinfo {pages} {166401} (\bibinfo {year}
  {2022}{\natexlab{a}})}\BibitemShut {NoStop}%
\bibitem [{\citenamefont {Teng}\ \emph {et~al.}(2022)\citenamefont {Teng},
  \citenamefont {Chen}, \citenamefont {Ye}, \citenamefont {Rosenberg},
  \citenamefont {Liu}, \citenamefont {Yin}, \citenamefont {Jiang},
  \citenamefont {Oh}, \citenamefont {Hasan}, \citenamefont {Neubauer},
  \citenamefont {Gao}, \citenamefont {Xie}, \citenamefont {Hashimoto},
  \citenamefont {Lu}, \citenamefont {Jozwiak}, \citenamefont {Bostwick},
  \citenamefont {Rotenberg}, \citenamefont {Birgeneau}, \citenamefont {Chu},
  \citenamefont {Yi},\ and\ \citenamefont {Dai}}]{Teng2022}%
  \BibitemOpen
  \bibfield  {author} {\bibinfo {author} {\bibfnamefont {Xiaokun}\ \bibnamefont
  {Teng}}, \bibinfo {author} {\bibfnamefont {Lebing}\ \bibnamefont {Chen}},
  \bibinfo {author} {\bibfnamefont {Feng}\ \bibnamefont {Ye}}, \bibinfo
  {author} {\bibfnamefont {Elliott}\ \bibnamefont {Rosenberg}}, \bibinfo
  {author} {\bibfnamefont {Zhaoyu}\ \bibnamefont {Liu}}, \bibinfo {author}
  {\bibfnamefont {Jia-Xin}\ \bibnamefont {Yin}}, \bibinfo {author}
  {\bibfnamefont {Yu-Xiao}\ \bibnamefont {Jiang}}, \bibinfo {author}
  {\bibfnamefont {Ji~Seop}\ \bibnamefont {Oh}}, \bibinfo {author}
  {\bibfnamefont {M.~Zahid}\ \bibnamefont {Hasan}}, \bibinfo {author}
  {\bibfnamefont {Kelly~J.}\ \bibnamefont {Neubauer}}, \bibinfo {author}
  {\bibfnamefont {Bin}\ \bibnamefont {Gao}}, \bibinfo {author} {\bibfnamefont
  {Yaofeng}\ \bibnamefont {Xie}}, \bibinfo {author} {\bibfnamefont {Makoto}\
  \bibnamefont {Hashimoto}}, \bibinfo {author} {\bibfnamefont {Donghui}\
  \bibnamefont {Lu}}, \bibinfo {author} {\bibfnamefont {Chris}\ \bibnamefont
  {Jozwiak}}, \bibinfo {author} {\bibfnamefont {Aaron}\ \bibnamefont
  {Bostwick}}, \bibinfo {author} {\bibfnamefont {Eli}\ \bibnamefont
  {Rotenberg}}, \bibinfo {author} {\bibfnamefont {Robert~J.}\ \bibnamefont
  {Birgeneau}}, \bibinfo {author} {\bibfnamefont {Jiun-Haw}\ \bibnamefont
  {Chu}}, \bibinfo {author} {\bibfnamefont {Ming}\ \bibnamefont {Yi}}, \ and\
  \bibinfo {author} {\bibfnamefont {Pengcheng}\ \bibnamefont {Dai}},\
  }\bibfield  {title} {\enquote {\bibinfo {title} {Discovery of charge density
  wave in a kagome lattice antiferromagnet},}\ }\href {\doibase
  10.1038/s41586-022-05034-z} {\bibfield  {journal} {\bibinfo  {journal}
  {Nature}\ }\textbf {\bibinfo {volume} {609}},\ \bibinfo {pages} {490--495}
  (\bibinfo {year} {2022})}\BibitemShut {NoStop}%
\bibitem [{\citenamefont {Teng}\ \emph {et~al.}(2023)\citenamefont {Teng},
  \citenamefont {Oh}, \citenamefont {Tan}, \citenamefont {Chen}, \citenamefont
  {Huang}, \citenamefont {Gao}, \citenamefont {Yin}, \citenamefont {Chu},
  \citenamefont {Hashimoto}, \citenamefont {Lu}, \citenamefont {Jozwiak},
  \citenamefont {Bostwick}, \citenamefont {Rotenberg}, \citenamefont
  {Granroth}, \citenamefont {Yan}, \citenamefont {Birgeneau}, \citenamefont
  {Dai},\ and\ \citenamefont {Yi}}]{Teng2023}%
  \BibitemOpen
  \bibfield  {author} {\bibinfo {author} {\bibfnamefont {Xiaokun}\ \bibnamefont
  {Teng}}, \bibinfo {author} {\bibfnamefont {Ji~Seop}\ \bibnamefont {Oh}},
  \bibinfo {author} {\bibfnamefont {Hengxin}\ \bibnamefont {Tan}}, \bibinfo
  {author} {\bibfnamefont {Lebing}\ \bibnamefont {Chen}}, \bibinfo {author}
  {\bibfnamefont {Jianwei}\ \bibnamefont {Huang}}, \bibinfo {author}
  {\bibfnamefont {Bin}\ \bibnamefont {Gao}}, \bibinfo {author} {\bibfnamefont
  {Jia-Xin}\ \bibnamefont {Yin}}, \bibinfo {author} {\bibfnamefont {Jiun-Haw}\
  \bibnamefont {Chu}}, \bibinfo {author} {\bibfnamefont {Makoto}\ \bibnamefont
  {Hashimoto}}, \bibinfo {author} {\bibfnamefont {Donghui}\ \bibnamefont {Lu}},
  \bibinfo {author} {\bibfnamefont {Chris}\ \bibnamefont {Jozwiak}}, \bibinfo
  {author} {\bibfnamefont {Aaron}\ \bibnamefont {Bostwick}}, \bibinfo {author}
  {\bibfnamefont {Eli}\ \bibnamefont {Rotenberg}}, \bibinfo {author}
  {\bibfnamefont {Garrett~E.}\ \bibnamefont {Granroth}}, \bibinfo {author}
  {\bibfnamefont {Binghai}\ \bibnamefont {Yan}}, \bibinfo {author}
  {\bibfnamefont {Robert~J.}\ \bibnamefont {Birgeneau}}, \bibinfo {author}
  {\bibfnamefont {Pengcheng}\ \bibnamefont {Dai}}, \ and\ \bibinfo {author}
  {\bibfnamefont {Ming}\ \bibnamefont {Yi}},\ }\bibfield  {title} {\enquote
  {\bibinfo {title} {Magnetism and charge density wave order in kagome
  \ce{FeGe}},}\ }\href {\doibase 10.1038/s41567-023-01985-w} {\bibfield
  {journal} {\bibinfo  {journal} {Nat. Phys.}\ }\textbf {\bibinfo {volume}
  {19}},\ \bibinfo {pages} {814--822} (\bibinfo {year} {2023})}\BibitemShut
  {NoStop}%
\bibitem [{\citenamefont {Han}\ \emph {et~al.}(2024)\citenamefont {Han},
  \citenamefont {Li}, \citenamefont {Tang}, \citenamefont {Wang}, \citenamefont
  {Zhang}, \citenamefont {Diao}, \citenamefont {Zhao}, \citenamefont {Sun},
  \citenamefont {Tian}, \citenamefont {Breese}, \citenamefont {Cai},
  \citenamefont {Milosevic}, \citenamefont {Qi}, \citenamefont {Wee},\ and\
  \citenamefont {Yin}}]{Han2024}%
  \BibitemOpen
  \bibfield  {author} {\bibinfo {author} {\bibfnamefont {Shulun}\ \bibnamefont
  {Han}}, \bibinfo {author} {\bibfnamefont {Linyang}\ \bibnamefont {Li}},
  \bibinfo {author} {\bibfnamefont {Chi~Sin}\ \bibnamefont {Tang}}, \bibinfo
  {author} {\bibfnamefont {Qi}~\bibnamefont {Wang}}, \bibinfo {author}
  {\bibfnamefont {Lingfeng}\ \bibnamefont {Zhang}}, \bibinfo {author}
  {\bibfnamefont {Caozheng}\ \bibnamefont {Diao}}, \bibinfo {author}
  {\bibfnamefont {Mingwen}\ \bibnamefont {Zhao}}, \bibinfo {author}
  {\bibfnamefont {Shuo}\ \bibnamefont {Sun}}, \bibinfo {author} {\bibfnamefont
  {Lijun}\ \bibnamefont {Tian}}, \bibinfo {author} {\bibfnamefont {Mark B.~H.}\
  \bibnamefont {Breese}}, \bibinfo {author} {\bibfnamefont {Chuanbing}\
  \bibnamefont {Cai}}, \bibinfo {author} {\bibfnamefont {Milorad~V.}\
  \bibnamefont {Milosevic}}, \bibinfo {author} {\bibfnamefont {Yanpeng}\
  \bibnamefont {Qi}}, \bibinfo {author} {\bibfnamefont {Andrew T.~S.}\
  \bibnamefont {Wee}}, \ and\ \bibinfo {author} {\bibfnamefont {Xinmao}\
  \bibnamefont {Yin}},\ }\href {https://arxiv.org/abs/2407.01076} {\enquote
  {\bibinfo {title} {Orbital origin of magnetic moment enhancement induced by
  charge density wave in kagome \ce{FeGe}},}\ } (\bibinfo {year} {2024}),\
  \Eprint {http://arxiv.org/abs/2407.01076} {arXiv:2407.01076
  [cond-mat.str-el]} \BibitemShut {NoStop}%
\bibitem [{\citenamefont {Teng}\ \emph {et~al.}(2024)\citenamefont {Teng},
  \citenamefont {Tam}, \citenamefont {Chen}, \citenamefont {Tan}, \citenamefont
  {Xie}, \citenamefont {Gao}, \citenamefont {Granroth}, \citenamefont {Ivanov},
  \citenamefont {Bourges}, \citenamefont {Yan}, \citenamefont {Yi},\ and\
  \citenamefont {Dai}}]{Teng2024}%
  \BibitemOpen
  \bibfield  {author} {\bibinfo {author} {\bibfnamefont {Xiaokun}\ \bibnamefont
  {Teng}}, \bibinfo {author} {\bibfnamefont {David~W.}\ \bibnamefont {Tam}},
  \bibinfo {author} {\bibfnamefont {Lebing}\ \bibnamefont {Chen}}, \bibinfo
  {author} {\bibfnamefont {Hengxin}\ \bibnamefont {Tan}}, \bibinfo {author}
  {\bibfnamefont {Yaofeng}\ \bibnamefont {Xie}}, \bibinfo {author}
  {\bibfnamefont {Bin}\ \bibnamefont {Gao}}, \bibinfo {author} {\bibfnamefont
  {Garrett~E.}\ \bibnamefont {Granroth}}, \bibinfo {author} {\bibfnamefont
  {Alexandre}\ \bibnamefont {Ivanov}}, \bibinfo {author} {\bibfnamefont
  {Philippe}\ \bibnamefont {Bourges}}, \bibinfo {author} {\bibfnamefont
  {Binghai}\ \bibnamefont {Yan}}, \bibinfo {author} {\bibfnamefont {Ming}\
  \bibnamefont {Yi}}, \ and\ \bibinfo {author} {\bibfnamefont {Pengcheng}\
  \bibnamefont {Dai}},\ }\href@noop {} {\enquote {\bibinfo {title}
  {Spin-charge-lattice coupling across the charge density wave transition in a
  kagome lattice antiferromagnet},}\ } (\bibinfo {year} {2024}),\ \Eprint
  {http://arxiv.org/abs/2404.04459} {arXiv:2404.04459 [cond-mat.str-el]}
  \BibitemShut {NoStop}%
\bibitem [{\citenamefont {Wu}\ \emph {et~al.}(2021{\natexlab{b}})\citenamefont
  {Wu}, \citenamefont {Schwemmer}, \citenamefont {M\"uller}, \citenamefont
  {Consiglio}, \citenamefont {Sangiovanni}, \citenamefont {Di~Sante},
  \citenamefont {Iqbal}, \citenamefont {Hanke}, \citenamefont {Schnyder},
  \citenamefont {Denner}, \citenamefont {Fischer}, \citenamefont {Neupert},\
  and\ \citenamefont {Thomale}}]{Wu2021}%
  \BibitemOpen
  \bibfield  {author} {\bibinfo {author} {\bibfnamefont {Xianxin}\ \bibnamefont
  {Wu}}, \bibinfo {author} {\bibfnamefont {Tilman}\ \bibnamefont {Schwemmer}},
  \bibinfo {author} {\bibfnamefont {Tobias}\ \bibnamefont {M\"uller}}, \bibinfo
  {author} {\bibfnamefont {Armando}\ \bibnamefont {Consiglio}}, \bibinfo
  {author} {\bibfnamefont {Giorgio}\ \bibnamefont {Sangiovanni}}, \bibinfo
  {author} {\bibfnamefont {Domenico}\ \bibnamefont {Di~Sante}}, \bibinfo
  {author} {\bibfnamefont {Yasir}\ \bibnamefont {Iqbal}}, \bibinfo {author}
  {\bibfnamefont {Werner}\ \bibnamefont {Hanke}}, \bibinfo {author}
  {\bibfnamefont {Andreas~P.}\ \bibnamefont {Schnyder}}, \bibinfo {author}
  {\bibfnamefont {M.~Michael}\ \bibnamefont {Denner}}, \bibinfo {author}
  {\bibfnamefont {Mark~H.}\ \bibnamefont {Fischer}}, \bibinfo {author}
  {\bibfnamefont {Titus}\ \bibnamefont {Neupert}}, \ and\ \bibinfo {author}
  {\bibfnamefont {Ronny}\ \bibnamefont {Thomale}},\ }\bibfield  {title}
  {\enquote {\bibinfo {title} {Nature of unconventional pairing in the kagome
  superconductors {$A{\mathrm{V}}_{3}{\mathrm{Sb}}_{5}$
  ($A=\mathrm{K},\mathrm{Rb},\mathrm{Cs}$)}},}\ }\href {\doibase
  10.1103/PhysRevLett.127.177001} {\bibfield  {journal} {\bibinfo  {journal}
  {Phys. Rev. Lett.}\ }\textbf {\bibinfo {volume} {127}},\ \bibinfo {pages}
  {177001} (\bibinfo {year} {2021}{\natexlab{b}})}\BibitemShut {NoStop}%
\bibitem [{\citenamefont {Li}\ \emph {et~al.}(2024)\citenamefont {Li},
  \citenamefont {Kim},\ and\ \citenamefont {Kee}}]{Li2024a}%
  \BibitemOpen
  \bibfield  {author} {\bibinfo {author} {\bibfnamefont {Heqiu}\ \bibnamefont
  {Li}}, \bibinfo {author} {\bibfnamefont {Yong~Baek}\ \bibnamefont {Kim}}, \
  and\ \bibinfo {author} {\bibfnamefont {Hae-Young}\ \bibnamefont {Kee}},\
  }\bibfield  {title} {\enquote {\bibinfo {title} {Intertwined van {Hove}
  singularities as a mechanism for loop current order in kagome metals},}\
  }\href {\doibase 10.1103/PhysRevLett.132.146501} {\bibfield  {journal}
  {\bibinfo  {journal} {Phys. Rev. Lett.}\ }\textbf {\bibinfo {volume} {132}},\
  \bibinfo {pages} {146501} (\bibinfo {year} {2024})}\BibitemShut {NoStop}%
\bibitem [{\citenamefont {Scammell}\ \emph {et~al.}(2023)\citenamefont
  {Scammell}, \citenamefont {Ingham}, \citenamefont {Li},\ and\ \citenamefont
  {Sushkov}}]{Scammell2023}%
  \BibitemOpen
  \bibfield  {author} {\bibinfo {author} {\bibfnamefont {Harley~D.}\
  \bibnamefont {Scammell}}, \bibinfo {author} {\bibfnamefont {Julian}\
  \bibnamefont {Ingham}}, \bibinfo {author} {\bibfnamefont {Tommy}\
  \bibnamefont {Li}}, \ and\ \bibinfo {author} {\bibfnamefont {Oleg~P}\
  \bibnamefont {Sushkov}},\ }\bibfield  {title} {\enquote {\bibinfo {title}
  {Chiral excitonic order from twofold van hove singularities in kagome
  metals},}\ }\href {https://www.nature.com/articles/s41467-023-35987-2}
  {\bibfield  {journal} {\bibinfo  {journal} {Nat. Commun.}\ }\textbf {\bibinfo
  {volume} {14}},\ \bibinfo {pages} {605} (\bibinfo {year} {2023})}\BibitemShut
  {NoStop}%
\bibitem [{\citenamefont {Ingham}\ \emph {et~al.}(2024)\citenamefont {Ingham},
  \citenamefont {Consiglio}, \citenamefont {di~Sante}, \citenamefont
  {Thomale},\ and\ \citenamefont {Scammell}}]{Ingham2024}%
  \BibitemOpen
  \bibfield  {author} {\bibinfo {author} {\bibfnamefont {Julian}\ \bibnamefont
  {Ingham}}, \bibinfo {author} {\bibfnamefont {Armando}\ \bibnamefont
  {Consiglio}}, \bibinfo {author} {\bibfnamefont {Domenico}\ \bibnamefont
  {di~Sante}}, \bibinfo {author} {\bibfnamefont {Ronny}\ \bibnamefont
  {Thomale}}, \ and\ \bibinfo {author} {\bibfnamefont {Harley~D.}\ \bibnamefont
  {Scammell}},\ }\bibfield  {title} {\enquote {\bibinfo {title} {Theory of
  excitonic order in scv $ \_6 $ sn $ \_6$},}\ }\href
  {https://arxiv.org/abs/2410.16365} {\bibfield  {journal} {\bibinfo  {journal}
  {arXiv preprint arXiv:2410.16365}\ } (\bibinfo {year} {2024})}\BibitemShut
  {NoStop}%
\bibitem [{\citenamefont {Yin}\ \emph {et~al.}(2022{\natexlab{b}})\citenamefont
  {Yin}, \citenamefont {Lian},\ and\ \citenamefont {Hasan}}]{Yin2022}%
  \BibitemOpen
  \bibfield  {author} {\bibinfo {author} {\bibfnamefont {Jia-Xin}\ \bibnamefont
  {Yin}}, \bibinfo {author} {\bibfnamefont {Biao}\ \bibnamefont {Lian}}, \ and\
  \bibinfo {author} {\bibfnamefont {M.~Zahid}\ \bibnamefont {Hasan}},\
  }\bibfield  {title} {\enquote {\bibinfo {title} {Topological kagome magnets
  and superconductors},}\ }\href {\doibase 10.1038/s41586-022-05516-0}
  {\bibfield  {journal} {\bibinfo  {journal} {Nature}\ }\textbf {\bibinfo
  {volume} {612}},\ \bibinfo {pages} {647--657} (\bibinfo {year}
  {2022}{\natexlab{b}})}\BibitemShut {NoStop}%
\bibitem [{\citenamefont {Winter}()}]{Pauling}%
  \BibitemOpen
  \bibfield  {author} {\bibinfo {author} {\bibfnamefont {Mark~J.}\ \bibnamefont
  {Winter}},\ }\href@noop {} {\enquote {\bibinfo {title} {Webelements: The
  periodic table on the www},}\ }\bibinfo {howpublished}
  {\url{https://winter.group.shef.ac.uk/webelements/chlorine/electronegativity.html}},\
  \bibinfo {note} {accessed: 16.12.23}\BibitemShut {NoStop}%
\bibitem [{\citenamefont {Streltsov}\ and\ \citenamefont
  {Khomskii}(2017)}]{Streltsov_2017}%
  \BibitemOpen
  \bibfield  {author} {\bibinfo {author} {\bibfnamefont {S~V}\ \bibnamefont
  {Streltsov}}\ and\ \bibinfo {author} {\bibfnamefont {D~I}\ \bibnamefont
  {Khomskii}},\ }\bibfield  {title} {\enquote {\bibinfo {title} {Orbital
  physics in transition metal compounds: new trends},}\ }\href {\doibase
  10.3367/UFNe.2017.08.038196} {\bibfield  {journal} {\bibinfo  {journal}
  {Physics-Uspekhi}\ }\textbf {\bibinfo {volume} {60}},\ \bibinfo {pages}
  {1121} (\bibinfo {year} {2017})}\BibitemShut {NoStop}%
\bibitem [{\citenamefont {Consiglio}\ \emph {et~al.}(2022)\citenamefont
  {Consiglio}, \citenamefont {Schwemmer}, \citenamefont {Wu}, \citenamefont
  {Hanke}, \citenamefont {Neupert}, \citenamefont {Thomale}, \citenamefont
  {Sangiovanni},\ and\ \citenamefont {Di~Sante}}]{PhysRevB.105.165146}%
  \BibitemOpen
  \bibfield  {author} {\bibinfo {author} {\bibfnamefont {Armando}\ \bibnamefont
  {Consiglio}}, \bibinfo {author} {\bibfnamefont {Tilman}\ \bibnamefont
  {Schwemmer}}, \bibinfo {author} {\bibfnamefont {Xianxin}\ \bibnamefont {Wu}},
  \bibinfo {author} {\bibfnamefont {Werner}\ \bibnamefont {Hanke}}, \bibinfo
  {author} {\bibfnamefont {Titus}\ \bibnamefont {Neupert}}, \bibinfo {author}
  {\bibfnamefont {Ronny}\ \bibnamefont {Thomale}}, \bibinfo {author}
  {\bibfnamefont {Giorgio}\ \bibnamefont {Sangiovanni}}, \ and\ \bibinfo
  {author} {\bibfnamefont {Domenico}\ \bibnamefont {Di~Sante}},\ }\bibfield
  {title} {\enquote {\bibinfo {title} {Van {Hove} tuning of
  {$A{\mathrm{V}}_{3}{\mathrm{Sb}}_{5}$} kagome metals under pressure and
  strain},}\ }\href {\doibase 10.1103/PhysRevB.105.165146} {\bibfield
  {journal} {\bibinfo  {journal} {Phys. Rev. B}\ }\textbf {\bibinfo {volume}
  {105}},\ \bibinfo {pages} {165146} (\bibinfo {year} {2022})}\BibitemShut
  {NoStop}%
\bibitem [{\citenamefont {LaBollita}\ and\ \citenamefont
  {Botana}(2021)}]{PhysRevB.104.205129}%
  \BibitemOpen
  \bibfield  {author} {\bibinfo {author} {\bibfnamefont {Harrison}\
  \bibnamefont {LaBollita}}\ and\ \bibinfo {author} {\bibfnamefont {Antia~S.}\
  \bibnamefont {Botana}},\ }\bibfield  {title} {\enquote {\bibinfo {title}
  {Tuning the van {Hove} singularities in {$A{\mathrm{V}}_{3}{\mathrm{Sb}}_{5}$
  $(A=\mathrm{K},\mathrm{Rb},\mathrm{Cs})$} via pressure and doping},}\ }\href
  {\doibase 10.1103/PhysRevB.104.205129} {\bibfield  {journal} {\bibinfo
  {journal} {Phys. Rev. B}\ }\textbf {\bibinfo {volume} {104}},\ \bibinfo
  {pages} {205129} (\bibinfo {year} {2021})}\BibitemShut {NoStop}%
\bibitem [{\citenamefont {Lin}\ \emph {et~al.}(2024)\citenamefont {Lin},
  \citenamefont {Consiglio}, \citenamefont {Forslund}, \citenamefont {Kuspert},
  \citenamefont {Denner}, \citenamefont {Lei}, \citenamefont {Louat},
  \citenamefont {Watson}, \citenamefont {Kim}, \citenamefont {Cacho},
  \citenamefont {Carbone}, \citenamefont {Leandersson}, \citenamefont {Polley},
  \citenamefont {Balasubramanian}, \citenamefont {Sante}, \citenamefont
  {Thomale}, \citenamefont {Guguchia}, \citenamefont {Sangiovanni},
  \citenamefont {Neupert},\ and\ \citenamefont {Chang}}]{lin2024giant}%
  \BibitemOpen
  \bibfield  {author} {\bibinfo {author} {\bibfnamefont {Chun}\ \bibnamefont
  {Lin}}, \bibinfo {author} {\bibfnamefont {Armando}\ \bibnamefont
  {Consiglio}}, \bibinfo {author} {\bibfnamefont {Ola~Kenji}\ \bibnamefont
  {Forslund}}, \bibinfo {author} {\bibfnamefont {Julia}\ \bibnamefont
  {Kuspert}}, \bibinfo {author} {\bibfnamefont {M.~Michael}\ \bibnamefont
  {Denner}}, \bibinfo {author} {\bibfnamefont {Hechang}\ \bibnamefont {Lei}},
  \bibinfo {author} {\bibfnamefont {Alex}\ \bibnamefont {Louat}}, \bibinfo
  {author} {\bibfnamefont {Matthew~D.}\ \bibnamefont {Watson}}, \bibinfo
  {author} {\bibfnamefont {Timur~K.}\ \bibnamefont {Kim}}, \bibinfo {author}
  {\bibfnamefont {Cephise}\ \bibnamefont {Cacho}}, \bibinfo {author}
  {\bibfnamefont {Dina}\ \bibnamefont {Carbone}}, \bibinfo {author}
  {\bibfnamefont {Mats}\ \bibnamefont {Leandersson}}, \bibinfo {author}
  {\bibfnamefont {Craig}\ \bibnamefont {Polley}}, \bibinfo {author}
  {\bibfnamefont {Thiagarajan}\ \bibnamefont {Balasubramanian}}, \bibinfo
  {author} {\bibfnamefont {Domenico~Di}\ \bibnamefont {Sante}}, \bibinfo
  {author} {\bibfnamefont {Ronny}\ \bibnamefont {Thomale}}, \bibinfo {author}
  {\bibfnamefont {Zurab}\ \bibnamefont {Guguchia}}, \bibinfo {author}
  {\bibfnamefont {Giorgio}\ \bibnamefont {Sangiovanni}}, \bibinfo {author}
  {\bibfnamefont {Titus}\ \bibnamefont {Neupert}}, \ and\ \bibinfo {author}
  {\bibfnamefont {Johan}\ \bibnamefont {Chang}},\ }\href@noop {} {\enquote
  {\bibinfo {title} {Giant strain response of charge modulation and singularity
  in a kagome superconductor},}\ } (\bibinfo {year} {2024}),\ \Eprint
  {http://arxiv.org/abs/2402.16089} {arXiv:2402.16089 [cond-mat.mtrl-sci]}
  \BibitemShut {NoStop}%
\bibitem [{\citenamefont {Tuniz}\ \emph {et~al.}(2024)\citenamefont {Tuniz},
  \citenamefont {Consiglio}, \citenamefont {Pokharel}, \citenamefont
  {Parmigiani}, \citenamefont {Neupert}, \citenamefont {Thomale}, \citenamefont
  {Sangiovanni}, \citenamefont {Wilson}, \citenamefont {Vobornik},
  \citenamefont {Salvador}, \citenamefont {Cilento}, \citenamefont {Sante},\
  and\ \citenamefont {Mazzola}}]{tuniz2024straininduced}%
  \BibitemOpen
  \bibfield  {author} {\bibinfo {author} {\bibfnamefont {Manuel}\ \bibnamefont
  {Tuniz}}, \bibinfo {author} {\bibfnamefont {Armando}\ \bibnamefont
  {Consiglio}}, \bibinfo {author} {\bibfnamefont {Ganesh}\ \bibnamefont
  {Pokharel}}, \bibinfo {author} {\bibfnamefont {Fulvio}\ \bibnamefont
  {Parmigiani}}, \bibinfo {author} {\bibfnamefont {Titus}\ \bibnamefont
  {Neupert}}, \bibinfo {author} {\bibfnamefont {Ronny}\ \bibnamefont
  {Thomale}}, \bibinfo {author} {\bibfnamefont {Giorgio}\ \bibnamefont
  {Sangiovanni}}, \bibinfo {author} {\bibfnamefont {Stephen~D.}\ \bibnamefont
  {Wilson}}, \bibinfo {author} {\bibfnamefont {Ivana}\ \bibnamefont
  {Vobornik}}, \bibinfo {author} {\bibfnamefont {Federico}\ \bibnamefont
  {Salvador}}, \bibinfo {author} {\bibfnamefont {Federico}\ \bibnamefont
  {Cilento}}, \bibinfo {author} {\bibfnamefont {Domenico~Di}\ \bibnamefont
  {Sante}}, \ and\ \bibinfo {author} {\bibfnamefont {Federico}\ \bibnamefont
  {Mazzola}},\ }\href@noop {} {\enquote {\bibinfo {title} {Strain-induced
  enhancement of the charge-density-wave in the kagome metal
  {ScV$_6$Sn$_6$}},}\ } (\bibinfo {year} {2024}),\ \Eprint
  {http://arxiv.org/abs/2403.18046} {arXiv:2403.18046 [cond-mat.str-el]}
  \BibitemShut {NoStop}%
\bibitem [{\citenamefont {Ranalli}\ \emph {et~al.}(2023)\citenamefont
  {Ranalli}, \citenamefont {Verdi}, \citenamefont {Monacelli}, \citenamefont
  {Kresse}, \citenamefont {Calandra},\ and\ \citenamefont {Franchini}}]{KTaO3}%
  \BibitemOpen
  \bibfield  {author} {\bibinfo {author} {\bibfnamefont {Luigi}\ \bibnamefont
  {Ranalli}}, \bibinfo {author} {\bibfnamefont {Carla}\ \bibnamefont {Verdi}},
  \bibinfo {author} {\bibfnamefont {Lorenzo}\ \bibnamefont {Monacelli}},
  \bibinfo {author} {\bibfnamefont {Georg}\ \bibnamefont {Kresse}}, \bibinfo
  {author} {\bibfnamefont {Matteo}\ \bibnamefont {Calandra}}, \ and\ \bibinfo
  {author} {\bibfnamefont {Cesare}\ \bibnamefont {Franchini}},\ }\bibfield
  {title} {\enquote {\bibinfo {title} {Temperature‐dependent anharmonic
  phonons in quantum paraelectric ktao3 by first principles and
  machine-learned force fields},}\ }\href {\doibase 10.1002/qute.202200131}
  {\bibfield  {journal} {\bibinfo  {journal} {Advanced Quantum Technologies}\
  }\textbf {\bibinfo {volume} {6}} (\bibinfo {year} {2023}),\
  10.1002/qute.202200131}\BibitemShut {NoStop}%
\bibitem [{\citenamefont {Kresse}\ and\ \citenamefont
  {Joubert}(1999)}]{PhysRevB.59.1758}%
  \BibitemOpen
  \bibfield  {author} {\bibinfo {author} {\bibfnamefont {G.}~\bibnamefont
  {Kresse}}\ and\ \bibinfo {author} {\bibfnamefont {D.}~\bibnamefont
  {Joubert}},\ }\bibfield  {title} {\enquote {\bibinfo {title} {From ultrasoft
  pseudopotentials to the projector augmented-wave method},}\ }\href {\doibase
  10.1103/PhysRevB.59.1758} {\bibfield  {journal} {\bibinfo  {journal} {Phys.
  Rev. B}\ }\textbf {\bibinfo {volume} {59}},\ \bibinfo {pages} {1758--1775}
  (\bibinfo {year} {1999})}\BibitemShut {NoStop}%
\bibitem [{\citenamefont {Kresse}\ and\ \citenamefont
  {Furthm\"uller}(1996)}]{PhysRevB.54.11169}%
  \BibitemOpen
  \bibfield  {author} {\bibinfo {author} {\bibfnamefont {G.}~\bibnamefont
  {Kresse}}\ and\ \bibinfo {author} {\bibfnamefont {J.}~\bibnamefont
  {Furthm\"uller}},\ }\bibfield  {title} {\enquote {\bibinfo {title} {Efficient
  iterative schemes for ab initio total-energy calculations using a plane-wave
  basis set},}\ }\href {\doibase 10.1103/PhysRevB.54.11169} {\bibfield
  {journal} {\bibinfo  {journal} {Phys. Rev. B}\ }\textbf {\bibinfo {volume}
  {54}},\ \bibinfo {pages} {11169--11186} (\bibinfo {year} {1996})}\BibitemShut
  {NoStop}%
\bibitem [{\citenamefont {Perdew}\ \emph {et~al.}(1996)\citenamefont {Perdew},
  \citenamefont {Burke},\ and\ \citenamefont
  {Ernzerhof}}]{PhysRevLett.77.3865}%
  \BibitemOpen
  \bibfield  {author} {\bibinfo {author} {\bibfnamefont {John~P.}\ \bibnamefont
  {Perdew}}, \bibinfo {author} {\bibfnamefont {Kieron}\ \bibnamefont {Burke}},
  \ and\ \bibinfo {author} {\bibfnamefont {Matthias}\ \bibnamefont
  {Ernzerhof}},\ }\bibfield  {title} {\enquote {\bibinfo {title} {Generalized
  gradient approximation made simple},}\ }\href {\doibase
  10.1103/PhysRevLett.77.3865} {\bibfield  {journal} {\bibinfo  {journal}
  {Phys. Rev. Lett.}\ }\textbf {\bibinfo {volume} {77}},\ \bibinfo {pages}
  {3865--3868} (\bibinfo {year} {1996})}\BibitemShut {NoStop}%
\bibitem [{\citenamefont {Wang}\ \emph {et~al.}(2021)\citenamefont {Wang},
  \citenamefont {Xu}, \citenamefont {Liu}, \citenamefont {Tang},\ and\
  \citenamefont {Geng}}]{WANG2021108033}%
  \BibitemOpen
  \bibfield  {author} {\bibinfo {author} {\bibfnamefont {Vei}\ \bibnamefont
  {Wang}}, \bibinfo {author} {\bibfnamefont {Nan}\ \bibnamefont {Xu}}, \bibinfo
  {author} {\bibfnamefont {Jin-Cheng}\ \bibnamefont {Liu}}, \bibinfo {author}
  {\bibfnamefont {Gang}\ \bibnamefont {Tang}}, \ and\ \bibinfo {author}
  {\bibfnamefont {Wen-Tong}\ \bibnamefont {Geng}},\ }\bibfield  {title}
  {\enquote {\bibinfo {title} {Vaspkit: A user-friendly interface facilitating
  high-throughput computing and analysis using vasp code},}\ }\href {\doibase
  https://doi.org/10.1016/j.cpc.2021.108033} {\bibfield  {journal} {\bibinfo
  {journal} {Computer Physics Communications}\ }\textbf {\bibinfo {volume}
  {267}},\ \bibinfo {pages} {108033} (\bibinfo {year} {2021})}\BibitemShut
  {NoStop}%
\bibitem [{\citenamefont {Momma}\ and\ \citenamefont {Izumi}(2008)}]{VESTA}%
  \BibitemOpen
  \bibfield  {author} {\bibinfo {author} {\bibfnamefont {Koichi}\ \bibnamefont
  {Momma}}\ and\ \bibinfo {author} {\bibfnamefont {Fujio}\ \bibnamefont
  {Izumi}},\ }\bibfield  {title} {\enquote {\bibinfo {title} {Vesta: a
  three-dimensional visualization system for electronic and structural
  analysis},}\ }\href {\doibase https://doi.org/10.1107/S0021889808012016}
  {\bibfield  {journal} {\bibinfo  {journal} {Journal of Applied
  Crystallography}\ }\textbf {\bibinfo {volume} {41}},\ \bibinfo {pages}
  {653--658} (\bibinfo {year} {2008})}\BibitemShut {NoStop}%
\bibitem [{\citenamefont {Mostofi}\ \emph {et~al.}(2008)\citenamefont
  {Mostofi}, \citenamefont {Yates}, \citenamefont {Lee}, \citenamefont {Souza},
  \citenamefont {Vanderbilt},\ and\ \citenamefont {Marzari}}]{MOSTOFI2008685}%
  \BibitemOpen
  \bibfield  {author} {\bibinfo {author} {\bibfnamefont {Arash~A.}\
  \bibnamefont {Mostofi}}, \bibinfo {author} {\bibfnamefont {Jonathan~R.}\
  \bibnamefont {Yates}}, \bibinfo {author} {\bibfnamefont {Young-Su}\
  \bibnamefont {Lee}}, \bibinfo {author} {\bibfnamefont {Ivo}\ \bibnamefont
  {Souza}}, \bibinfo {author} {\bibfnamefont {David}\ \bibnamefont
  {Vanderbilt}}, \ and\ \bibinfo {author} {\bibfnamefont {Nicola}\ \bibnamefont
  {Marzari}},\ }\bibfield  {title} {\enquote {\bibinfo {title} {wannier90: A
  tool for obtaining maximally-localised wannier functions},}\ }\href {\doibase
  https://doi.org/10.1016/j.cpc.2007.11.016} {\bibfield  {journal} {\bibinfo
  {journal} {Computer Physics Communications}\ }\textbf {\bibinfo {volume}
  {178}},\ \bibinfo {pages} {685--699} (\bibinfo {year} {2008})}\BibitemShut
  {NoStop}%
\bibitem [{nom()}]{nomad}%
  \BibitemOpen
  \href@noop {} {\enquote {\bibinfo {title} {The calculation data are
  accessable on zenodo under \url{ https://doi.org/10.5281/zenodo.17483257}.}}\
  }\BibitemShut {NoStop}%
\bibitem [{\citenamefont {Kaltak}(2015)}]{kaltak2015}%
  \BibitemOpen
  \bibfield  {author} {\bibinfo {author} {\bibfnamefont {Merzuk}\ \bibnamefont
  {Kaltak}},\ }\emph {\bibinfo {title} {Merging GW with DMFT}},\ \href
  {\doibase 10.25365/thesis.38099} {Ph.D. thesis},\ \bibinfo  {school}
  {Universit{\"a}t Wien} (\bibinfo {year} {2015})\BibitemShut {NoStop}%
\bibitem [{\citenamefont {Honerkamp}(2012)}]{Honerkamp2012e}%
  \BibitemOpen
  \bibfield  {author} {\bibinfo {author} {\bibfnamefont {Carsten}\ \bibnamefont
  {Honerkamp}},\ }\bibfield  {title} {\enquote {\bibinfo {title} {Effective
  interactions in multiband systems from constrained summations},}\ }\href
  {\doibase 10.1103/PhysRevB.85.195129} {\bibfield  {journal} {\bibinfo
  {journal} {Phys. Rev. B}\ }\textbf {\bibinfo {volume} {85}},\ \bibinfo
  {pages} {195129} (\bibinfo {year} {2012})}\BibitemShut {NoStop}%
\bibitem [{\citenamefont {Profe}\ \emph {et~al.}(2025)\citenamefont {Profe},
  \citenamefont {Vu\v{c}i\v{c}evi\'c}, \citenamefont {Stavropoulos},
  \citenamefont {R\"osner}, \citenamefont {Valent\'i},\ and\ \citenamefont
  {Klebl}}]{Profe2025e}%
  \BibitemOpen
  \bibfield  {author} {\bibinfo {author} {\bibfnamefont {Jonas~B.}\
  \bibnamefont {Profe}}, \bibinfo {author} {\bibfnamefont {Jak\v{s}a}\
  \bibnamefont {Vu\v{c}i\v{c}evi\'c}}, \bibinfo {author} {\bibfnamefont
  {P.~Peter}\ \bibnamefont {Stavropoulos}}, \bibinfo {author} {\bibfnamefont
  {Malte}\ \bibnamefont {R\"osner}}, \bibinfo {author} {\bibfnamefont {Roser}\
  \bibnamefont {Valent\'i}}, \ and\ \bibinfo {author} {\bibfnamefont {Lennart}\
  \bibnamefont {Klebl}},\ }\href {https://arxiv.org/abs/2507.16916} {\enquote
  {\bibinfo {title} {Exact downfolding and its perturbative approximation},}\ }
  (\bibinfo {year} {2025}),\ \Eprint {http://arxiv.org/abs/2507.16916}
  {arXiv:2507.16916 [cond-mat.str-el]} \BibitemShut {NoStop}%
\bibitem [{\citenamefont {Kaltak}\ \emph {et~al.}(2025)\citenamefont {Kaltak},
  \citenamefont {Hampel}, \citenamefont {Schlipf}, \citenamefont {Reddy},
  \citenamefont {Kim},\ and\ \citenamefont {Kresse}}]{Kaltak2025c}%
  \BibitemOpen
  \bibfield  {author} {\bibinfo {author} {\bibfnamefont {Merzuk}\ \bibnamefont
  {Kaltak}}, \bibinfo {author} {\bibfnamefont {Alexander}\ \bibnamefont
  {Hampel}}, \bibinfo {author} {\bibfnamefont {Martin}\ \bibnamefont
  {Schlipf}}, \bibinfo {author} {\bibfnamefont {Indukuru~Ramesh}\ \bibnamefont
  {Reddy}}, \bibinfo {author} {\bibfnamefont {Bongae}\ \bibnamefont {Kim}}, \
  and\ \bibinfo {author} {\bibfnamefont {Georg}\ \bibnamefont {Kresse}},\
  }\href {https://arxiv.org/abs/2508.15368} {\enquote {\bibinfo {title}
  {Constrained random phase approximation: the spectral method},}\ } (\bibinfo
  {year} {2025}),\ \Eprint {http://arxiv.org/abs/2508.15368} {arXiv:2508.15368
  [cond-mat.str-el]} \BibitemShut {NoStop}%
\bibitem [{\citenamefont {Lichtenstein}\ \emph {et~al.}(2017)\citenamefont
  {Lichtenstein}, \citenamefont {{S\'anchez de la Pe\~na}}, \citenamefont
  {Rohe}, \citenamefont {{Di Napoli}}, \citenamefont {Honerkamp},\ and\
  \citenamefont {Maier}}]{Lichtenstein2017h}%
  \BibitemOpen
  \bibfield  {author} {\bibinfo {author} {\bibfnamefont {J.}~\bibnamefont
  {Lichtenstein}}, \bibinfo {author} {\bibfnamefont {D.}~\bibnamefont
  {{S\'anchez de la Pe\~na}}}, \bibinfo {author} {\bibfnamefont
  {D.}~\bibnamefont {Rohe}}, \bibinfo {author} {\bibfnamefont {E.}~\bibnamefont
  {{Di Napoli}}}, \bibinfo {author} {\bibfnamefont {C.}~\bibnamefont
  {Honerkamp}}, \ and\ \bibinfo {author} {\bibfnamefont {S.A.}\ \bibnamefont
  {Maier}},\ }\bibfield  {title} {\enquote {\bibinfo {title} {High-performance
  functional renormalization group calculations for interacting fermions},}\
  }\href {\doibase https://doi.org/10.1016/j.cpc.2016.12.013} {\bibfield
  {journal} {\bibinfo  {journal} {Computer Physics Communications}\ }\textbf
  {\bibinfo {volume} {213}},\ \bibinfo {pages} {100--110} (\bibinfo {year}
  {2017})}\BibitemShut {NoStop}%
\bibitem [{\citenamefont {Beyer}\ \emph {et~al.}(2022)\citenamefont {Beyer},
  \citenamefont {Profe},\ and\ \citenamefont {Klebl}}]{Beyer2022r}%
  \BibitemOpen
  \bibfield  {author} {\bibinfo {author} {\bibfnamefont {Jacob}\ \bibnamefont
  {Beyer}}, \bibinfo {author} {\bibfnamefont {Jonas~B.}\ \bibnamefont {Profe}},
  \ and\ \bibinfo {author} {\bibfnamefont {Lennart}\ \bibnamefont {Klebl}},\
  }\bibfield  {title} {\enquote {\bibinfo {title} {Reference results for the
  momentum space functional renormalization group},}\ }\href {\doibase
  10.1140/epjb/s10051-022-00323-y} {\bibfield  {journal} {\bibinfo  {journal}
  {The European Physical Journal B}\ }\textbf {\bibinfo {volume} {95}}
  (\bibinfo {year} {2022}),\ 10.1140/epjb/s10051-022-00323-y}\BibitemShut
  {NoStop}%
\bibitem [{\citenamefont {Profe}\ and\ \citenamefont
  {Kennes}(2022)}]{Profe2022t}%
  \BibitemOpen
  \bibfield  {author} {\bibinfo {author} {\bibfnamefont {Jonas~B.}\
  \bibnamefont {Profe}}\ and\ \bibinfo {author} {\bibfnamefont {Dante~M.}\
  \bibnamefont {Kennes}},\ }\bibfield  {title} {\enquote {\bibinfo {title}
  {Tu$^2$frg: a scalable approach for truncated unity functional
  renormalization group in generic fermionic models},}\ }\href {\doibase
  10.1140/epjb/s10051-022-00316-x} {\bibfield  {journal} {\bibinfo  {journal}
  {The European Physical Journal B}\ }\textbf {\bibinfo {volume} {95}},\
  \bibinfo {pages} {60} (\bibinfo {year} {2022})}\BibitemShut {NoStop}%
\bibitem [{\citenamefont {Park}\ \emph {et~al.}(2021)\citenamefont {Park},
  \citenamefont {Ye},\ and\ \citenamefont {Balents}}]{Park2021}%
  \BibitemOpen
  \bibfield  {author} {\bibinfo {author} {\bibfnamefont {Takamori}\
  \bibnamefont {Park}}, \bibinfo {author} {\bibfnamefont {Mengxing}\
  \bibnamefont {Ye}}, \ and\ \bibinfo {author} {\bibfnamefont {Leon}\
  \bibnamefont {Balents}},\ }\bibfield  {title} {\enquote {\bibinfo {title}
  {Electronic instabilities of kagome metals: Saddle points and {Landau}
  theory},}\ }\href {\doibase 10.1103/PhysRevB.104.035142} {\bibfield
  {journal} {\bibinfo  {journal} {Phys. Rev. B}\ }\textbf {\bibinfo {volume}
  {104}},\ \bibinfo {pages} {035142} (\bibinfo {year} {2021})}\BibitemShut
  {NoStop}%
\bibitem [{\citenamefont {Nandkishore}\ \emph {et~al.}(2012)\citenamefont
  {Nandkishore}, \citenamefont {Chern},\ and\ \citenamefont
  {Chubukov}}]{Nandkishore2012}%
  \BibitemOpen
  \bibfield  {author} {\bibinfo {author} {\bibfnamefont {Rahul}\ \bibnamefont
  {Nandkishore}}, \bibinfo {author} {\bibfnamefont {Gia-Wei}\ \bibnamefont
  {Chern}}, \ and\ \bibinfo {author} {\bibfnamefont {Andrey~V.}\ \bibnamefont
  {Chubukov}},\ }\bibfield  {title} {\enquote {\bibinfo {title} {Itinerant
  half-metal spin-density-wave state on the hexagonal lattice},}\ }\href
  {\doibase 10.1103/PhysRevLett.108.227204} {\bibfield  {journal} {\bibinfo
  {journal} {Phys. Rev. Lett.}\ }\textbf {\bibinfo {volume} {108}},\ \bibinfo
  {pages} {227204} (\bibinfo {year} {2012})}\BibitemShut {NoStop}%
\bibitem [{\citenamefont {Xu}\ \emph {et~al.}(2023)\citenamefont {Xu},
  \citenamefont {Wu}, \citenamefont {Zhi}, \citenamefont {Cao}, \citenamefont
  {Dai}, \citenamefont {Cao}, \citenamefont {Wang},\ and\ \citenamefont
  {Lin}}]{Xu2023}%
  \BibitemOpen
  \bibfield  {author} {\bibinfo {author} {\bibfnamefont {Chenchao}\
  \bibnamefont {Xu}}, \bibinfo {author} {\bibfnamefont {Siqi}\ \bibnamefont
  {Wu}}, \bibinfo {author} {\bibfnamefont {Guo-Xiang}\ \bibnamefont {Zhi}},
  \bibinfo {author} {\bibfnamefont {Guanghan}\ \bibnamefont {Cao}}, \bibinfo
  {author} {\bibfnamefont {Jianhui}\ \bibnamefont {Dai}}, \bibinfo {author}
  {\bibfnamefont {Chao}\ \bibnamefont {Cao}}, \bibinfo {author} {\bibfnamefont
  {Xiaoqun}\ \bibnamefont {Wang}}, \ and\ \bibinfo {author} {\bibfnamefont
  {Hai-Qing}\ \bibnamefont {Lin}},\ }\href {https://arxiv.org/abs/2309.14812}
  {\enquote {\bibinfo {title} {Frustrated altermagnetism and charge density
  wave in kagome superconductor {CsCr}$_3${Sb}$_5$},}\ } (\bibinfo {year}
  {2023}),\ \Eprint {http://arxiv.org/abs/2309.14812} {arXiv:2309.14812
  [cond-mat.supr-con]} \BibitemShut {NoStop}%
\bibitem [{\citenamefont {Messio}\ \emph {et~al.}(2011)\citenamefont {Messio},
  \citenamefont {Lhuillier},\ and\ \citenamefont {Misguich}}]{Messio2011}%
  \BibitemOpen
  \bibfield  {author} {\bibinfo {author} {\bibfnamefont {L.}~\bibnamefont
  {Messio}}, \bibinfo {author} {\bibfnamefont {C.}~\bibnamefont {Lhuillier}}, \
  and\ \bibinfo {author} {\bibfnamefont {G.}~\bibnamefont {Misguich}},\
  }\bibfield  {title} {\enquote {\bibinfo {title} {Lattice symmetries and
  regular magnetic orders in classical frustrated antiferromagnets},}\ }\href
  {\doibase 10.1103/PhysRevB.83.184401} {\bibfield  {journal} {\bibinfo
  {journal} {Phys. Rev. B}\ }\textbf {\bibinfo {volume} {83}},\ \bibinfo
  {pages} {184401} (\bibinfo {year} {2011})}\BibitemShut {NoStop}%
\end{thebibliography}
\end{document}